\documentclass{aa}  

\usepackage{graphicx}

\usepackage{natbib}

\usepackage[varg]{txfonts}
\newcommand{\fmipscobeam}{$f_{\rm 24~\mu m,CObeam}$}
\newcommand{\kms}{\nobreak{km~s$^{-1}$}} 
\newcommand{\mi}{$\,\mu$m} 
 
\newcommand{\htwopah}{H$_2$/7.7$\mu$m} 
\newcommand{\htwo}{H$_2$} 
\newcommand{\lb}{$\,L_{\rm B}$}
\newcommand{\lk}{$\,L_{\rm K}$}

\newcommand{\lsun}{$\,L_{\sun}$}
\newcommand{\lsunk}{$\,L_{\sun,k}$}

\newcommand{\lmipscobeam}{$L_{\rm 24~\mu m,CObeam}$}
\newcommand{\msun}{$\,M_{\sun}$}
\newcommand{\mhtwo}{$M_{\rm H_2}$}
\newcommand{\mhtwotot}{$M_{\rm H_2, tot}$}
\newcommand{\mhtwocenter}{$M_{\rm H_2, 0}$}
\newcommand{\mhtwomapped}{$M_{\rm H_2, mapped}$}
\newcommand{\mhtwowarm}{$M_{\rm H_2, warm}$}

\newcommand{\spitzer}{{\it Spitzer Space Observatory}}
\newcommand{\ropt}{r$_{\rm 25}$}
\newcommand{\taudep}{$\tau_{\rm dep}$}

\begin{document}
   \title{CO in Hickson Compact Group galaxies with enhanced warm \htwo\ emission: Evidence for galaxy evolution?}
    \titlerunning{CO in Hickson Compact Group galaxies with enhanced warm \htwo\ emission}


   \author{U. Lisenfeld
          \inst{1,2}
          \and
          P. N. Appleton\inst{3}\  
          \and
          M. E. Cluver\inst{4,5}
          \and
          P. Guillard\inst{6,7}
          \and
          K. Alatalo\inst{3}
          \and
          P. Ogle\inst{8}
          }

   \institute{
	      Departamento de F\'isica Te\'orica y del Cosmos, Universidad de Granada, Spain\\
              \email{ute@ugr.es}
         \and
         Instituto Universitario Carlos I de F\'isica Te\'orica y Computacional, Facultad de Ciencias, 18071, Granada, Spain 
         \and
             NASA Herschel Science Center, California Institute of Technology, Pasadena, CA 91125, USA
	\and
     Department of Astronomy, University of Cape Town, Private
Bag X3, Rondebosch, 7701, South Africa
        \and  ARC Super Science Fellow, Australian Astronomical Observatory,
PO Box 915, North Ryde, NSW 1670, Australia
      \and
         Institut d'Astrophysique Spatiale, CNRS, UMR 8617, Universit\'e Paris-Sud, B\^at. 121,  91405 Orsay Cedex, France
                 \and
        Spitzer Science Center, IPAC, California Institute of Technology, Pasadena, CA 91125, USA
         \and
             NASA Extragalactic Database, IPAC, California Institute of Technology, Pasadena, CA 91125
             }

   \date{Received February 13, 2014; accepted July 9, 2014}

 
  \abstract
   {Galaxies in Hickson Compact Groups (HCGs) are believed to experience 
 morphological transformations from blue,
   star-forming galaxies to red, early-type galaxies. 
   Galaxies with a high ratio between the luminosities of the warm  \htwo\  to the 7.7\mi\ PAH emission (so-called {\it Molecular Hydrogen
Emission Galaxies},  MOHEGs) are  predominantly  in an intermediate
   phase, the {\it green valley}. Their enhanced \htwo\ emission suggests that the molecular gas is affected in the 
 transition.}
   {We study the properties of the molecular gas traced by CO  in galaxies 
    in HCGs with measured  warm \htwo\ emission  in order to 
   look for evidence of the perturbations affecting the warm \htwo\ 
   in the  kinematics, morphology and mass of the molecular gas.
      }
   {{We observed  the CO(1-0) emission  of 20 galaxies in HCGs and complemented our sample with
   11 CO(1-0) spectra  from the literature.
   Most of the galaxies have measured}  warm \htwo\ emission, 
   and 14 of them  are classified as MOHEGs. 
   We mapped some of these galaxies in order to search for extra-galactic CO emission.
   We analyzed the molecular gas mass derived from CO(1-0), \mhtwo, and its kinematics, and then compared it to the mass
   of the warm molecular gas, the stellar mass and star formation rate (SFR). }
   {Our results are the following. 
   (i)  The mass ratio between the CO-derived and the warm \htwo\ molecular gas is in the same range as found
   for  field galaxies. 
   (ii) Some of the galaxies, mostly MOHEGs, have very broad CO linewidths of up to 1000 \kms\
    in the central pointing. The line shapes are irregular and show various components.
  (iii)    In the mapped objects we found  asymmetric distributions of the cold molecular gas.
     (iv) The star formation efficiency (= SFR/\mhtwo) of galaxies in HCGs is very similar to isolated galaxies. No significant difference between 
   MOHEGs and non-MOHEGs {or  between early-type and spiral galaxies} has been found. 
   In a few objects 
   the  SFE is significantly lower, {indicating}
   the presence of molecular gas that is not actively forming stars.
      (v) The molecular gas masses, \mhtwo , and ratios \mhtwo/\lk\  are lower 
   in MOHEGs (predominantly early-types)  than in
   non-MOHEGs (predominantly spirals). This trend  
   remains when
   comparing MOHEGs and non-MOHEGs of the same morphological type.
   }
   { We found differences in the molecular gas properties of MOHEGs  that support
   the view that they have suffered (or are presently suffering)  perturbations of the molecular gas, as well as
   a decrease in the molecular gas
   content and  {associated} SFR.  
   Higher  resolution observations of the molecular gas are needed to shed light on the nature of these
   perturbations and their cause.
    }


   \maketitle
%

\section{Introduction}

Compact groups represent the highest density enhancements outside clusters.
   Galaxies within them have relatively low velocity dispersions \citep[median dispersion $\sim$ 200 \kms,][]{1992ApJ...399..353H} which
   makes compact groups ideal objects to study the influence of the environment and interaction
   on galaxy properties and morphological evolution. 
   \citet{1982ApJ...255..382H} selected a uniform sample of 100 nearby compact groups (Hickson Compact Groups,
   hereafter HCGs) consisting of four to eight members which have been extensively studied since then.
   
   HCGs show a high ratio  of early-type to spiral galaxies   \citep[the spiral fraction is about 50\%  compared to  that of field galaxies,][]{1982ApJ...255..382H} which suggests
   that morphological transformations from late-type to early-type galaxies are taking places. 
      It has been proposed \citep[e.g.][]{2000Sci...288.1617Q, 2011MNRAS.415.1783B},
   that  S0 galaxies in HCGs might be stripped spirals. 
   %
   %
   It is, however, still completely unclear what the dominant mechanism for such a morphological
   evolution could be. Different scenarios  for the transformation of spirals into S0s in groups and clusters have been proposed, 
   which include ram pressure stripping
   \citep[e.g.][] {1972ApJ...176....1G,1980ApJ...241..928F,1999MNRAS.308..947A}, 
tidal encounters 
\citep[e.g.][] {1985A&A...144..115I,2011MNRAS.415.1783B},
truncation of gas replenishment 
\citep[e.g.][] {1980ApJ...237..692L,2002ApJ...577..651B},
minor and unequal-mass
merging \citep[e.g.][] {1998ApJ...502L.133B} and
harassment combined with interaction with intergalactic gas \citep{2013ApJ...765...93C}.

Observations of the atomic hydrogen (HI) have provided further evidence for strong interactions. Many HCGs show
a  considerable amount of intergalactic HI and at the same time an  HI deficiency,
   both globally for the groups as well as individually for their members  \citep{2001A&A...377..812V}.
    The fate of this missing gas is still unclear, even though deep single-dish
   observations discovered a  low surface brightness HI component, missed by interferometric observations,
   which can reduce, but not completely eliminate, the HI deficiency  \citep{2010ApJ...710..385B}.
   Recent {\it Herschel} observations \citep{2014arXiv1402.1470B} of 28 HCGs  give further support for the presence
   of intergalactic material,
  discovering intergalactic dust in 4 groups and finding that a fraction of 15-20\% of the 250 \mi\ emission resides outside of the
   main body of late-type and lenticular galaxies.
   Observations of the CO emission of galaxies in HCGs 
showed a much less dramatic picture.
\citet{1998A&A...330...37L} found a slight enhancement
of \mhtwo/\lb\ of galaxies in HCGs compared to a sample of field and interacting galaxies.
On the contrary, \citet{1998ApJ...497...89V} found no evidence for an enhancement of 
\mhtwo\ in HCGs compared to a sample of isolated galaxies, whereas \citet{2007ggnu.conf..349V}
found tentative evidence of a depressed molecular gas content
in HI deficient galaxies in HCGs.  \citet{2012A&A...540A..96M}
solved these apparent contradictions by showing that there is an enhancement of
$\sim$ 0.2 dex in \mhtwo/\lk\ 
 compared to isolated galaxies. They found  tentative evidence that this enhancement was mainly in groups
in an early evolutionary phase, whereas in galaxies in more evolved groups the molecular
gas mass was the same as in isolated galaxies. They speculated on an evolutionary sequence
in which the molecular gas mass was enhanced in early stages due to tidal interaction and
later \mhtwo\ decreased either due to gas depletion or environmental effects.

Further evidence for evolution in HCGs came from \citet{2007AJ....134.1522J} who found  {an infrared} ``gap", i.e. a seemingly underpopulated region,
in the \spitzer\ IRAC color diagram in the region separating actively starforming,
blue galaxies from red galaxies dominated by an evolved stellar population.
This gap is not found in comparison samples (e.g. the  {\it Spitzer Infrared Nearby Galaxies Survey}  (SINGS) sample).
The low density of galaxies in this ``gap"
suggests that the evolutionary phase corresponding to the
gap is a very short transition. The existence of an underpopulated region was confirmed in larger samples
\citep{2010AJ....140.1254W, 2012AJ....143...69W} but reduced in extension to a smaller area 
in the mid-IR colors \citep{2012AJ....143...69W}. 

\citet{2013ApJ...765...93C}  studied a sample of Hickson Compact Groups (HCGs) with
intermediate HI deficiencies and identified a number of galaxies with
enhanced warm \htwo\ emission, well above the level expected from star formation (SF) and
indicative of shock-excitation, classifying them as co-called {\it Molecular Hydrogen
Emission Galaxies} (MOHEGs) \citep{2010ApJ...724.1193O}.
Interestingly, most of these galaxies have IRAC colors in the range of the previously found 
gap \citep{2007AJ....134.1522J, 2010AJ....140.1254W}.
Furthermore, based on their extinction-corrected optical colors, these
galaxies fall predominantly in the ``green valley" between blue, starforming galaxies
and red-sequence objects. 
%
\citet{2013ApJ...765...93C}  conclude that the increased warm \htwo\ emission is most likely due to shock-excitation.
They propose that these galaxies might be in a special phase of their evolution in which
interaction with material from the IGM  -- most likely previously stripped gas -- 
causes viscous stripping and shock-excitation of the molecular gas.
This event could be important for their further evolution and might even trigger their
evolution from the blue cloud to the red sequence.

Enhanced \htwo\ emission had previously been found in one of the most spectacular
HCGs, Stephan's Quintet (HCG~92) \citep{2006ApJ...639L..51A,2010ApJ...710..248C} where it originated in
an intergalactic region dominated by  gas interaction producing large scale shocks.
\citet{2009A&A...502..515G} showed that \htwo\ can form in the dense, post-shock gas and that its luminosity can be powered
by the dissipation of  turbulent energy after the energy injected on large scales by the galaxy collision  cascades
on smaller scales.
 Not only warm \htwo\  gas,  emitting in the mid-infrared, but also molecular gas traced by the
CO(1-0) line was found abundantly  in this region, with velocities showing that
it is present both in the pre- and in the post-shock gas \citep{2012ApJ...747...95G}. This component of the
molecular gas carries a significant fraction of the total kinetic energy.

{All these results suggest that the molecular gas plays an important role in processes that are relevant to galaxy
evolution. The warm molecular gas seems to be a tracer for galaxies in a transitionary phase. The observations of Stephan's Quintet show
that the warm and the cold (CO traced)  molecular gas\footnote{
Throughout this paper we call the molecular gas derived from the CO(1-0) emission the ``cold" molecular gas,  in contrast to
the warm molecular gas derived from the \htwo\ emission,  and we denote it as \mhtwo\, \mhtwocenter, or \mhtwotot, whereas we
call the warm molecular gas explicitly \mhtwowarm . {We note, however, that CO(1-0) can  trace relatively
warm \htwo\ gas and that, without additional measurements, we cannot determine the physical temperture of the CO(1-0)
emitting gas.    Therefore, the above  naming has to be understood as a pure convention without
meaning  real physical tempertures.}}
 are closely related. So far no study has focussed on the 
molecular gas content 
 in galaxies in HCGs that have been selected based on their warm \htwo\ emission.
In order to address this issue, in the present paper, we present and analyze CO observations for a sample of galaxies in HCGs with existing warm \htwo\ data.
We chose a subsample of the galaxies
studied by  \citet{2013ApJ...765...93C}  which includes both galaxies with
an enhanced warm \htwo\ emission (MOHEGs) and galaxies in which the \htwo\ emission can be explained by SF.
The goal of our work  is to characterize the 
properties of the  CO-traced ("cold")  molecular gas
in a sample with information on the luminosity  and mass
of the warm molecular gas in order to study the relation between both emissions and look for evidence of
differences between MOHEGs and non-MOHEGs that could help to better understand the mechanisms responsible for 
galaxy evolution in HCGs. 
}



\section{The sample}

We selected our sample from the galaxies studied by \citet{2013ApJ...765...93C} who 
carried out a {\it Spitzer} mid-infrared spectroscopic study of 74 galaxies located in 23
HCGs. Their sample selected groups with intermediate HI deficiencies because
those are expected to be most likely in an active phase of transformation. 
{Tab.~\ref{tab:tab_group_properties}  lists the groups included in our study together with  some general group properties.}
Furthermore, only groups with visible signs of tidal interaction in two or more group
members were included in their sample. Their study allowed to measure the ratio of the
warm \htwo\ luminosity (summed over the 0-0S(0) to 0-0S(3) lines) to the
7.7~\mi\ PAH luminosity, hereafter referred to as \htwopah.

\begin{table*}[h!]
\caption{Basic data of the groups}
\label{tab:tab_group_properties}
\begin{tabular}{lccccccccc}
\noalign{\smallskip} \hline \noalign{\medskip}
Group & $z$\tablefootmark{a} &$\Theta$\tablefootmark{b} & {log($\sigma_{\rm v})$}\tablefootmark{c} &  {log($R$) }\tablefootmark{d} & log($L_{\rm B})$\tablefootmark{e} & log($M_{\rm HI})$\tablefootmark{f} &
HI deficiency\tablefootmark{f} & log($M_{\rm star})$\tablefootmark{g} & log($M_{\rm dyn})$\tablefootmark{h}  \\
   & & [arcmin] & [\kms] & [kpc] & [W] &[\msun]  &  & [\msun]  & [\msun]  \\
\noalign{\smallskip} \hline \noalign{\medskip}
   HCG   6   &  0.0379   &   1.60   &    2.40   &    1.54   &   37.58   &    9.71   &    0.33   &   11.30   &   12.63  \\
HCG  15   &  0.0228   &   7.70   &    2.63   &    2.03   &   37.63   &    9.59   &    0.46   &   11.04   &   13.19  \\
HCG  25   &  0.0212   &   6.40   &    1.79   &    1.82   &   37.38   &$>$   10.15   &$<$    0.02   &   10.96   &   11.11  \\
HCG  31   &  0.0135   &   0.90   &    1.75   &    1.04   &   37.38   &   10.36   &    0.18   &      --   &      --  \\
HCG  40   &  0.0223   &   1.70   &    2.17   &    1.32   &   37.66   &    9.84   &    0.29   &   11.56   &   12.27  \\
HCG  44   &  0.0046   &  16.40   &    2.13   &    1.72   &   37.12   &    8.98   &    0.76   &   10.75   &   12.68  \\
HCG  47   &  0.0317   &   2.30   &    1.63   &    1.70   &   37.44   &    9.95   &    0.28   &   11.24   &   11.60  \\
HCG  55   &  0.0526   &   0.90   &    2.33   &    1.42   &   37.71   &      --   &      --   &   11.39   &   12.49  \\
HCG  56   &  0.0270   &   2.10   &    2.23   &    1.47   &   37.47   &    9.38   &    0.73   &   10.92   &   12.78  \\
HCG  57   &  0.0304   &   5.50   &    2.43   &    2.00   &   37.98   &    9.73   &    0.86   &   11.70   &   12.55  \\
HCG  67   &  0.0245   &   3.30   &    2.32   &    1.83   &   37.92   &    9.69   &    0.62   &   11.43   &   12.13  \\
HCG  68   &  0.0080   &   9.20   &    2.19   &    1.66   &   37.54   &$>$    9.77   &$<$    0.12   &   11.29   &   12.80  \\
HCG  79   &  0.0145   &   1.30   &    2.14   &    0.97   &   37.17   &    9.64   &    0.08   &   10.80   &   11.68  \\
HCG  82   &  0.0362   &   3.10   &    2.79   &    1.99   &   37.89   &$<$    9.71   &$>$    0.76   &   11.54   &   13.01  \\
HCG  91   &  0.0238   &   5.20   &    2.26   &    1.86   &   37.89   &   10.38   &    0.20   &   11.13   &   12.64  \\
HCG  95   &  0.0396   &   1.50   &    2.49   &    1.62   &   37.80   &$>$   10.12   &$<$    0.21   &   11.22   &   12.67  \\
HCG  96   &  0.0292   &   2.30   &    2.12   &    1.62   &   37.76   &$>$   10.22   &$<$    0.17   &      --   &      --  \\
HCG 100   &  0.0178   &   3.60   &    1.95   &    1.72   &   37.24   &$>$    9.97   &$<$    0.28   &   10.79   &   11.13  \\

\noalign{\smallskip} \hline \noalign{\medskip}
\end{tabular}
\tablefoot{
\tablefoottext{a}{Redshift from NED.}
\tablefoottext{b}{Angular size \citep{1992ApJ...399..353H}.}
\tablefoottext{c}{Radial velocity dispersion \citep{1992ApJ...399..353H}.}
\tablefoottext{d}{Median projected distance from \citet{1992ApJ...399..353H}, adapted to our distance.}
\tablefoottext{e}{Blue luminosity, from \citet{1992ApJ...399..353H}, adapted to our distance.}
\tablefoottext{f}{Total HI mass, adapted to our distance, and HI deficiency (defined as log(M(HI))$_{predicted}$-log(M(HI))$_{observed}$, from \citet{2010ApJ...710..385B},
if available, or otherwise from \citet{2001A&A...377..812V}.}
\tablefoottext{g}{Total stellar mass, from \citet{2011A&A...533A.142B}, adapted to our distance.}
\tablefoottext{h}{Dynamical mass, from \citet{2011A&A...533A.142B}, adapted to our distance.}
}

\end{table*}

We observed 20 of these galaxies with the IRAM 30m telescope, selecting  preferentially 
 those galaxies with  
a high \htwopah\ ratio. This ratio can be used as an indicator of whether the warm \htwo\
emission is due to UV excitation  (\htwopah $\sim 0.01$ or below) or due to dissipation of turbulent
kinetic energy  (\htwopah $\gtrsim 0.01$)  \citep{2012ApJ...747...95G}.
We adopt the definition of  Ogle et al. (2010) that a    {\it Molecular Hydrogen Emission Galaxy}
(MOHEG)  has \htwopah $> 0.04$, which is also used
in \citet{2013ApJ...765...93C}.  
For 11 additional galaxies 
CO data from the literature is available, mostly  from the IRAM 30m telescope. 
Thus, our complete sample includes 31 galaxies in 18 groups and spans a large
range of HI depletion ranging from 
$0  \le  \log(M({\rm HI})_{\rm pred}) - \log({M({\rm HI})_{\rm obs}}) \le 0.97$.
In Tab.~\ref{tab:tab_general_data_paper} we list the galaxies in the sample together with some general properties.

\begin{table}[h!]
\caption{Basic data of the sample galaxies}
\label{tab:tab_general_data_paper}
\begin{tabular}{lccccc}
\noalign{\smallskip} \hline \noalign{\medskip}
Galaxy & $D$\tablefootmark{a} & T(RC3)\tablefootmark{b} &  $D_{\rm 25}$\tablefootmark{c}  & $i$\tablefootmark{d} & log(\lk)\tablefootmark{e}  \\
       & [Mpc] &   &   [\arcmin] & [$^\circ$] & \lsunk \\
\noalign{\smallskip} \hline \noalign{\medskip}
    HCG  6b   &  155.65   & 1   &    0.87   &  51   &   11.09  \\
HCG  6c   &  155.65   &-3   &    0.91   &  90   &       -  \\
HCG 15a   &   93.64   &-2   &    1.29   &  74   &   11.17  \\
HCG 15d   &   93.64   &-3   &    0.91   &  44   &   10.72  \\
HCG 25b   &   87.07   & 1   &    0.95   &  56   &   11.06  \\
HCG 25f   &   87.07   &-2   &    0.43   &  56   &   10.11  \\
HCG31ac   &   55.44   & 9   &    1.00   &  80   &   10.08  \\
HCG 40b   &   91.58   &-3   &    0.89   &  53   &   10.93  \\
HCG 40c   &   91.58   & 2   &    1.55   &  90   &   11.09  \\
HCG 40d   &   91.58   & 0   &    0.91   &  81   &   10.92  \\
HCG 44a   &   18.89   & 1   &    3.63   &  87   &   10.84  \\
HCG 44d   &   18.89   & 5   &    2.24   &  74   &    9.35  \\
HCG 47a   &  130.19   & 2   &    1.02   &  66   &   11.19  \\
HCG 47d   &  130.19   & 7   &    0.50   &  32   &   10.51  \\
HCG 55c   &  216.02   & 1   &    0.62   &  54   &   11.24  \\
HCG 56b   &  110.88   & 0   &    0.76   &  62   &   11.07  \\
HCG 56c   &  110.88   &-2   &    0.81   &  48   &   10.78  \\
HCG 57a   &  124.85   & 2   &    1.62   &  90   &   11.54  \\
HCG 57d   &  124.85   & 3   &    0.55   &  36   &   10.85  \\
HCG 67b   &  100.62   & 3   &    1.07   &  88   &   11.03  \\
HCG 68a   &   32.85   &-2   &    2.40   &  80   &   10.98  \\
HCG 68b   &   32.85   &-2   &    3.02   &  90   &       -  \\
HCG 68c   &   32.85   & 4   &    2.69   &  54   &   10.92  \\
HCG 79a   &   59.55   & 0   &    1.78   &  16   &   10.57  \\
HCG 79b   &   59.55   &-2   &    2.29   &  67   &   10.55  \\
HCG 82b   &  148.67   &-2   &    0.89   &  55   &   11.27  \\
HCG 82c   &  148.67   & 8   &    0.72   &  90   &   10.97  \\
HCG 91a   &   97.74   & 5   &    1.95   &  47   &   11.58  \\
HCG 95c   &  162.63   & 9   &    0.74   &  78   &   10.92  \\
HCG 96a   &  119.92   & 4   &    1.12   &   0   &       -  \\
HCG100a   &   73.10   & 0   &    0.87   &  60   &   11.13  \\

\noalign{\smallskip} \hline \noalign{\medskip}
\end{tabular}
\tablefoot{
\tablefoottext{a}{Distance, based on  redshifts from  the NASA/IPAC Extragalactic Database (NED) listed in \citet{2013ApJ...765...93C} and adopting a Hubble constant
of 73 \kms Mpc$^{-1}$.}
\tablefoottext{b}{Morphological type taken from the LEDA  database, following the RC3 classification \citep{1991trcb.book.....D}.}
\tablefoottext{c}{Optical major diameter at the 25 mag \rm{arcsec$^{- 2}$} isophot taken from LEDA.}
\tablefoottext{d}{Inclination taken from LEDA.}
\tablefoottext{e}{Decimal logarithm of the luminosity in the K-band in units of the solar luminosity in the 
$K_{\rm S}$-band ($L_{K,\odot} = 5.0735 \times 10^{32}$ erg s$^{-1}$).
We calculated the $K_{\rm S}$ luminosity, ${\it L_{\rm K}}$, from the total (extrapolated) $K_{\rm S}$ flux, $f_{\rm K}$, as ${\it L_{\rm K}}$ = $\nu f_{\rm K}(\nu)$ (where $\nu$ is the frequency of the K-band, 1.38$\times$10$^{14}$ Hz).
The fluxes in  the $K_{\rm S}$ (2.17 $\mu$m) band were taken from \citet{2011A&A...533A.142B}
who obtained them partly from their own observations and partly from the 2MASS Extended Source Catalog \citep{2000AJ....119.2498J}. }
}
\end{table}


\section{The data }

\subsection{CO observations and reduction }

We observed the CO(1--0) and CO(2--1)  lines at 115 and 230\,GHz
between June and September 2011 with the IRAM 30m
telescope on Pico Veleta.  
We used the EMIR dual-polarization receivers, with the
autocorrelator WILMA as  backend. This setup yields a resolution
of 2 MHz and a bandwidth of 3.7 GHz. All observations
were performed in wobbler-switching mode, with a throw in azimuth
between 150 and 240~\arcsec , chosen 
carefully to avoid any other
 galaxies in the group  at the off-position. 
  The telescope pointing was checked on a nearby
quasar about every 90 minutes. The average pointing offset was 
 3-4~\arcsec . 
 
All observations were carried out in good weather conditions,
the mean system temperatures being 200 K for the CO(1-0) and 290 K for the
CO(2-1) transition on the $T_{\rm A}^*$ scale.
All CO spectra and intensities are
presented on the main beam temperature scale ($T_{\rm mb}$) which is
defined as $T_{\rm mb} = (F_{\rm eff}/B_{\rm eff})\times T_{\rm A}^*$.
 The main beam efficiencies
are 0.77 (115 GHz), and 0.58 (230 GHz), with
half-power beam widths of about 22\arcsec\ (115 GHz),
and 11\arcsec\ (230 GHz).

Most of the galaxies were only observed at their central position,
but some were mapped (HCG~25b, HCG~40c, HCG~91a and  HCG~57a).
{We selected a galaxy for mapping if its central CO emission was bright and 
its size was considerably larger than the IRAM CO(1-0) beam. A further criterion was
the presence of strong warm \htwo\ emission either in the centre or at an off-centre position.}
In Fig.~\ref{fig:map_co_25b}-\ref{fig:map_co_91a} the observed positions are overlaid over
IRAC three-color images of the galaxies.
 In two of these objects, one of  the mapped
positions coincides very accurately with the central position of a neighbouring
galaxy. This is the case for the position (-3, -33) in HGC~25b which coincides
exactly with the central position in HCG~25f and the position (15, 15) in HCG~57a
which is 1\arcsec\ away from the central position in  HGC~57d.
In the following we identify these positions with the corresponding
neighboring galaxies HCG~25f and HCG~57a. 
In the case of HCG~31ac, the CO spectrum taken from the 
literature \citep{1998A&A...330...37L} 
corresponds
to the position of HCG~31c. At the position of HCG~31a only
an upper limit was observed \citep{2012A&A...540A..96M}. Since the objects
HCG~31a and HCG~31c are in the process of merging and
very close to each other (distance of the centres 16\arcsec ),
we follow the convention suggested in Gallagher et al.  (2010) to 
consider both objects as a  merger blend and call it HCG~31ac.

The data were reduced in the standard way using the CLASS software
in the GILDAS package\footnote{http://www.iram.fr/IRAMFR/GILDAS}.
We selected the observations with
a good quality (taken
during satisfactory weather conditions and showing a flat baseline),
averaged the spectra over the individual
positions and subtracted a baseline which was a 
constant continuum level for the CO(1--0) spectra and a linear
baseline for the CO(2--1) data. 

In order to check the relative calibration, we observed the central position
of HCG~68c, HCG~91a and  HCG~96a,  which exhibit strong CO(1-0) and CO(2-1)
lines, on different days. The agreement between the
velocity integrated line intensities of CO(1-0), $I_{\rm CO(1-0)}$,
is better than 10\% most of the time, with the exception of one day  where the disagreement
is 30\%.
The reason for this poorer calibration agreement is not entirely clear, possibly it was
caused by a  slight pointing offset in one of the observations.
The agreement for the CO(2-1) line intensities, $I_{\rm CO(2-1)}$, is between 15 and 30\%.
Based on these measurements we adopt a (probably conservative) calibration error
of 20\% for $I_{\rm CO(1-0)}$ and 30\% for $I_{\rm CO(2-1)}$.

\subsection{{\it Spitzer} data}

Our  analysis made use of \spitzer\  IRAC and MIPS data.
We took the total  IRAC fluxes of the galaxies from \citet{2013ApJ...765...93C} 
(their Tab.~4), which were obtained with aperture photometry covering
the entire galaxies,  and applied  the  k-correction derived by these
authors. We furthermore took the ratio
between \htwo\ luminosity and the luminosity in the 7.7 \mi\ band,
\htwopah, from these authors  (their  Tab. 2).

We took the total MIPS 24 \mi\  fluxes from \citet[][their Tab. 1]{2011A&A...533A.142B}.
A few objects  are missing in their sample (HCG 31ac and 
HCG 96a) or no 24~\mi\ data is given (HCG~25f).  We did not attempt to derive the total
fluxes for these objects because either the emission is very
weak so that no reliable total flux can be derived, 
or the galaxies are so close to a neighboring object that no separate total emission can be 
derived. 

In addition, we derived the 24~\mi\ flux within the
IRAM CO(1-0) beam directly from the MIPS images 
in order to locally compare this SF tracer to  the molecular
gas mass.
%
%
To achieve this, we multiplied the  MIPS 24 \mi\ image with the IRAM beam pattern
(approximated as a normalized Gaussian beam) placed
at the position where the CO beam was pointed, and measured the total 
24~\mi\ emission of this map. We used a FWHM of 21~\arcsec\  in order
to achieve, after convolution with the 24~\mi\  map (spatial resolution 5.9\arcsec),  a total
FWHM of 22\arcsec which corresponds to the FWHM of the IRAM beam at the
frequency of our observations.

Some of our objects habor AGNs 
\citep[HCG~47a, 56B, 57a, 91A, 95C and 96A, from Tab. 3 in][]{2013ApJ...765...93C}  
and their 24\mi\ flux might be  
affected by this.
In order to avoid a contribution of the AGN to the 24~\mi\  emission which
would produce an overestimate of the SFR (see Sect.~\ref{sec:sfe}) 
we  inspected the 24~\mi\ images of these galaxies  and searched for strong, 
centrally concentrated emission indicative of an  AGN.
HCG~56b  and HCG~96a  show intense, point-like 24 \mi\ emission
in the centre. The central emission in HCG~56b is so bright that the value of $f_{\rm 24 \mu m, CO beam}$ for 
the neighboring galaxy, HCG~56c, is strongly affected.
We therefore excluded all three objects from 
 the  analysis  of the SFR and star formation efficiency (SFE).
 In the other objects no indication of a contribution of the
 AGN to the 24~\mi\  was noticeable.
 
The fluxes are listed in Tab.~\ref{tab:spitzer}. 
%
%
%
For some objects the flux within the CO beam is larger than the total 
flux. This is because our value for
$f_{\rm 24 \mu m, CO beam}$ can  include  the emission of
nearby objects if they are  strong and close-by, whereas \citet{2011A&A...533A.142B}
have attempted to separate the flux for the two.

\begin{table}
\caption{{\it Spitzer} data}
\label{tab:spitzer}
\resizebox{\linewidth}{!}{%
\begin{tabular}{lllll}
\noalign{\smallskip} \hline \noalign{\medskip}
Galaxy  &  $f_{5.8}/f_{3.6}$\tablefootmark{a} & \htwopah\tablefootmark{b} & $f_{24\mu m,tot}$\tablefootmark{c} & $f_{24\mu m,CObeam}$\tablefootmark{d} \\
    &      & & [mJy] & [mJy] \\
\noalign{\smallskip} \hline \noalign{\medskip}
    HCG  6b   &   -0.25   &   0.198   &    3.80 $\pm$    0.19   &    2.53 $\pm$    0.08  \\
HCG  6c   &   -0.38   &       -   &    0.80 $\pm$    0.04   &    1.32 $\pm$    0.08  \\
HCG 15a   &   -0.29   &   0.044   &    5.60 $\pm$    0.28   &    4.20 $\pm$    0.09  \\
HCG 15d   &   -0.30   &   0.112   &    2.70 $\pm$    0.14   &    3.72 $\pm$    0.10  \\
HCG 25b   &   -0.24   &   0.098   &    4.70 $\pm$    0.24   &    5.69 $\pm$    0.07  \\
HCG 25f   &   -0.36   &       -   &       -   &    0.70 $\pm$    0.07  \\
HCG31ac   &    0.32   &   0.002   &       -   &  307.68 $\pm$    0.14  \\
HCG 40b   &   -0.34   &   0.083   &    3.00 $\pm$    0.15   &    9.07 $\pm$    0.07  \\
HCG 40c   &    0.09   &   0.010   &   73.40 $\pm$    3.67   &   35.08 $\pm$    0.07  \\
HCG 40d   &    0.06   &   0.013   &   88.10 $\pm$    4.41   &   67.71 $\pm$    0.07  \\
HCG 44a   &   -0.20   &   0.042$^{(1)}$ &  246.00 $\pm$   12.30   &   73.85 $\pm$    0.06  \\
HCG 44d   &    0.19   &   0.021   &   79.90 $\pm$    4.00   &   19.80 $\pm$    0.06  \\
HCG 47a   &   -0.01   &   0.035   &   87.00 $\pm$    4.35   &   38.00 $\pm$    0.13  \\
HCG 47d   &   -0.19   &       -   &    4.50 $\pm$    0.23   &    4.70 $\pm$    0.13  \\
HCG 55c   &   -0.01   &   0.018   &    7.70 $\pm$    0.39   &    6.13 $\pm$    0.07  \\
HCG 56b   &    0.16   &   0.053   &  185.00 $\pm$    9.25   &  134.56 $\pm$    0.07  \\
HCG 56c   &   -0.32   &   0.037   &    0.90 $\pm$    0.05   &   10.39 $\pm$    0.07  \\
HCG 57a   &   -0.27   &   0.174   &   13.80 $\pm$    0.69   &    9.63 $\pm$    0.05  \\
HCG 57d   &    0.03   &       -   &   24.00 $\pm$    1.20   &   15.50 $\pm$    0.05  \\
HCG 67b   &    0.02   &   0.006   &   84.80 $\pm$    4.24   &   32.65 $\pm$    0.07  \\
HCG 68a   &   -0.35   &   0.741   &   21.30 $\pm$    1.07   &   14.63 $\pm$    0.05  \\
HCG 68b   &   -0.35   &   0.073   &    9.80 $\pm$    0.49   &    7.79 $\pm$    0.05  \\
HCG 68c   &    0.01   &   0.044   &  403.60 $\pm$   20.18   &  122.81 $\pm$    0.06  \\
HCG 79a   &   -0.06   &   0.011   &   23.30 $\pm$    1.17   &   17.25 $\pm$    0.05  \\
HCG 79b   &   -0.18   &   0.010   &   33.70 $\pm$    1.69   &   27.95 $\pm$    0.05  \\
HCG 82b   &   -0.31   &   0.045   &    5.80 $\pm$    0.29   &    3.76 $\pm$    0.04  \\
HCG 82c   &    0.41   &   0.014   &   90.60 $\pm$    4.53   &   63.30 $\pm$    0.04  \\
HCG 91a   &    0.09   &   0.030   &  351.00 $\pm$   17.55   &  184.46 $\pm$    0.06  \\
HCG 95a   &   -0.08   &   0.071   &   19.80 $\pm$    0.99   &   15.17 $\pm$    0.07  \\
HCG 96c   &    0.37   &   0.007   &       -   &  473.06 $\pm$    1.40  \\
HCG100a   &   -0.05   &   0.013   &  153.00 $\pm$    7.65   &  102.47 $\pm$    0.08  \\

\noalign{\smallskip} \hline \noalign{\medskip}
\end{tabular}
}%
\tablefoot{
\tablefoottext{1}{Value taken from \citet{2007ApJ...669..959R}}
\tablefoottext{a}{IRAC flux ratio.}
\tablefoottext{b}{Ratio between luminosity of the total (sum of S(0)-S(3) lines) warm H$_2$ emission.
and the 7.7~\mi\ PAH feature. If \htwopah $>$ 0.04 the galaxy is classified as a MOHEG (Ogle et al. 2010).}
\tablefoottext{c}{Total 24~\mi\  flux of the galaxy and its error, taken  from \citet{2011A&A...533A.142B}.}
\tablefoottext{d}{24~\mi\ flux within
the IRAM CO(1-0) beam (see text) and its photometric error, derived from the
rms of the background. In addition, a calibration error of 10\% 
error \citep{2007PASP..119..994E} has to be taken into account.}
}
\end{table}

\subsection{Properties of the warm molecular gas from {\it Spitzer} observations.}
\label{sect:prop_warm_gas}

The properties of the warm molecular gas were derived from the
strength and extent of the detected pure-rotational 0-0S(0) to 0-0S(5)  lines of molecular
hydrogen from the observations of \citet{2013ApJ...765...93C}  where the full
details of the observations are given. 
Tab.~\ref{tab:tab_mh2_warm} lists the derived properties of the warm H$_2$ gas
for each galaxy. 
In order to
measure the masses and temperature of the warm molecular hydrogen we
constructed excitation diagrams 
\citep[see for example][]{2002A&A...389..374R}. The
excitation diagrams plot the column density of H$_2$, N,  in the
upper level of each transition, normalized by its statistical weight, g,
versus the upper level energy 
for each rotational H$_2$ line
observed by {\it Spitzer}. 
The diagrams can be used to estimate the total column density
and temperature of the warm H$_2$ within the slit for each galaxy
observed. Gas with a constant temperature would appear as a linear
distribution of points in the excitation diagram after correcting for
an assumed ortho-to-para ratio. 

\begin{figure}[h!]
\centerline{
\includegraphics[width=13.cm]{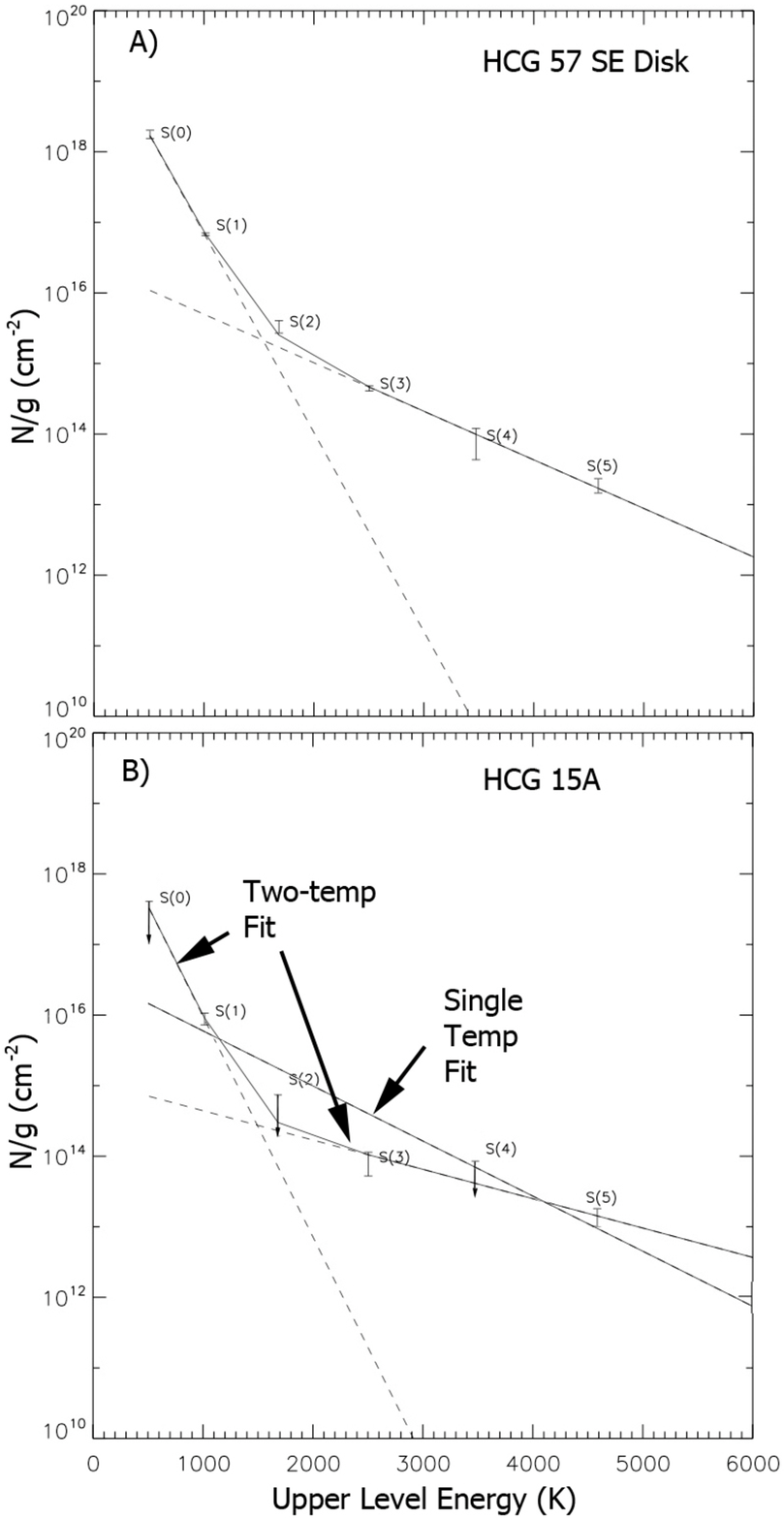}
}
\caption{Examples of the temperature fit to the excitation diagram showing
the column density of the upper level of each transition normalized by its statistical weight versus
the upper level energy (see text for further
discussion of these figure.) {\it Top:}  Case where
 H$_2$ is detected  at all  levels S(0) - S(5).
{\it Bottom:} Case where several upper limits are present.
}
\label{fig:excitation_diagram}
\end{figure}

Fig.~\ref{fig:excitation_diagram} (A)
 shows an example (from HCG~57a) 
where we have detected all 6 rotational lines and have the best constraints 
on the \htwo\ temperatures and column densities. 
A problem stems from the sparse sampling of the excitation
diagram in some slit positions. Fig.~\ref{fig:excitation_diagram} (B) (HCG~15a)  shows an examples where 
we did not detect some of the transitions and only have upper limits.
In this case, if only the detected points are used,
the slope of the fit is shallow leading to a higher temperature for the gas,
and a low mass. However, if we assume the 0-0S(0) line (at 28$\mu$m) is detected
we can perform a two-temperature fit and determine an effective upper limit for
the H$_2$ mass, since pushing the detection down to lower values would increase
the temperature of the low-temperature component, decreasing its mass. In Tab.~\ref{tab:tab_mh2_warm},
we have indicated cases where we have used the upper limit to the detection of 0-0S(0)
as an upper-limit to the mass using this technique. 
%

We have assumed a thermal equilibrium value (LTE) for the
ortho-to-para ratio of the H$_2$. For temperatures above 300K this
ratio is 3, but varies to lower values of ortho-to-para ratio in LTE for lower
temperatures \citep[see][]{2000A&A...356.1010W}. Thus we interactively fit
the excitation diagram points, allowing for lower values of ortho-to-para ratio as we home-in on a solution. 
In some cases we do not have enough points on the excitation
diagram to detect an offset between the ortho- and para-transitions that
would result from non-LTE conditions. In those cases we assume LTE
conditions, and accept an unknown uncertainty in the
derived H$_2$ masses. This may be justified, since in those cases where we have
enough points to notice an offset between the odd and even transitions, we 
do not see any systematic deviations from LTE.

Warm H$_2$ masses were estimated for each galaxy by multiplying the derived column densities
with the linear extent of the emission.
{Most objects were covered by a combination of perpendicular slits, an 10.2\arcsec\ wide
Long-Low (LL) slit and a 3.6\arcsec\ wide Short-Low (SL) slit.  We assumed for the extent of the
\htwo\ emission the linear extent in the LL slit times its slit width of 10.2\arcsec. 
In those cases where also SL measurements were available, we could directly measure the
 perpendicular extension and applied a correction factor  $f_1 = $ SL extension (in arcsec)/10.2\arcsec\
 if the extension in the SL slit was found to be different from the 10.2\arcsec\ width of the perpendicular LL slit.
In order to compare our warm H$_2$ results  
to the cold H$_2$ masses derived from the IRAM
telescope,  we applied a second correction factor, a beam correction factor, $f_2$.
This factor depends on the assumed distribution of warm H$_2$,
varying from the case where all the warm H$_2$ lies within the 22\arcsec\ $\times$ 22\arcsec\
 IRAM beam (no correction), to cases where the H$_2$ is
extended in one or (rarely) both dimensions relative to the IRAM
beam. Approximate H$_2$ dimensions were obtained from the extent of the H$_2$
in both the LL and SL slits \citep[tabulated by][]{2013ApJ...765...93C}.
The total correction factor was then  $f_{\it corr} = f_1\times f_2$. It 
varies from unity in most cases, to the largest correction of
0.36 for HCG~44a, where the H$_2$ emission is highly elongated
along the direction of the LL
slit compared with the 22\arcsec\  IRAM beam.
}

\begin{table*}
\caption{Warm molecular gas mass}
\label{tab:tab_mh2_warm}
\resizebox{\linewidth}{!}{%
\begin{tabular}{lccccccccccc}
\noalign{\smallskip} \hline \noalign{\medskip}
Galaxy  &  Lines\tablefootmark{a} & $T_{\rm 1}$\tablefootmark{b}  & OPR$_{\rm 1}$\tablefootmark{c}  &  log($M_{\rm 1}$)\tablefootmark{d} & $T_{\rm 2}$\tablefootmark{b}   & OPR$_{\rm 2}$\tablefootmark{c}  &  log($M_{\rm 2}$)\tablefootmark{d}  & log($M_{\rm 1+2}$) & \htwo\ Dimensions\tablefootmark{e} & f$_{\rm corr}$\tablefootmark{f} & log(\mhtwowarm)\tablefootmark{g}  \\
   &      &  [K] & & [\msun]  & [K] & & [\msun]  & [\msun]  &  [\arcsec$^2$] & &  [\msun]  \\
\noalign{\smallskip} \hline \noalign{\medskip}
    HGC 6b	        & 0123(45) 		        &163     & 2.6   & 7.91       & 792     &3      &5.3         &7.91         &15.3 x 5.4     &1.00  &7.91      \\
HGC 15a	        & (0)135	             &139     & 2.4   & 7.69       & 1044    & 3     & 4.46       & $<$ 7.69   &30.6 x 14.4	   &1.01  & $<$ 7.70 \\
HGC 15d	          & 0123(4)5			&150     &2.5   & 7.78	      &630	&3	&5.30        &7.78	   &25.5 x 7.2     &0.86  &7.72	   \\  
HGC 25b   	&	(0)1	   	        &173	 &2.7	&7.39	      &-	&-	&-	     &$<$ 7.39	   &20.4 x [10.2]    &1.00  &$<$ 7.39  \\  
HGC 31ac	&	(0)1			&268	 &3.0	&5.02	      &-        &-	&-	     &$<$ 5.02     &10.2 x [10.2]	   &1.00  &$<$ 5.02  \\	  
HGC 40b		&(0)1(2)34(5)			&142     &2.4	&7.71	      &911	&3	&4.11	     &$<$ 7.71     &20.4 x 3.6	   &1.00  &$<$ 7.71 \\	  
HGC 40c 	&01				&157     &2.6	&8.60	      &-	&-	&-	     &8.60	   &40.8 x [10.2]	   &0.54  &8.33     \\	  
HGC 40d		&0123(4)5			&144 	 &2.4	&8.26	      &482	&3	&6.18        &8.26	   &20.4 x 9.0	   &1.00  &8.26     \\	  
HGC 44a 	&01			        &158     &2.6	&7.53         &-        &-	&-	     &7.53	   &61.2 x [10.2]    &0.36  &7.08     \\	  
HGC 44d		&01(234)5			&141	 &2.4	&7.18	      &577	&3	&5.40	     &7.18	   &51.0 x [10.2]    &0.43  &6.81     \\	 
HGC 47a   	&  (0)12(3)4(5)			&145     &2.4	&8.60         &935	&3	&5.85	     &$<$ 8.60	   &35.7 x [10.2]	   &0.62  &$<$ 8.39 \\	 
HGC 47d		&(0)12(3)4(5)		&100     &1.6	&8.81	      &799	&3	&6.40	     &$<$ 8.81     &22 x 22	   &1.00  &$<$ 8.81 \\	 
HGC 55c	&	01234(5)			&158     &2.6	&7.99	      &1031	&3	&5.30	     &7.99	   &10.2 x 7.2	   &1.00  &7.99      \\	 
HGC 56b	&	(0)12(34)5			&-	 &-	&-	      &540	&3	&3.22	     &$<$ 6.32	   &35.7 x [10.2]	   &0.62  &$<$ 6.11 \\	 
HGC 56c		&(0)1(2)3(4)5			&154	 &2.5	&7.54	      &779	&3	&5.15	     &$<$ 7.54	   &15.3 x 10.8    &1.00  &$<$ 7.54 \\	 
HGC 56e		&(0)1(2)3(4)5			&113     &1.9	&8.44	      &1365	&3	&3.96	     &$<$ 8.44	   &22 x 22	   &1.00  &$<$ 8.44 \\	 
HGC 57a		&01	               &206	 &2.9	&8.25	      &-	&-      &-	     &8.25	   &40.8 x [10.2]    &0.54  &7.98 \\	 
HGC 67b		&01234(5)			&141	 &2.4	&6.81	      &1169	&3	&3.58	     &6.81	   &25.5 x [10.2]	   &0.86  &6.75 \\	 
HGC 68a		&(0)123(45)			&100	 &1.6	&7.18	      &429	&3	&5.51	     &$<$ 7.19	   &25.5 x 10.8	   &0.86  &$<$ 7.12 \\ 
HGC 68b		&(0)1(2)3(45)			&140	 &2.4	&6.81	      &1169	&3	&3.60  	     &$<$ 6.81	   &20.4 x 10.8    &1.00  &$<$ 6.81 \\	 
HGC 68c		&0123(45)			&100	 &1.6	&7.89	      &415	&3	&6.00 	     &7.90	   &71.4 x 68.4    &0.52  &7.62 \\	 
HGC 79a		&0123(4)(5)			&123	 &2.1	&7.77         &400	&3	&5.83	     &7.77	   &20.4 x 10.8	   &1.00  &7.77 \\	 
HGC 79b		&0123(4)5			&193	 &2.8	&6.70	      &-	&-	&-	     &6.70	   &15.3 x 7.2     &1.00  &6.70 \\	 
HGC 82b		&01(2)3(45)			&115	 &1.9	&8.28	      &373	&3	&6.28	     &8.29	   &15.3 x 3.6	   &1.00  &8.29 \\	 
HGC 82c		&0123(4)5			&156	 &2.6	&8.64	      &652	&3	&6.04	     &8.64	   &10.8 x 12.6	   &1.24  &8.74 \\	 
HGC 91a		&0123(4)5			&151     &2.5	&8.48	      &1058	&3	&5.18	     &8.48	   &45.9 x [10.2]	   &0.48  &8.16 \\	 
HGC 95c		&0123(4)5			&140	 &2.4	&8.18	      &605	&3	&6.15	     &8.19	   &20.4 x 19.8	   &1.94  &8.48 \\	 
HGC 96a		&0123(4)5			&100     &1.6	&9.58	      &-	&-	&-	     &-	           &9.58 x 45.9	   &0.48  &9.26 \\	 
HGC 100a 	&01234(5)			&194	 &2.8	&7.47	      &1057	&3	&4.83	     &7.47	   &35.7 x 10.8	   &0.62  &7.26 \\	

\noalign{\smallskip} \hline \noalign{\medskip}
\end{tabular}
}%
\tablefoot{
No measurements for HCG 6c, HCG 25f and HCG 57d are available.
\tablefoottext{a}{Number of the observed lines, $n$ denotes the transition 0-0S($n$). 
A parenthesis indicates that the line was not detected so that only an upper limit could be measured.}
\tablefoottext{b} {Temperture of coolest component  ($T_{\rm 1}$), and the second, hotter  ($T_{\rm 2}$) component derived from the fit to the excitation diagrams.}
\tablefoottext{c} {Ortho-to-para ratio determined from the fit.}
\tablefoottext{d}{Logarithm of the mass of each component determined over the extracted area.  This mass is an upper limit if
the lowest line, 0-0S(0) is not detected.}
\tablefoottext{e}{Dimension of the warm \htwo\ emission, measured where possible in orthogonal LL and SL IRS slits. If no SL slit measurement was
available, the width of the LL slit was assumed and the width is given in parenthesis. }
\tablefoottext{f}{Correction factor, which corrects IRS observations for the geometry
of the extraction area relative to the LL slit width and  to the IRAM beam (see text for a more
detailed explanation). }
\tablefoottext{g}{Logarithm of the mass of the warm molecular gas within the IRAM 30m CO(1-0) beam. If the S(0) line is an upper limit,
the mass is an upper limit (see text).}
}

\end{table*}

The assumption and limitations in the data produce, in some cases, a considerable uncertainty
in the derived warm molecular gas mass. This uncertainty is difficult to quantify
precisely, especially in the cases where a correction factor was applied, but we can estimate it. 
The error of the   $H_2$ line fluxes is about 10\%, and the error in the mass determination
for those cases where no correction factor was applied about 20\%. We estimate the 
error introduced by the correction factor to be $\sim$ 30\% on average so that in these cases the
total error of the warm \htwo\ mass is about 40\%.

\section{Results}

\subsection{CO spectra, integrated intensities and kinematics}

In Appendix A we show the spectra of the detections and tentative detection of CO(1-0) and CO(2-1).
For each spectrum, we integrated the intensity along the velocity interval where emission is detected. For nondetections we set an upper limit as

\begin{equation}
I_{\rm CO} < 3 \times rms \times \sqrt{\delta \rm{V} \ \Delta V},
\end{equation}

\noindent where $\delta \rm{V}$ is the channel width, $\Delta$V the total line width and $rms$ the root mean square noise.
When the source is detected in only one transition, this line width is used to calculate the upper limit in the other transition.
 In case of nondetection of both lines,  the mean value
of all  detections was used as an estimate for the linewidth 
($\Delta$V = 470  \kms\  for CO(1-0) and $\Delta$V = 410  \kms\ for CO(2-1)). 
The results of our CO(1-0) and CO(2-1) observations are listed in Table~\ref{tab:co}.


\begin{table*}
\caption{Integrated CO intensities}
\label{tab:co}
\begin{tabular}{lllllllll}
\noalign{\smallskip} \hline \noalign{\medskip}
Galaxy & RA offset & DEC offset &  rms\tablefootmark{a} &  $I_{\rm CO(1-0)}$ & $\rm \Delta \rm{V}_{CO(1-0)}$\tablefootmark{b} & rms\tablefootmark{a} & $I_{\rm CO(2-1)}$\tablefootmark{c}  & $\rm \Delta \rm{V}_{CO(2-1)}$\tablefootmark{b}  \\
  &  [\arcsec ] &  [\arcsec ] & [mK] & [K \kms]  & [\kms] & [mK] & [K \kms]  & [\kms] \\
\noalign{\smallskip} \hline \noalign{\medskip}
    HCG 6b   &   0   &   0   &    2.98   &    1.13 $\pm$    0.30*   &  620   &    3.98   &    1.78 $\pm$    0.40*   &  620  \\
HCG 6b   &   0   &  15   &    2.99   &             $<$    2.37   &    -   &    4.77   &             $<$    3.66   &    -  \\
HCG 6b   &   0   & -15   &    3.12   &    0.58 $\pm$    0.16*   &  160   &    3.96   &   $<$    1.81   &    -  \\
HCG 6b   &  15   &   0   &    3.04   &             $<$    2.41   &    -   &    4.25   &             $<$    3.26   &    -  \\
HCG 6b   & -15   &   0   &    2.99   &             $<$    2.37   &    -   &    4.87   &             $<$    3.74   &    -  \\
HCG 6c   &   0   &   0   &    2.48   &    0.64 $\pm$    0.15   &  220   &    3.64   &   $<$    1.96   &    -  \\
HCG15d   &   0   &   0   &    2.25   &             $<$    1.77   &    -   &    3.82   &             $<$    2.91   &    -  \\
HCG15d   &   0   &  15   &    3.46   &             $<$    2.73   &    -   &    4.94   &             $<$    3.77   &    -  \\
HCG25b   &   0   &   0   &    2.85   &    3.37 $\pm$    0.26   &  520   &    3.80   &    5.01 $\pm$    0.35   &  540  \\
HCG25b   &   0   &  -7   &    2.52   &    3.56 $\pm$    0.25   &  600   &    4.48   &    5.28 $\pm$    0.42   &  560  \\
HCG25b   &   0   &  15   &    3.33   &             $<$    2.62   &    -   &    4.52   &             $<$    3.45   &    -  \\
HCG25b   &   0   & -15   &    2.03   &    0.51 $\pm$    0.10   &  160   &    2.34   &    1.08 $\pm$    0.21   &  520  \\
HCG25b   &   0   & -23   &    2.42   &             $<$    1.91   &    -   &    4.14   &             $<$    3.16   &    -  \\
HCG25b   &  15   &   0   &    1.48   &    1.15 $\pm$    0.14   &  600   &    2.44   &    0.53 $\pm$    0.15*   &  240  \\
HCG25b   &  15   & -15   &    2.62   &    0.70 $\pm$    0.19*   &  330   &    3.43   &   $<$    2.23   &    -  \\
HCG25b$^{(1)}$ &  -3   & -33   &    2.80   &             $<$    2.21   &    -   &    3.87   &             $<$    2.96   &    -  \\
HCG25b   & -15   &   0   &    2.53   &             $<$    2.00   &    -   &    3.81   &             $<$    2.90   &    -  \\
HCG25b   & -15   & -15   &    4.18   &             $<$    3.30   &    -   &    9.62   &             $<$    7.34   &    -  \\
HCG40c   &  -3   &   1   &    5.69   &   10.20 $\pm$    0.59   &  680   &    8.64   &   11.83 $\pm$    0.81   &  540  \\
HCG40c   &   4   &   8   &    4.28   &    6.25 $\pm$    0.39   &  520   &    5.42   &    2.33 $\pm$    0.47   &  460  \\
HCG40c   &   4   &  -6   &    5.83   &    8.24 $\pm$    0.53   &  520   &    7.81   &    8.48 $\pm$    0.73   &  550  \\
HCG40c   &   4   & -21   &    4.04   &    0.54 $\pm$    0.15*   &   90   &    5.89   &   $<$    2.01   &    -  \\
HCG40c   &  34   & -36   &    4.48   &    1.18 $\pm$    0.35*   &  380   &    6.68   &   $<$    4.69   &    -  \\
HCG40c   & -11   &  -6   &    3.20   &    5.18 $\pm$    0.28   &  490   &    4.80   &    4.01 $\pm$    0.42   &  480  \\
HCG40c   & -14   &   7   &    5.75   &    7.14 $\pm$    0.50   &  480   &    7.55   &    6.36 $\pm$    0.52   &  290  \\
HCG40c   & -18   & -14   &    2.79   &             $<$    2.20   &    -   &    3.68   &             $<$    2.81   &    -  \\
HCG47d   &   0   &   0   &    2.07   &    1.09 $\pm$    0.13   &  260   &        - & - & - \\
HCG55c   &   0   &   0   &    1.23   &    1.54 $\pm$    0.13   &  730   &    3.33   &    2.36 $\pm$    0.34   &  640  \\
HCG56b   &   0   &   0   &    2.69   &             $<$    2.12   &    -   &    4.33   &             $<$    3.31   &    -  \\
HCG56c   &   0   &   0   &    3.57   &             $<$    2.81   &    -   &    6.36   &             $<$    4.85   &    -  \\
HCG57a   &   0   &   0   &    2.41   &    4.57 $\pm$    0.28   &  840   &        - & - & - \\
HCG57a   &   0   &  15   &    2.00   &    2.70 $\pm$    0.22   &  720   &    3.58   &    1.63 $\pm$    0.32   &  490  \\
HCG57a   &   0   & -15   &    1.96   &    3.19 $\pm$    0.21   &  740   &    3.26   &   $<$    3.19   &    -  \\
HCG57a   &  15   &   0   &    1.91   &    2.68 $\pm$    0.22   &  810   &        - & - & - \\
HCG57a$^{(2)}$ &  15   &  15   &    2.61   &    2.06 $\pm$    0.19   &  340   &    4.85   &    2.70 $\pm$    0.39   &  400  \\
HCG57a   &  15   & -15   &    1.56   &    0.99 $\pm$    0.13   &  440   &    2.89   &    1.01 $\pm$    0.23   &  380  \\
HCG57a   & -15   &   0   &    2.30   &    3.90 $\pm$    0.26   &  820   &        - & - & - \\
HCG57a   & -15   &  15   &    2.03   &    0.60 $\pm$    0.15   &  360   &    3.97   &   $<$    2.71   &    -  \\
HCG68a   &   0   &   0   &    1.92   &    2.59 $\pm$    0.23   &  960   &    2.90   &    5.10 $\pm$    0.35   &  920  \\
HCG68b   &   0   &   0   &    2.57   &             $<$    2.01   &    -   &    5.73   &             $<$    4.34   &    -  \\
HCG68c   &   0   &   0   &    4.97   &    8.67 $\pm$    0.36   &  340   &    8.38   &   12.65 $\pm$    0.65   &  380  \\
HCG82b   &   0   &   0   &    1.83   &    1.13 $\pm$    0.17   &  500   &    3.59   &   $<$    2.91   &    -  \\
HCG91a   &   0   &   0   &    6.39   &   15.81 $\pm$    0.58   &  520   &    8.36   &   14.13 $\pm$    0.83   &  620  \\
HCG91a   &   0   &  15   &    9.07   &   14.20 $\pm$    0.75   &  430   &   13.32   &   10.69 $\pm$    0.71   &  180  \\
HCG91a   &   0   & -15   &    6.81   &    7.62 $\pm$    0.44   &  270   &   10.21   &    5.28 $\pm$    0.58   &  200  \\
HCG91a   &  15   &   0   &    7.50   &   12.85 $\pm$    0.72   &  580   &    8.99   &   11.25 $\pm$    0.79   &  490  \\
HCG91a   &  15   &  15   &    9.51   &    9.21 $\pm$    0.74   &  380   &   27.79   &   $<$   19.48   &    -  \\
HCG91a   &  15   & -15   &    8.79   &    5.55 $\pm$    0.59   &  290   &   10.88   &    4.17 $\pm$    0.73   &  280  \\
HCG91a   & -15   &   0   &    8.13   &    9.66 $\pm$    0.71   &  470   &   14.05   &   11.96 $\pm$    1.19   &  450  \\
HCG91a   & -15   &  15   &    8.43   &    9.10 $\pm$    0.67   &  390   &   16.29   &    6.18 $\pm$    1.11   &  290  \\
HCG91a   & -15   & -15   &    7.55   &    5.99 $\pm$    0.54   &  320   &   21.73   &   $<$   13.98   &    -  \\
HCG91a   &  30   & -30   &    3.80   &    0.79 $\pm$    0.17   &  130   &    6.73   &   $<$    2.76   &    -  \\
HCG95c   &   0   &   0   &    3.04   &    3.15 $\pm$    0.25   &  420   &    5.81   &    8.75 $\pm$    0.52   &  490  \\
HCG95c   &   0   &  15   &    3.44   &    2.37 $\pm$    0.34   &  600   &    4.89   &   $<$    4.34   &    -  \\
HCG96a   &   0   &   0   &    3.38   &   18.02 $\pm$    0.25   &  340   &    4.93   &   24.24 $\pm$    0.36   &  320  \\

\noalign{\smallskip} \hline \noalign{\medskip}
\end{tabular}
\tablefoot{
\tablefoottext{*}{Tentative detections.}
\tablefoottext{1}{Coincides with the position of HCG~25f.}
\tablefoottext{2}{Coincides (within 1\arcsec) with the position of HCG~57d.}
\tablefoottext{a}{Root-mean-square noise at a velocity resolution of 16 \kms.}
\tablefoottext{b}{Zero-level line width. The uncertainty is roughly given by the velocity resolution (16 \kms).}
\tablefoottext{c}{At some positions the bandwidth did not entirely cover the expected line range so that no value for $I_{\rm CO(2-1)}$
can be given.}
}
\end{table*}


 
{The zero-level line widths of the detected spectra range from $\sim$ 150 to 950 \kms . Most noticeable are
two MOHEGs, HCG~68a and HCG~57a, with very broad ($>$ 800 \kms ) lines. 
Also HCG 55c, a galaxy with enhanced \htwo\ emission, slightly below the MOHEG threshold,
has a broad spectrum of 730 \kms . }
Their spectra have an irregular profile with indications
of different components. A possible reason for  the broad lines could be a 
 projection effect, i.e. the presence of an additional molecular gas component not
 associated with galaxy rotation, e.g. infalling or outflowing molecular gas or
 gas displaced by tidal effects.
Higher resolution CO observations are necessary to resolve this issue.
In the case of HCG~57a, interferometric observations with CARMA  have been carried
out  \citep{alatalo_sub} and shown
that the position-velocity diagram is not well described by rotation alone, but has
an additional component which seems to trace outflowing or infalling gas.

{The average value for the line ratio is  $\log(I_{\rm CO(2-1)}/I_{\rm CO(1-0)}) = 0.00 \pm 0.07$ 
(corresponding to $I_{\rm CO(2-1)}/I_{\rm CO(1-0)} \sim 1.0$) for
the central position and a slightly lower value of $\log(I_{\rm CO(2-1)}/I_{\rm CO(1-0)}) = -0.07 \pm 0.05$
(corresponding to $I_{\rm CO(2-1)}/I_{\rm CO(1-0)} \sim 0.85$)  for the off-centre positions.
In the calculation of the mean values we  did not take into account  those objects/positions where
both lines are upper limits and we did not attempt to correct for the different beam sizes.
We did not find any correlations of this line ratio with other
parameters and do not discuss it further in this paper.}

\subsection{Molecular gas masses}

We calculate the molecular gas mass, ${\it M_{\rm H_{2}}}$,
from the CO(1-0) emission using the following equation:
\begin{equation}
M_{\rm H_{2}}=75 \times D^{2}I_{\rm CO(1-0)}\Omega
\label{eq:mh2}
\end{equation}

\noindent in \msun . Here,  $\Omega$ is the area covered by the observations in arcsec$^{2}$ (i.e. $\Omega$ = 1.13 $\theta^{2}$ for a single pointing with a Gaussian beam where $\theta$ is the HPBW), $D$ is the distance in Mpc and $I_{\rm CO(1-0)}$ is the velocity integrated line intensity in
K \kms . This equation assumes a CO-to-H$_2$ conversion factor 
 X=$N_{\rm H_{2}}/I_{\rm CO}$ = $2\times 10^{20}\rm cm^{-2}$ (K km s$^{-1})^{-1}$ \citep[e.g.] []{1986ApJ...309..326D}. 
No correction factor for the fraction of helium and other heavy metals is included. 

For most of the galaxies only an observation at the centre of the galaxy was obtained so that a 
 correction for emission outside of the beam is necessary in order to derive the total molecular gas mass.
We carried out this aperture correction in the same way as described in \citet{2011A&A...534A.102L}, 
assuming an exponential distribution of the CO emission:

\begin{equation}
I_{\rm CO}(r) = I_{0}\propto \exp(r/r_{\rm e}) .
\label{Ico_r}
\end{equation}

We adopt a scale length of $r_{\rm e}$ = 0.2$\times$$r_{\rm 25}$, where \ropt\  is the major optical isophotal radius at 25 mag arcsec$^{-2}$, 
following \citet{2011A&A...534A.102L}, who derived this scale length from different  studies 
 \citep{2001PASJ...53..757N, 2001ApJ...561..218R, 2008AJ....136.2782L}
 and from their own CO data.  These studies are based on spiral galaxies, however, the analysis of the spatial extent of CO in early-type galaxies in the 
 ALTAS$^{\rm 3D}$ survey showed that the relative CO extent, normalized to \ropt ,  of early-type and spiral galaxies is  the same
 \citep{2013MNRAS.429..534D}. 
We used this distribution to calculate the expected CO emission from the entire disk, taking the galaxy inclination into account
 \citep[see][for more details]{2011A&A...534A.102L}. This approach assumes that the distribution of the molecular gas in galaxies in HCGs is the
 same as in isolated  galaxies. 
 {Unfortunately, so far, no observational data exists that could confirm that this assumption is  correct for galaxies in this different environment.
 Indeed, there is no alternative to applying an aperture correction because the molecular gas mass measured
 in the central pointing represents a different fraction of the total molecular gas mass for each galaxy, depending on its diameter and inclination.
 Thus, a study of the central molecular gas mass  is only meaningful when
 we are able to compare it with a magnitude that is measured in the same area, as was done in this paper 
 for the comparison of the cold molecular gas mass to the SFR and
 the warm molecular gas mass.
 }

The resulting aperture correction factor, $f_{\rm aper}$, defined as the ratio between the molecular gas mass  observed in the central pointing, \mhtwocenter,
 and  the molecular gas mass
extrapolated to the entire disk, \mhtwotot,  lies between 1.15  and 4.85 with a mean value of 2.0. 
The values for the molecular gas mass in the central pointing, the extrapolated molecular gas mass and the aperture correction factor are listed in Table \ref{tab:mh2}.  
{In order to test whether biases might be   introduced by the aperture correction, 
we also carried out, as far as possible, all the analysis based on the extrapolated molecular gas masses only for those galaxies where the aperture correction factor
was small. In particular, we calculated the mean values for the total molecular gas, \mhtwotot, and the ratio \mhtwo/\lk, listed in Tab.~\ref{tab:mean_mhtwo_lk},
with the restriction of $f_{\rm aper}<1.6$ for the subsample of MOHEGs, 
non-MOHEGs, early type and spiral galaxies. The number of galaxies in the corresponding subsample was about half the total size which made a 
finer subdivision impossible.  The mean values of the restricted subsamples were in very good agreement with those of the total
sample. 
}

\begin{table}
\caption{Cold molecular gas mass}
\label{tab:mh2}
\begin{tabular}{lllll}
\noalign{\smallskip} \hline \noalign{\medskip}
Galaxy & Ref. &log(\mhtwocenter)\tablefootmark{a} & $f_{\rm aper}$\tablefootmark{b} & log(\mhtwotot)\tablefootmark{c} \\
  & &  [\msun]  &   &  [\msun]    \\
\noalign{\smallskip} \hline \noalign{\medskip}
    HCG   6b   &   1   &    9.05*   &    1.51   &    9.23*  \\
HCG   6c   &   1   &    8.80   &    1.44   &    8.96  \\
HCG  15a   &   2  &  $<$    8.18   &    1.89  &  $<$    8.45  \\
HCG  15d   &   1  &  $<$    8.32   &    1.78  &  $<$    8.57  \\
HCG  25b   &   1   &    9.02   &    1.60   &    9.23  \\
HCG  25f   &   1  &  $<$    8.36   &    1.15  &  $<$    8.42  \\
HCG 31ac   &   3   &    8.29   &    1.46   &    8.45  \\
HCG  40b   &   2   &    8.55   &    1.69   &    8.78  \\
HCG  40c   &   1   &    9.55   &    1.91   &    9.83  \\
HCG  40d   &   3   &    9.24   &    1.46   &    9.41  \\
HCG  44a   &   3   &    8.19   &    3.68   &    8.75  \\
HCG  44d   &   2   &    7.53   &    2.98   &    8.01  \\
HCG  47a   &   3   &    9.53   &    1.52   &    9.71  \\
HCG  47d   &   1   &    8.88   &    1.26   &    8.98  \\
HCG  55c   &   1   &    9.47   &    1.29   &    9.58  \\
HCG  56b   &   1  &  $<$    8.55   &    1.45  &  $<$    8.71  \\
HCG  56c   &   1  &  $<$    8.67   &    1.51  &  $<$    8.85  \\
HCG  57a   &   1   &    9.47   &    1.91   &    9.75  \\
HCG  57d   &   1   &    9.12   &    1.32   &    9.24  \\
HCG  67b   &   3   &    9.44   &    1.56   &    9.64  \\
HCG  68a   &   1   &    8.06   &    2.59   &    8.47  \\
HCG  68b   &   1  &  $<$    7.47   &    3.18  &  $<$    7.97  \\
HCG  68c   &   1   &    8.58   &    4.83   &    9.27  \\
HCG  79a   &   3  &  $<$    8.39   &    3.89  &  $<$    8.98  \\
HCG  79b   &   2   &    8.12   &    2.98   &    8.60  \\
HCG  82b   &   1   &    9.01   &    1.59   &    9.21  \\
HCG  82c   &   3   &    9.67   &    1.26   &    9.77  \\
HCG  91a   &   1   &    9.79   &    3.16   &   10.29  \\
HCG  95c   &   1   &    9.53   &    1.36   &    9.67  \\
HCG  96a   &   1   &   10.03   &    2.28   &   10.38  \\
HCG 100a   &   3   &    8.98   &    1.45   &    9.14  \\

\noalign{\smallskip} \hline \noalign{\medskip}
\end{tabular}
\tablefoot{
\tablefoottext{*}{Tentative detection.} 
\tablefoottext{a}{Logarithm of the cold H$_{2}$ mass (in solar masses) calculated from the observed  central $I_{\rm CO(1-0)}$.}
\tablefoottext{b}{Aperture correction factor, defined as \mhtwocenter/\mhtwotot.}
\tablefoottext{c}{Logarithm of the cold H$_{2}$ mass (in solar masses) extrapolated to the emission from the total disk (see text).}
}
\tablebib{
(1)  this work, (2) \citet{2012A&A...540A..96M},
(3) \citet{1998A&A...330...37L}, observed with the IRAM 30m telescope.
}
\end{table}

\begin{figure}
\centering
\includegraphics[width=8.cm]{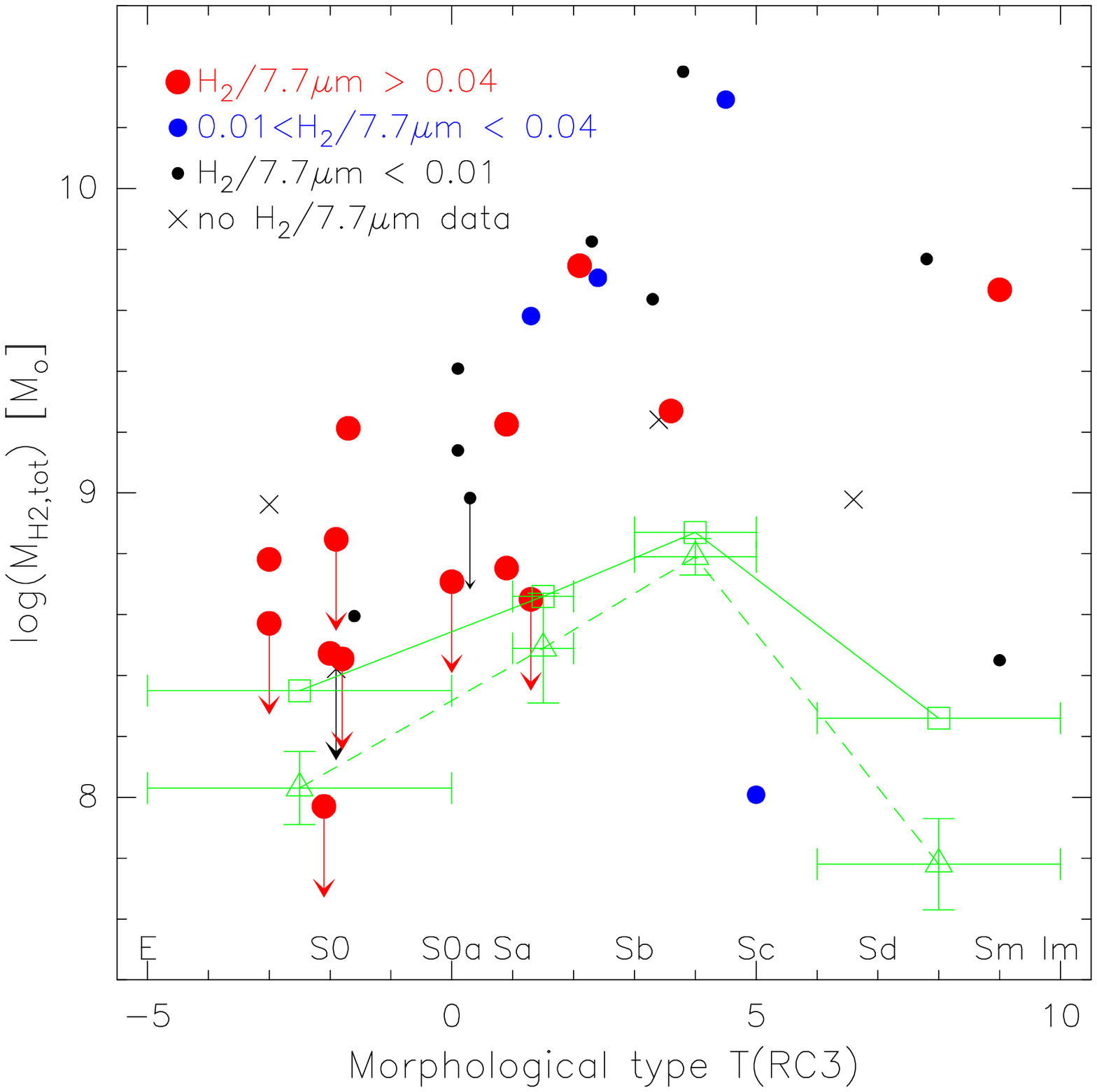}
\caption{Total extrapolated molecular gas mass, \mhtwotot,  as a function of
morphological type.  
Galaxies classified as MOHEGs (\htwopah $\ge 0.04$) are shown as large red circles, non-MOHEG galaxies 
with $0.01<$ \htwopah $< 0.04$ as smaller blue circles, galaxies with  \htwopah $< 0.01$ as black dots and galaxies with no 
warm \htwo\ data are black crosses.
The green symbols denote means (triangles) and medians (squares) for a sample of isolated galaxies \citep{2011A&A...534A.102L}. 
The mean, respectively median, are taken for the interval of morphological types
shown by the horizontal bar.
}
\label{fig:mh2_vs_hubble}
\end{figure}

In Fig.~\ref{fig:mh2_vs_hubble} we show the molecular gas mass as a function
of morphological type for our sample. 
Included in the figure are the mean and
median values for a sample of $\sim$ 200 isolated galaxies   \citep{2011A&A...534A.102L} from the
AMIGA ({\it Analysis of the Interstellar Medium of Isolated Galaxies}) project \citep{2005A&A...436..443V}. The molecular gas masses
were obtained in the same way as here, by an extrapolation from observations
of CO(1-0) from a central pointing.
In Tab.~\ref{tab:mean_mhtwo_lk}, the mean
values\footnote{  %
When calculating the mean values we take upper limits into account
by using the program ASURV  Rev, Astronomical Survival Analysis) Rev, 1.1
\citep{1992BAAS...24..839L} which is a generalized statistical package that implements the methods presented by \cite{1985ApJ...293..192F}.}
of the molecular gas
 for different subsamples are listed.

\begin{table*}
\caption{Mean values and their errors of \mhtwotot\ and \mhtwotot/\lk\ for different subsamples}
\begin{tabular}{llllll}
\noalign{\smallskip} \hline \noalign{\medskip}
Name &  $T$(RC3)\tablefootmark{a} &  log(\mhtwotot)  & $n/n_{\rm up}$\tablefootmark{b}  & log(\mhtwotot/\lk) & $n/n_{\rm up}$\tablefootmark{b}  \\
     &   & [\msun] & & [\msun/\lsunk] &  \\
\noalign{\smallskip} \hline \noalign{\medskip}
MOHEG & all $T$ & 8.72 $\pm$0.18 & 14/6 & -2.16 $\pm$ 0.13  & 13/5 \\
                &  $T \leq 0$   & 8.37 $\pm$0.17 & 8/5 & -2.45 $\pm$ 0.11  & 7/4 \\
                & $T\geq 1$  & 9.24 $\pm$0.16 & 6/1 & -1.78 $\pm$ 0.12   & 6/1 \\
\hline                
non-MOHEG & all $T$ &  9.32 $\pm$0.20 & 13/1 & -1.55 $\pm$ 0.08  & 12/1 \\
                & $T\leq 0$ & 8.94 $\pm$0.18 & 4/1 & -1.86 $\pm$ 0.10  & 4/1 \\
                & $T\geq 1$ & 9.52 $\pm$0.25 & 9/0 & -1.41 $\pm$ 0.06   & 8/0 \\
\hline                
 all               & $T\leq 0$ & 8.52 $\pm$0.15 & 14/7 & -2.26 $\pm$ 0.12 & 12/6 \\
                & $T\geq 1$ & 9.36 $\pm$0.16 & 17/1 & -1.57 $\pm$ 0.07   & 16/1 \\
\hline
AMIGA$^{c}$  & $T\leq 0$ & 8.03 $\pm$0.12 & 23/15 &  -2.32 $\pm$ 0.10 & 22/15 \\
                & $T\geq 1$ & 8.38 $\pm$0.09 & 150/64 & -1.76 $\pm$ 0.05   & 135/50 \\
\noalign{\smallskip} \hline \noalign{\medskip}
\end{tabular}
\tablefoot{{
\tablefoottext{a}{Morphological subsample.}
\tablefoottext{b}{Total number of galaxies in the subsample ($n$) and number of upper limits  for the considered magnitude ($n_{\rm up}$).}}
\tablefoottext{c}{AMIGA sample of isolated galaxies (data from Lisenfeld et al. 2011).
The mean values were calculated with ASURV.}
}
\label{tab:mean_mhtwo_lk}
\end{table*}

The mean molecular gas mass for MOHEGs is lower than the average
molecular gas mass for non-MOHEGs. This is partly due to the fact that
most MOHEGs are early-type galaxies which have lower average molecular gas masses
in all samples considered here. When comparing the mean
values of the molecular gas mass for the same group of morphological types, we find
for early-type galaxies that MOHEGs have a mean value lower by 0.57 dex (2.3$\sigma$) than non-MOHEGs, whereas
for spiral galaxies both groups have identical mean values within the errors.

The molecular gas masses of our sample cannot be directly compared to those
of the isolated galaxies because of the different mean distances of both samples.
Whereas our HCG  sample includes objects between 33 and 216 Mpc,
the sample of isolated galaxies is restricted to more nearby objects with distances up to 70 Mpc.
As a consequence, the galaxies in the HCG are on average brighter in all wavelength,
and, indeed, we find that the molecular gas masses of our sample are well above the values found  for isolated
galaxies.

However, we can compare relative properties. In both samples,  early-type galaxies have a lower mean molecular gas mass than spirals.
For isolated galaxies, the  difference
is 0.35 dex (2.3 $\sigma$), very similar to the difference found in non-MOHEG galaxies in HCGs of 0.58 dex  
(1.8 $\sigma$).
 For MOHEGs in HCGs, the difference is, however, much larger with spiral galaxies having
a mean molecular gas mass of 0.87 dex (3.7$\sigma$) higher than early-type galaxies.


In summary, we find  indications for a lower molecular gas mass  in
MOHEG early-type galaxies  compared to non-MOHEGs early types,
and for an unusually large differences (when compared  to non-MOHEGs or to isolated galaxies)
in molecular gas mass between early-type and spiral MOHEG galaxies.
In Sect.~\ref{sec:mhtwo_lk} we continue to investigate this issue
by using the distance-independent ratio \mhtwotot/\lk .

\subsection{Total mapped molecular gas mass}

We determined the total molecular gas mass for the mapped galaxies and compared it
to the value obtained from the extrapolation.
We calculated the total molecular gas mass by averaging the spectra over the positions
where CO was detected. The total molecular gas mass was then calculated from
eq.~\ref{eq:mh2} where $\Omega$ is the total area {covered by the beams}.
In Tab. \ref{tab:mapped_co} we list the total molecular gas mass and compare it to
the extrapolated value.  
%
%

The agreement is reasonable, with a maximum difference of a factor of 1.4 in the case
of HCG~25, giving support to the reliability of the aperture correction that was applied to the
galaxies.

\begin{table}
\caption{Mapped and extrapolated molecular gas mass}
\begin{tabular}{lllll}
\noalign{\smallskip} \hline \noalign{\medskip}
Name &  Area\tablefootmark{a} & \mhtwomapped\tablefootmark{b} & \mhtwotot\tablefootmark{c} & $\frac{M_{\rm H_2, extra}}{M_{\rm H_2, mapped}}$\\
     & [\arcsec$^2$] & [$10^9$ \msun] &[$10^9$ \msun] & \\
\noalign{\smallskip} \hline \noalign{\medskip}
HCG~25b & 1292  & 1.2  & 1.7 & 1.4\\
HCG~40c &  1547 & 6.4&  6.8 & 1.1 \\
HCG~57a &  2547 & 6.9 &  5.6& 0.8\\
HCG~91a & 3055   & 16.6& 19.5 & 1.2\\
\noalign{\smallskip} \hline \noalign{\medskip}
\end{tabular}
\tablefoot{
\tablefoottext{a}{Area covered by the beams over which the 
CO(1-0) intensity was averaged.}
\tablefoottext{b}{Molecular gas mass determined from the mapping.}
\tablefoottext{c}{Molecular gas mass determined from the extrapolation of the central value.}
}
\label{tab:mapped_co}
\end{table}

\subsection{Distribution of the molecular gas}

\begin{figure*}[h!]
\centerline{
\includegraphics[width=8.5cm]{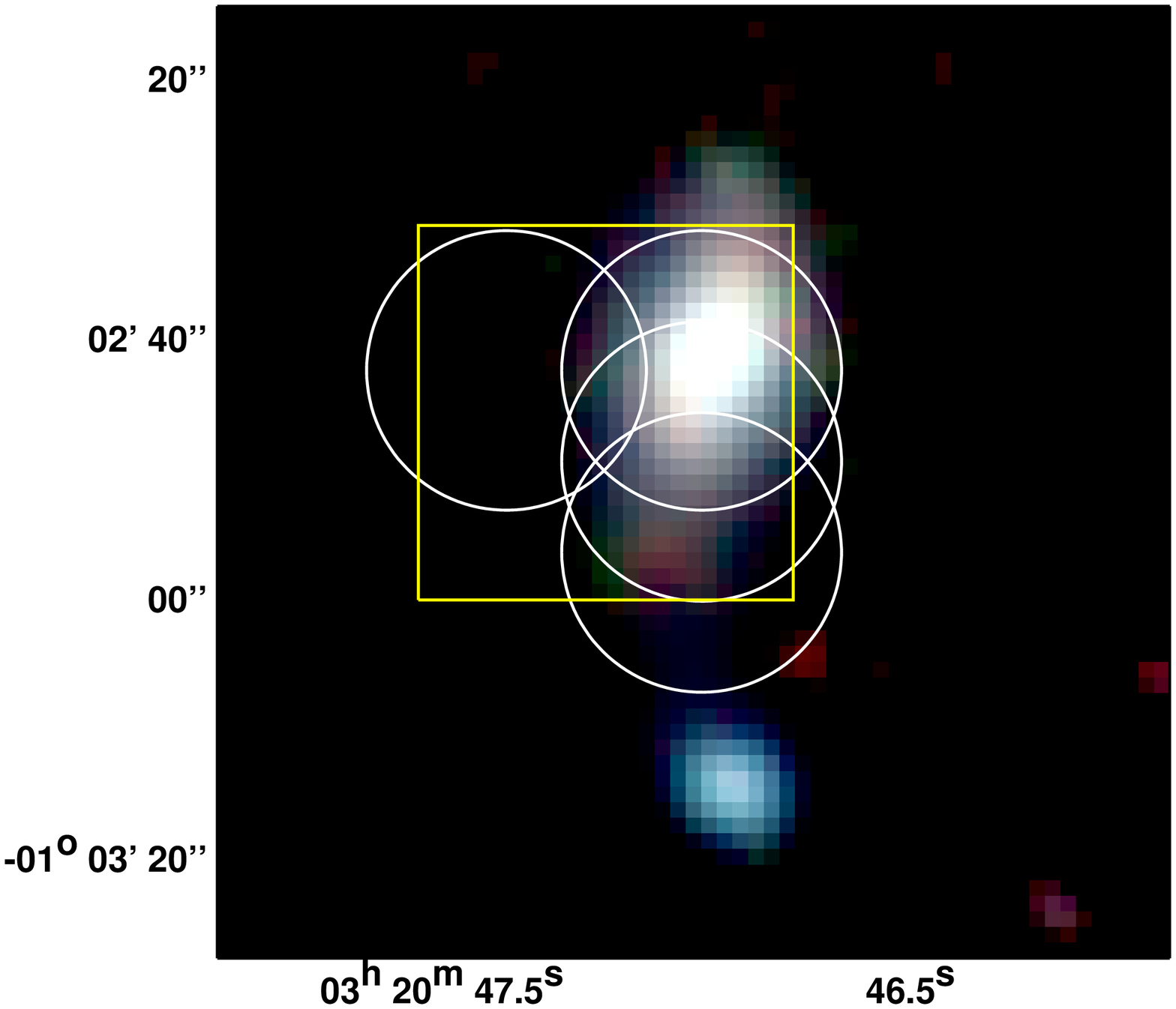}
\quad
\includegraphics[width=8.5cm]{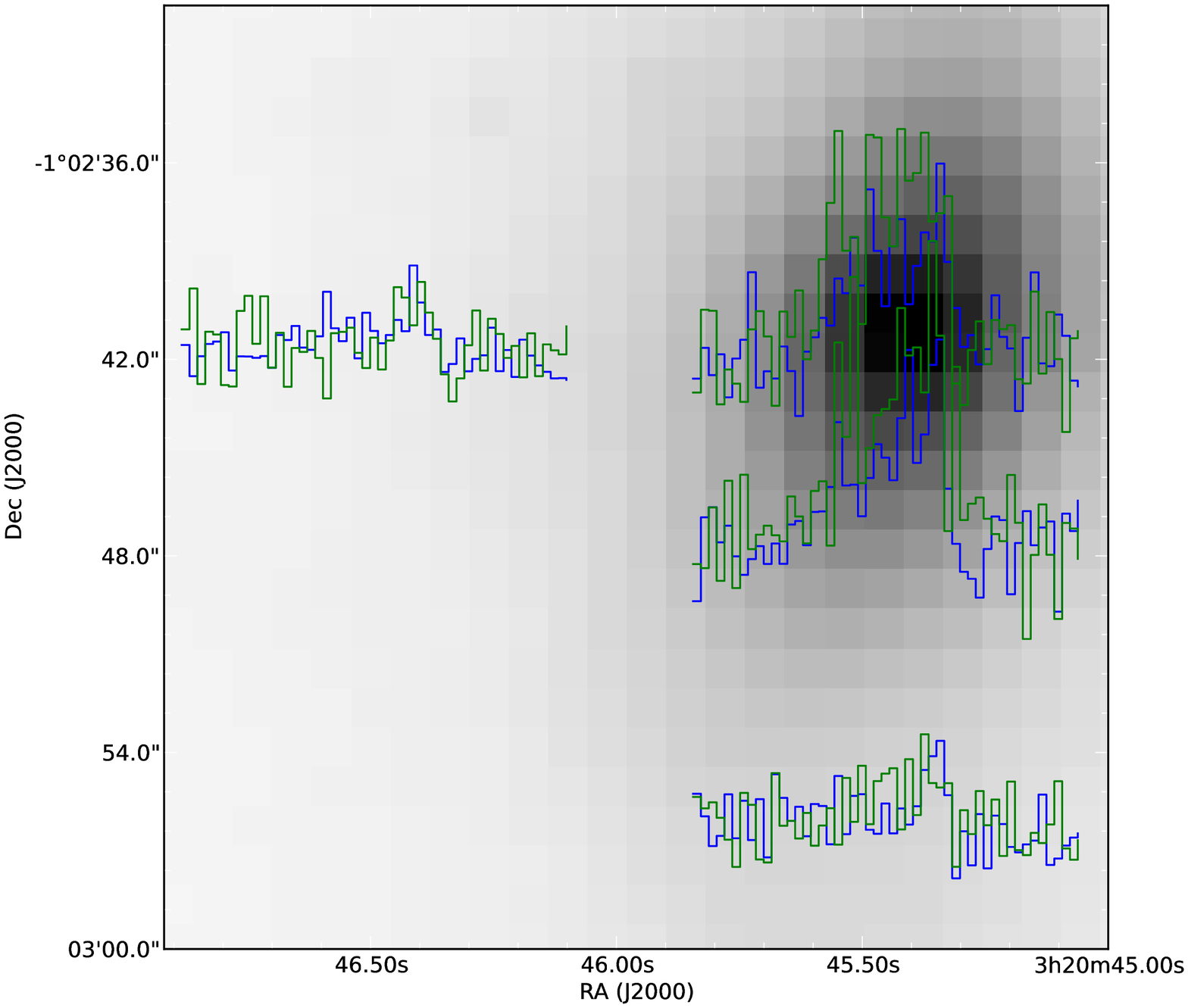}
}
\caption{{\it Left:}  Three-color
IRAC image (blue is the 3.6 \mi, green the 4.5\mi\ and red the  8\mi\ band) of HCG~25b. 
The circles represent the positions detected in CO(1-0). The size of the circles corresponds to
the FWHM size of the IRAM 30m beam at the frequency of CO(1-0).
The large yellow rectangle shows, for reference, the size of the zoomed image at the right.
{\it Right:}  The detected CO(1-0)  (blue) and CO(2-1) (green)  spectra  at these positions,   overlaid
over an IRAC~3.5\mi\  image.  The velocity range is from 5600 to 7200 \kms\ for all spectra.
}
\label{fig:map_co_25b}
\end{figure*}

\begin{figure*}[h!]
\centerline{
\includegraphics[width=8.5cm]{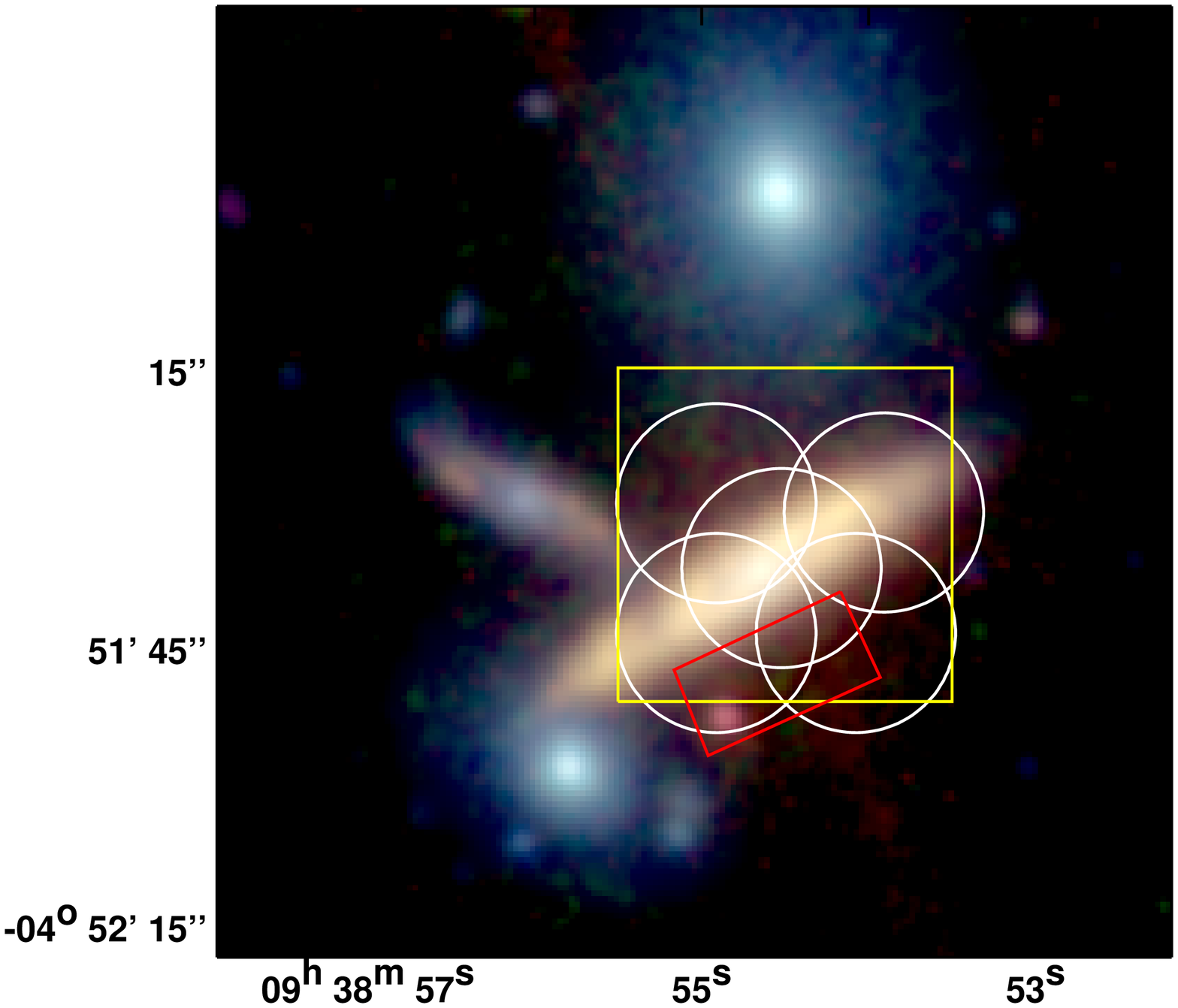}
\quad
\includegraphics[width=8.5cm]{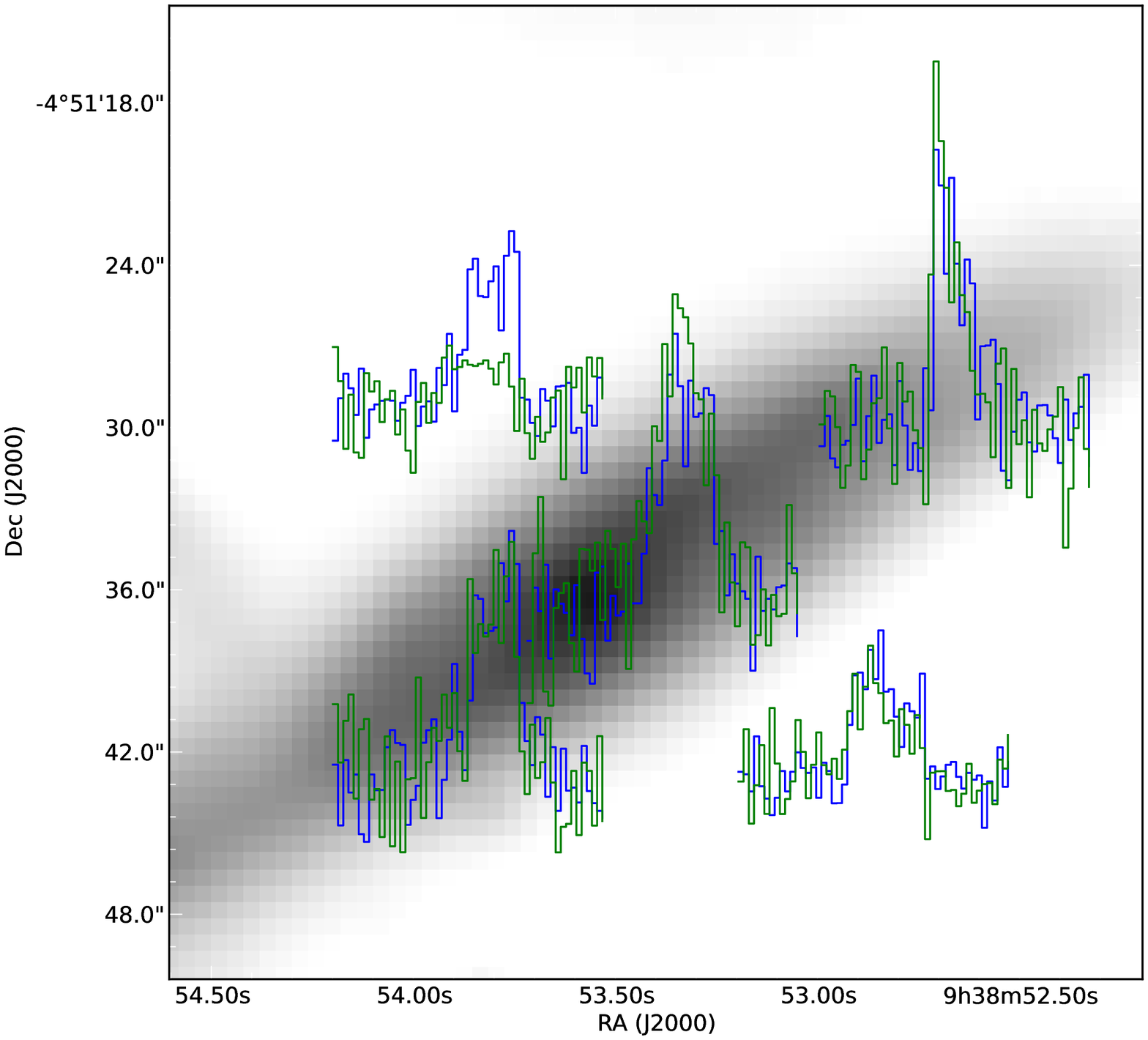}
}
\caption{{\it Left:}  Three-color
IRAC image (blue is the 3.6 \mi, green the 4.5\mi\ and red the  8\mi\ band) of HCG~40c. 
The circles represent the positions detected in CO(1-0). The size of the circles corresponds to
the FWHM size of the IRAM 30m beam at the frequency of CO(1-0).
The large yellow rectangle shows, for reference, the size of the zoomed image at the right.
The small red rectangle is the position where extragalactic, warm \htwo\ was detected by \citet{2013ApJ...765...93C}.
{\it Right:}  The detected CO(1-0)  (blue) and CO(2-1) (green)  spectra  at these positions,   overlaid
over an IRAC~3.5\mi\  image.  The velocity range is from 5500 to 7200 \kms\ for all spectra.
}
\label{fig:map_co_40c}
\end{figure*}

\begin{figure*}
\centerline{
\includegraphics[width=8.5cm]{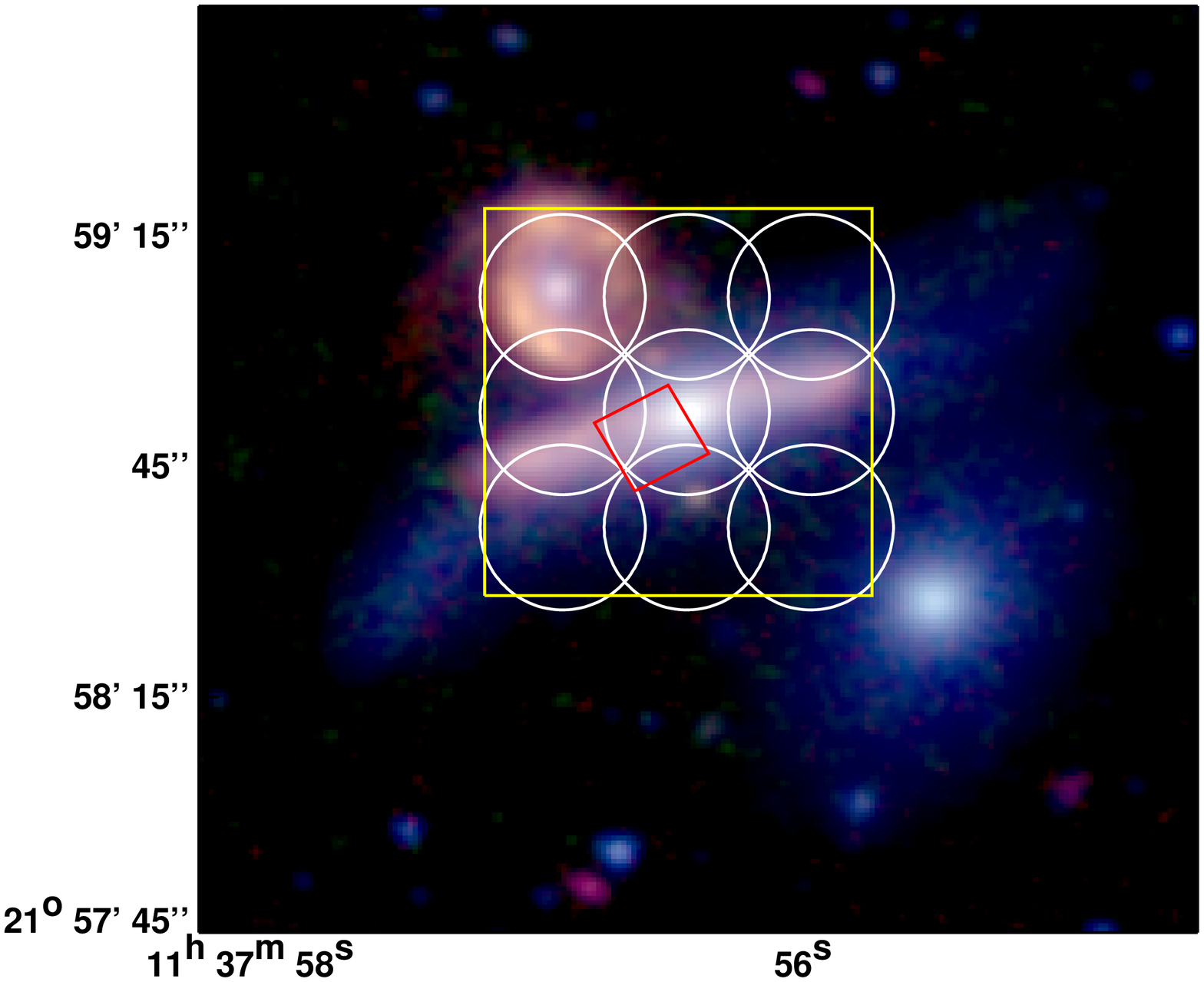}
\quad
\includegraphics[width=8.5cm]{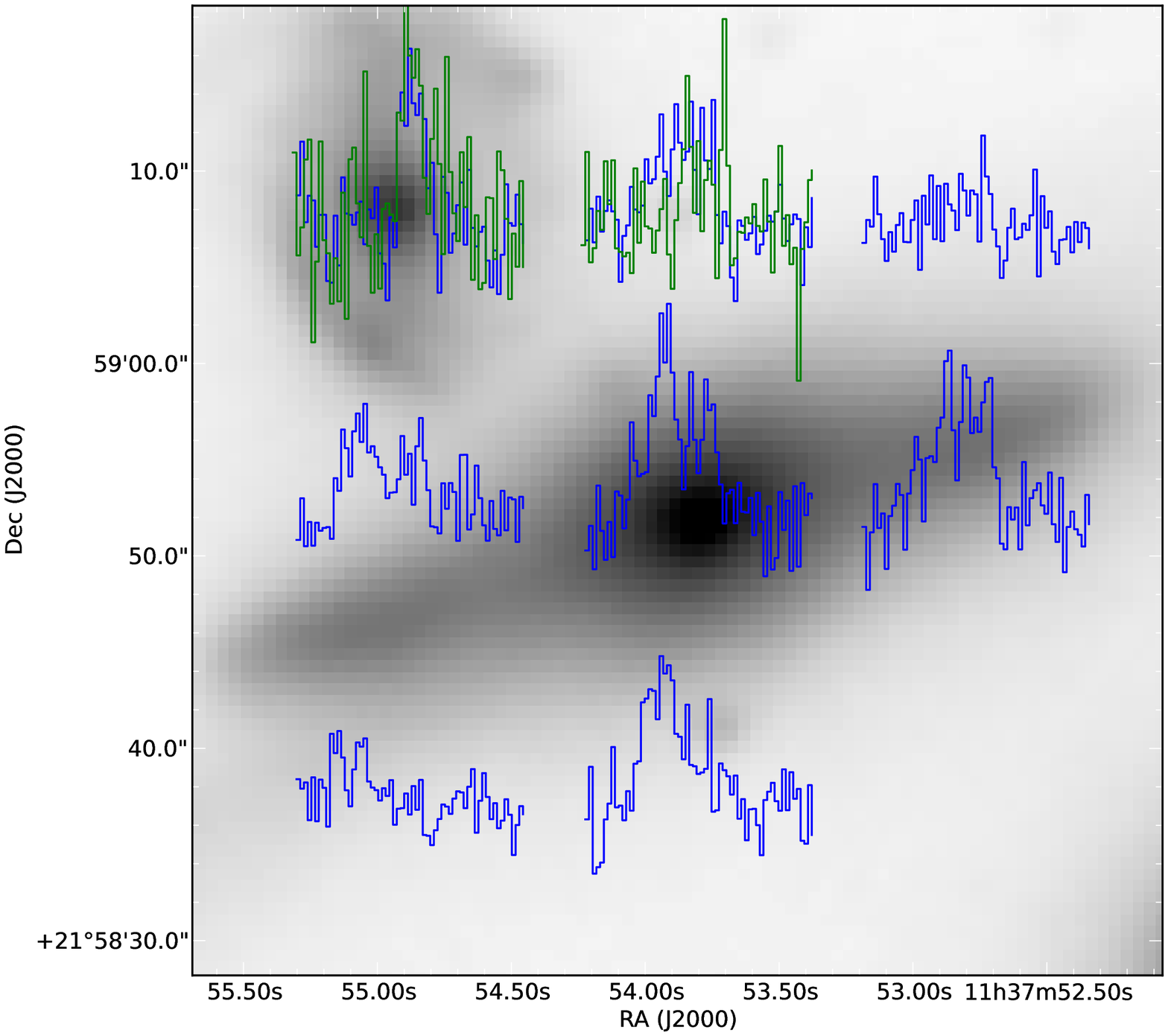}
}
\caption{{\it Left:}  Three-color
IRAC image (blue is the 3.6 \mi, green the 4.5\mi\ and red the  8\mi\ band) of HCG~57a (southern object) and 57d (northern object). 
The circles represent the positions detected in CO(1-0). The size of the circles corresponds to
the FWHM size of the IRAM 30m beam at the frequency of CO(1-0).
The large yellow rectangle shows, for reference, the size of the zoomed image at the right.
The small red rectangle is the position where extragalactic, warm \htwo\ was detected by \citet{2013ApJ...765...93C}.
{\it Right:}  The detected CO(1-0)  (blue) and CO(2-1) (green)  spectra  at these positions,   overlaid
over an IRAC~3.5\mi\  image.  The velocity range is from 8000 to 10000 \kms\ for all spectra.
}
\label{fig:map_co_57a}
\end{figure*}

\begin{figure*}
\centerline{
\includegraphics[width=8.5cm]{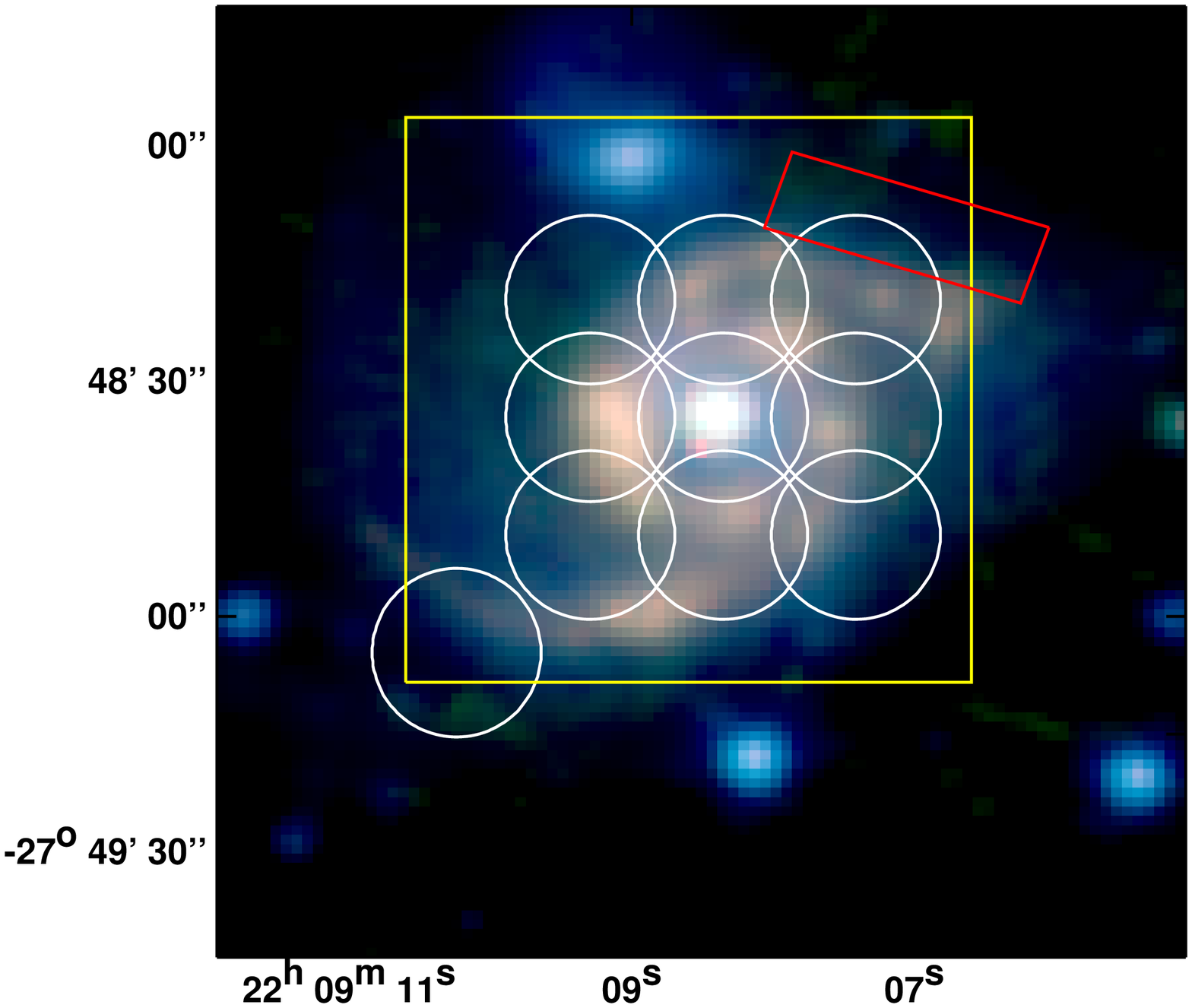}
\quad
\includegraphics[width=8.5cm]{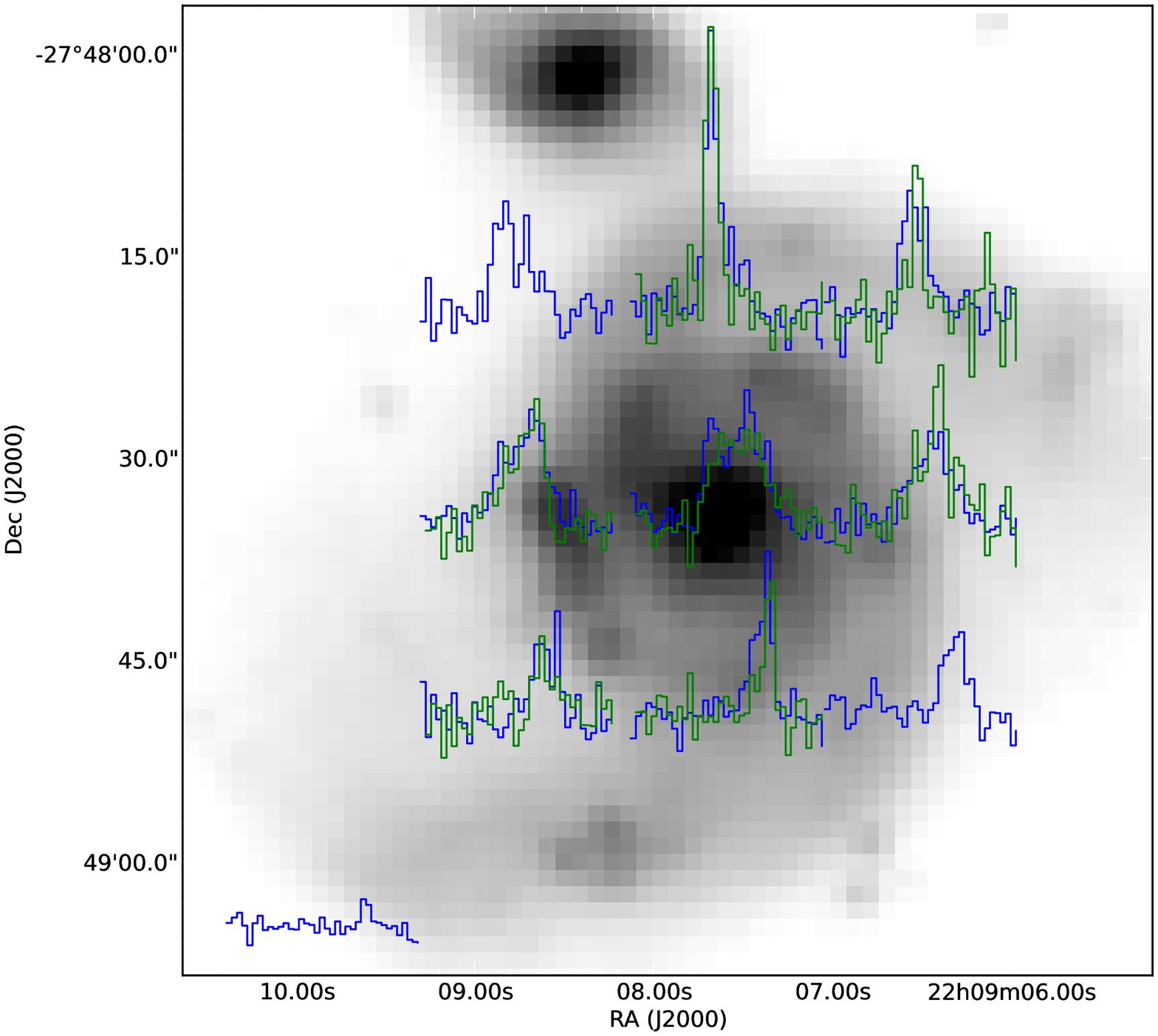}
}
\caption{{\it Left:}  Three-color
IRAC image (blue is the 3.6 \mi, green the 4.5\mi\ and red the  8\mi\ band) of HCG~91a. 
The circles represent the positions detected in CO(1-0). The size of the circles corresponds to
the FWHM size of the IRAM 30m beam at the frequency of CO(1-0).
The large yellow rectangle shows, for reference, the size of the zoomed image at the right.
The small red rectangle is the position where extragalactic, warm \htwo\ was detected by \citet{2013ApJ...765...93C}.
{\it Right:}  The detected CO(1-0)  (blue) and CO(2-1) (green)  spectra  at these positions,   overlaid
over an IRAC~3.5\mi\  image.  The velocity range is from 6300 to 7500 \kms\ for all spectra.
}
\label{fig:map_co_91a}
\end{figure*}

In Figs.~\ref{fig:map_co_25b} - \ref{fig:map_co_91a} we show the distribution of the spectra within and around
the mapped galaxies. 
In all cases, the CO distribution is extended and, at least slightly,  asymmetric.

The CO(1-0) distribution in HCG~25b (Fig.~\ref{fig:map_co_25b}) is concentrated mostly in a north-south direction.
Emission is also found towards the East (positons (15, 0) and tentatively at (15, -15)) which is the
side where the HI  shows more emission (Verdes-Montenegro et al. in prep).

The distribution of the CO emission in HCG~91a (Fig.~\ref{fig:map_co_91a}) is very asymmetric. There is a pronounced  maximum in the North,
towards HCG~91c
where the spectrum is not only stronger, but also very peaked  (position 0, 15).
 There is also considerable emission towards the NW  of the galaxy (position -15, 15), close
 to the place where \citet{2013ApJ...765...93C} found 
extragalactic   mid-infrared rotational \htwo\ emission (see red square in Fig.~\ref{fig:map_co_91a}).

In HCG~40c (Fig.~\ref{fig:map_co_40c}) we find some evidence for extraplanar CO. There is CO(1-0) and CO(2-1) emission detected at two positions
about 10.5\arcsec\ (corresponding to 4.6 kpc at the distance of HCG~40c) 
above the disk. 
The CO(1-0) integrated emission is very similar for both positions, of about half the central emission ($0.54 \pm 0.05$ for the NE
and $0.49\pm 0.04$ for the SW position). Since the
FWHM of the CO(1-0) is large compared to the offset, we still expect to detect this level of emission even
if the CO is very concentrated towards the disk of the galaxy. The 
CO(2-1) emission with its smaller beam is more sensitive to variations in spatial scales. Here, we found about twice as much 
emission in the SW position than in the NE. The emission in the SW 
is about $0.32\pm0.04$ times the emission of the central position, which is about 3 times the amount that would be expected if the
CO distribution is completely concentrated to the galactic plane. The off-plane CO(2-1) emission {could} be explained with an
exponential CO distribution with a scale-height of 5\arcsec , corresponding to  2.2 kpc at the distance of HCG 40.
This is an unusually large scale height and an indication for off-planar molecular gas.
{Interestingly,} the SW position overlaps with the place where \citet{2013ApJ...765...93C} found 
extragalactic  {warm}  \htwo\ emission (see red square in Fig.~\ref{fig:map_co_40c}).
{There might be a relation between both molecular gas emissions, but the poor
spatial resolution of our data does not allow us to draw any firm conclusions.}

\subsection{Ratio between warm and cold molecular gas mass}

\begin{figure}
\centering
\includegraphics[width=8.cm]{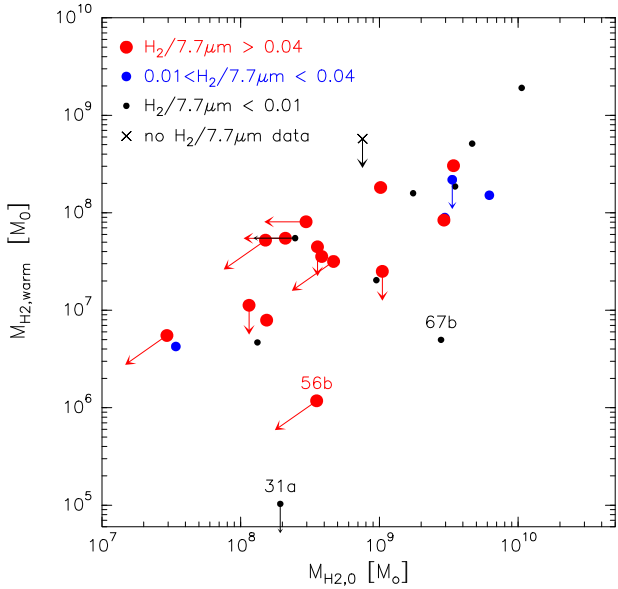}
\caption{Warm vs. cold molecular gas masses, both for the central pointing.
The green line shows the relation \mhtwowarm = 0.1 \mhtwocenter.
}
\label{fig:warm_cold_molecular_gas}
\end{figure}

Fig.~\ref{fig:warm_cold_molecular_gas} shows the relation between
the warm and cold molecular gas mass in the centres of the galaxies.
The cold molecular gas mass corresponds to the mass measured
in the central pointing and the warm molecular gas mass is corrected
 if the dimension of its emission is larger than the IRAM beam
(see Sect. ~\ref{sect:prop_warm_gas}).
There is a correlation between the warm and cold molecular gas mass,
albeit with a large scatter, with
the warm molecular gas mass being $\sim$ 1-2 orders of magnitude lower than the cold
molecular gas mass.  Part of the scatter might be due to uncertainties 
arising from the correction of the warm molecular gas mass.

\begin{figure}
\centering
\includegraphics[width=8.cm]{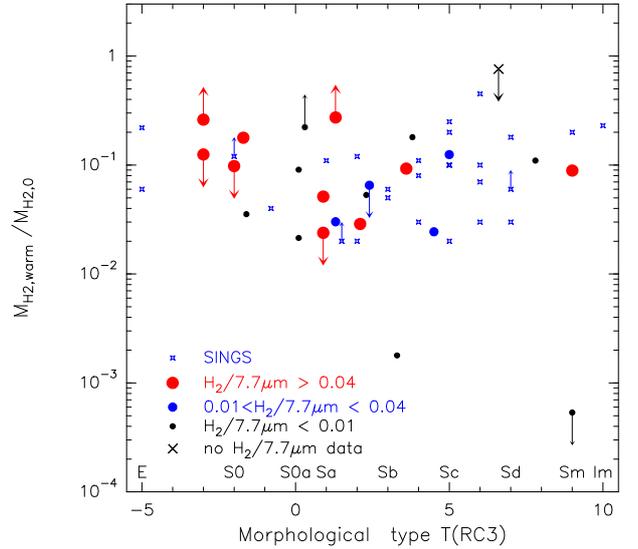}
\caption{Warm-to-cold molecular gas mass ratio as a function of morphological type.
}
\label{fig:warm_over_cold_vs_hubble}
\end{figure}

We tested the relation between the warm-to-cold molecular gas mass ratio
 and the morphological type (Fig.~\ref{fig:warm_over_cold_vs_hubble}), 
and   \htwopah\ (not shown here) and
no correlation with either of these parameters was found.
In this figure we  include 31 star-forming galaxies from the SINGS sample whose
warm molecular gas properties were studied in 
\citet{2007ApJ...669..959R}. The range of \mhtwowarm/\mhtwo\ of the SINGS galaxies
coincides very well  with the range of our sample.


\subsection{SF rates and efficiencies}
\label{sec:sfe}

We chose for our analysis the 24\mi\ luminosity as the SF tracer, after, as explained in Sect. 3.2,  excluding two galaxies where the central
24~\mi\ flux seemed to be affected by an AGN.
We calculated the SF rate (SFR) within the IRAM pointing from the 24\mi\ luminosity, 
\lmipscobeam = $\nu$ \fmipscobeam $\times 4 \pi D^2$, measured within the 
CO beam as described in Sect. 3.2,
adopting the  relation derived by 
\citet{2007ApJ...666..870C} 

\begin{equation}
SFR_{\rm 24~\mu m} (M_\sun\, {\rm yr}^{-1})= 1.27 \, 10^{-38}  L_{\rm 24~\mu m, CObeam}^{0.885}  ({\rm erg\, s^{-1}}).
\label{sfr}
\end{equation}
This relation assumes a \citet{2001MNRAS.322..231K}  Initial Mass Function (IMF)  
and was derived from a spatially resolved analysis of the HII regions in 33 galaxies
which is adequate for this study which deals with resolved regions
of the galaxies.  
We define the SFE as the ratio between SFR and molecular gas mass, 
$SFE =  SFR_{\rm 24~\mu m}$/\mhtwocenter , 
where 
 \mhtwocenter\ is the (cold) molecular gas mass
within the central pointing.

As a check, we carried out the entire analysis described in this section
also for the total SFE, derived from the total 24~\mi\  emission  and the extrapolated
molecular gas mass. For this case, we calculated the SFR 
from a relation derived for entire galaxies,
$SFR_{\rm 24~\mu m} (M_\sun\, {\rm yr}^{-1})= 8.1 \, 10^{-37}  L_{\rm 24~\mu m}^{0.848}  ({\rm erg\, s^{-1}})$
\citep{2008ApJ...686..155Z}. We obtained very similar values for the SFE (the mean values agree within
10\%) and  in general consistent results in comparison to the analysis described in the following.
Very similar values (within 10\%) and consistent trends were also obtained when calculating
the SFR from the 8~\mi\ emission, adopting the 
relation 
$SFR_{\rm 8~\mu m} (M_\sun\, {\rm yr}^{-1})= 1.2 \, 10^{-43}  L_{\rm 8~\mu m}  ({\rm erg\, s^{-1}})$
\citep{2008ApJ...686..155Z}, derived for entire galaxies.

\begin{figure}
\centering
\includegraphics[width=8.cm]{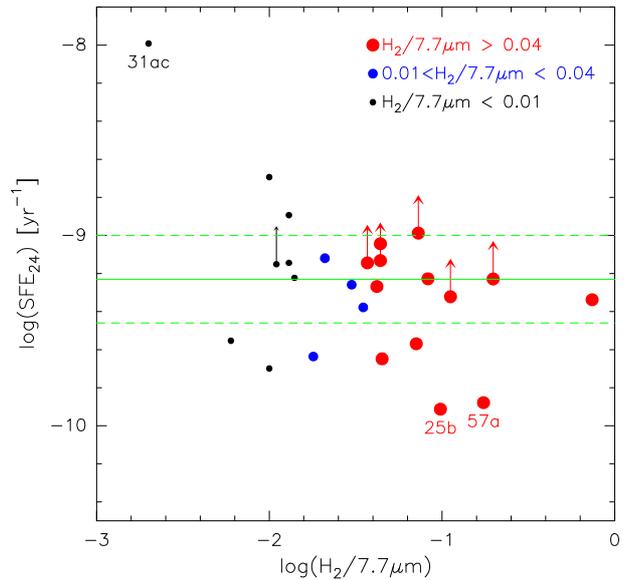}
\caption{The SFE within the central CO beam  as a function of
 \htwopah. The full green line shows the average value derived by \citet{2011ApJ...730L..13B}
 for a sample of spiral galaxies and its dispersion (dashed green line).
}
\label{fig:sfe_24_moheg}
\end{figure}
\begin{figure}
\centering
\includegraphics[width=8.cm]{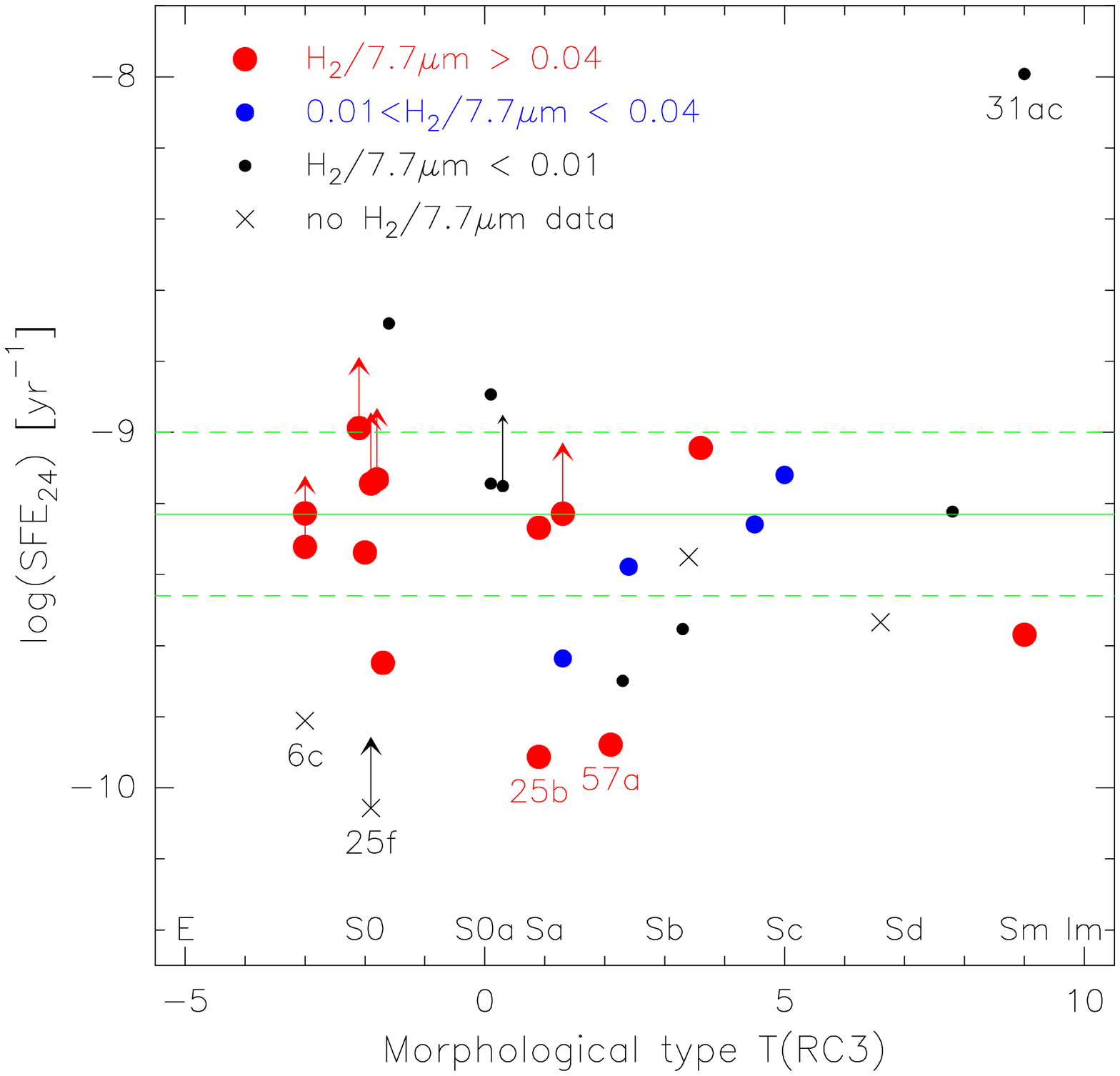}
\caption{The SFE  within the central CO beam   as a function of 
the mophological type.
The full green line shows the average value derived by \citet{2011ApJ...730L..13B}
 for a sample of spiral galaxies and its dispersion (dashed green line).
}
\label{fig:sfe_24_hubble}
\end{figure}

Figure~\ref{fig:sfe_24_moheg} shows the SFE as a function of 
\htwopah . No difference between MOHEGs and
 non-MOHEG galaxies is found. We neither find variations of the SFE
 with the IRAC color $f_{5.8}/f_{3.6}$ (not shown) or 
with the morphological type
(Fig.~\ref{fig:sfe_24_hubble}).  
It is interesting that  early-type galaxies have the same SFE as spiral
galaxies and suggests that they are
indeed  forming stars in a similar way.
This is in contrast to the results of  \citet{2013MNRAS.432.1914M} who found for early-type galaxies from
the ATLAS$^{\rm 3D}$ that they  form stars two to five times less efficiently than
a comparison sample of spiral galaxies. 
{The reason for this difference is unclear. We speculate that 
maybe} the different environment of HCGs -- gravitational interaction instead of the
relatively isolated position of ATLAS$^{\rm 3D}$  galaxies -- is the reason for
this more active SF compared to the molecular gas content.


The mean SFE of our sample is log(SFE) = $-9.2\pm 0.1$ yr$^{-1}$ which corresponds to a gas depletion time
\taudep = (SFE)$^{-1}$ of $1.6\times 10^9$  yr. This is very close to the  result of 
 \citet{2011ApJ...730L..13B}  who derived log(SFE) = -9.23, with a standard deviation of 0.23,
from  a spatially resolved analysis of 30 nearby galaxies from the HERACLES
survey  (the value is adjusted to the Kroupa IMF and no helium fraction).
In Figs.~\ref{fig:sfe_24_moheg} and \ref{fig:sfe_24_hubble} this value and its dispersion 
 are shown as green lines.

The most extreme outliers are HCG~31ac, HCG~25b and HCG~57a.
The very high SFE in the merger HCG~31ac (more than an order of magnitude higher than the mean value
of Bigiel et al.) is surprising. The  SFR derived from the 8\mi\ emission is 0.4 dex lower than that derived from the 24\mi\ for this
object, so that the discrepancy is less, but still considerable.
HCG~31ac is in a very intense  phase of SF and classified as a Wolf-Rayet galaxy
\citep{2002ApJ...575..747O}. No indication for the presence of an embedded AGN has been found
\citep{2002ApJ...575..747O}.
 Interferometric CO observations with OVRO \citep{1997ApJ...475L..21Y} showed that the brightest
CO emission is centred in the overlap region between galaxy A and C with  a mean surface density
of 30~\msun pc$^{-2}$, among the highest of our sample.
Given the high-level and spatially concentrated SF present in HCG~31ac, it might indeed be forming stars with
a higher SFE.


We also derived the SFEs for the off-centre positions in HCG~25b,
HCG~40b, HCG~57a and HCG~91a and were able to look for spatial 
variations within the galaxies. 
{Whereas in HCG~40b the position at the outskirts of the galaxy have similar values as 
in the centre, noticeable differences were found in the
other galaxies.} Lower SFEs  were  found in the South and Western side
of HCG~57a, the Northern side of HCG~ 91c, and the  Eastern position of HCG~25b.
In the case of HCG~91a, these positions are on the side where an interacting galaxy and luminous \htwo\ emission is found
(see Fig. ~\ref{fig:map_co_91a}). 
%
 {In HCG~57a the low SFE  and the spatial trend (lower SFE in the centre and West and higher SFE in the East)
 has been confirmed by  CARMA  observations \citep{alatalo_sub}. 
The higher resolution of their data allows the relation of  local SF properties 
 to the kinematical properties at those positions. The  low
SFEs in the centre and Western part  coincide with peculiar kinematics of  the CO with clear deviations from
a pure rotation which is most
likely due to interaction with HCG 57d. This coincidence suggests that  this interaction might also 
be the reason for suppression of the SF by kinematically disturbing the molecular gas and making gravitational collapse more
difficult. Alatalo et al. furthermore discuss a local decrease of the X-factor as an alternative
explanation and conclude that it is unlikely to solely explain the low SFEs.

Similarly to the case of HCG~57a, the low SFE at some positions of the other galaxies
might be due to a perturbation of the molecular
gas that makes gravitational collapse more inefficient and thereby supresses SF. This explanation is
suggestive especially in HCG~91a where the low SFE is towards the companion galaxies and  at positions
with enhanced \htwo\ emission indicating the presence of shocks.
However, we clearly need  higher resolution observations in order to draw any firm conclusion.}


\subsection{Ratio between \mhtwotot\  and \lk }
\label{sec:mhtwo_lk}

We examine the relation between the molecular gas mass
and the luminosity in the K-band, \lk, which is 
a reliable tracer for the stellar mass.
\citet{2011A&A...534A.102L} showed
 that a roughly linear relation exists
between the molecular gas mass and \lk\ for a sample of isolated galaxies so that the  ratio between both
quantities can be used as an indicator for the normalcy of the molecular gas content.
In this section we compare the values of \mhtwotot/\lk\ between MOHEG,
non-MOHEG and isolated galaxies. The mean values
 for the different ratios are
listed in Tab.~\ref{tab:mean_mhtwo_lk}.

\begin{figure}
\centering
\includegraphics[width=8.cm]{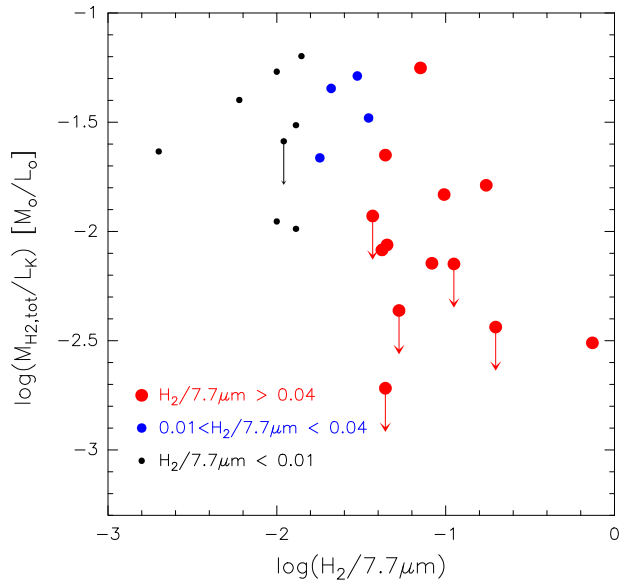}
\caption{Ratio between total extrapolated molecular gas mass, \mhtwotot, and 
luminosity in the K band, \lk,  as a function of \htwopah.
The color coding is as in Fig.~\ref{fig:mh2_vs_hubble}.
}
\label{fig:mh2_over_lk_vs_moheg}
\end{figure}

Fig.~\ref{fig:mh2_over_lk_vs_moheg} shows 
\mhtwotot/\lk\ as a function of \htwopah . MOHEG galaxies show considerably
lower values
of \mhtwotot/\lk\ than non-MOHEGs. 
Since we found no difference in the SFE between MOHEGs, non-MOHEGs and field galaxies,
this result is consistent with the findings of  \citet[][their Fig. 15]{2013ApJ...765...93C} that MOHEGs 
have the lowest specific SFRs (sSFR=SFR/\lk ).

\begin{figure}
\centering
\includegraphics[width=8.cm]{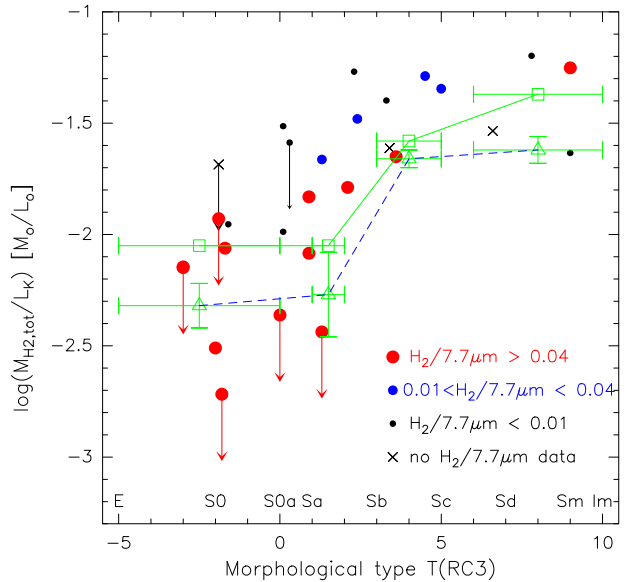}
\caption{Ratio between total extrapolated molecular gas mass, \mhtwotot, and luminosity in the K band, \lk, 
as a function of morphological type.
The green symbols denote means (triangles) and medians (squares) for a sample of isolated galaxies
 \citep{2011A&A...534A.102L}. 
The mean, respectively median, are taken for the interval of morphological types
shown by the horizontal bar.
}
\label{fig:mh2_over_lk_vs_hubble}
\end{figure}

Fig.~\ref{fig:mh2_over_lk_vs_hubble} presents  
\mhtwotot/\lk\ as a function of morphological type.
The difference in the mean values of \mhtwotot/\lk\  between MOHEG and
non-MOHEG galaxies  is at the same time a difference 
between early-type galaxies (predominatly MOHEGs) and 
spirals (predominatly non-MOHEGs).  
We can compare the mean values for
early-type galaxies separately for MOHEGs and non-MOHEGs and find  (see Tab.~\ref{tab:mean_mhtwo_lk})
that  \mhtwotot/\lk\ in
MOHEG early-type galaxies is   0.59 dex (4 $\sigma$) lower than in non-MOHEG
early-types. When only considering spiral galaxies, 
MOHEGs also have a lower \mhtwotot/\lk\  than non-MOHEGs, but the difference is
less (0.37 dex, corresponding to 2.8 $\sigma$).
Thus, there are indications, albeit for small subsamples  ($n <$ 10), that
MOHEG galaxies have a lower value of \mhtwo/\lk\ compared
to non-MOHEG galaxies, independent of their morphological type.

We include  in Fig.~\ref{fig:mh2_over_lk_vs_hubble}  the mean and median values for the sample of
isolated galaxies from \citet{2011A&A...534A.102L}, mean values  are
given in Tab.~\ref{tab:mean_mhtwo_lk}.
Spiral galaxies in HCGs have a mean  \mhtwotot/\lk\  which is 0.21 dex  (2.4$\sigma$) higher than the mean values
for isolated spiral galaxies, in agreement with the finding of Badenes-Martinez et al. (2011), who
obtained an identical mean value of \mhtwotot/\lk\ as we for a larger sample of HCG spiral galaxies. 
The difference is due to the non-MOHEG galaxies, whereas MOHEG spiral galaxies have,
within the errors, the same
\mhtwotot/\lk\  as  isolated galaxies.
Also for  early-type galaxies there is good agreement between the values
of isolated galaxies and MOHEGs whereas 
non-MOHEG galaxies tend to have a higher value by 0.46 dex (3.3 $\sigma$).

\begin{figure}
\centering
\includegraphics[width=8.cm]{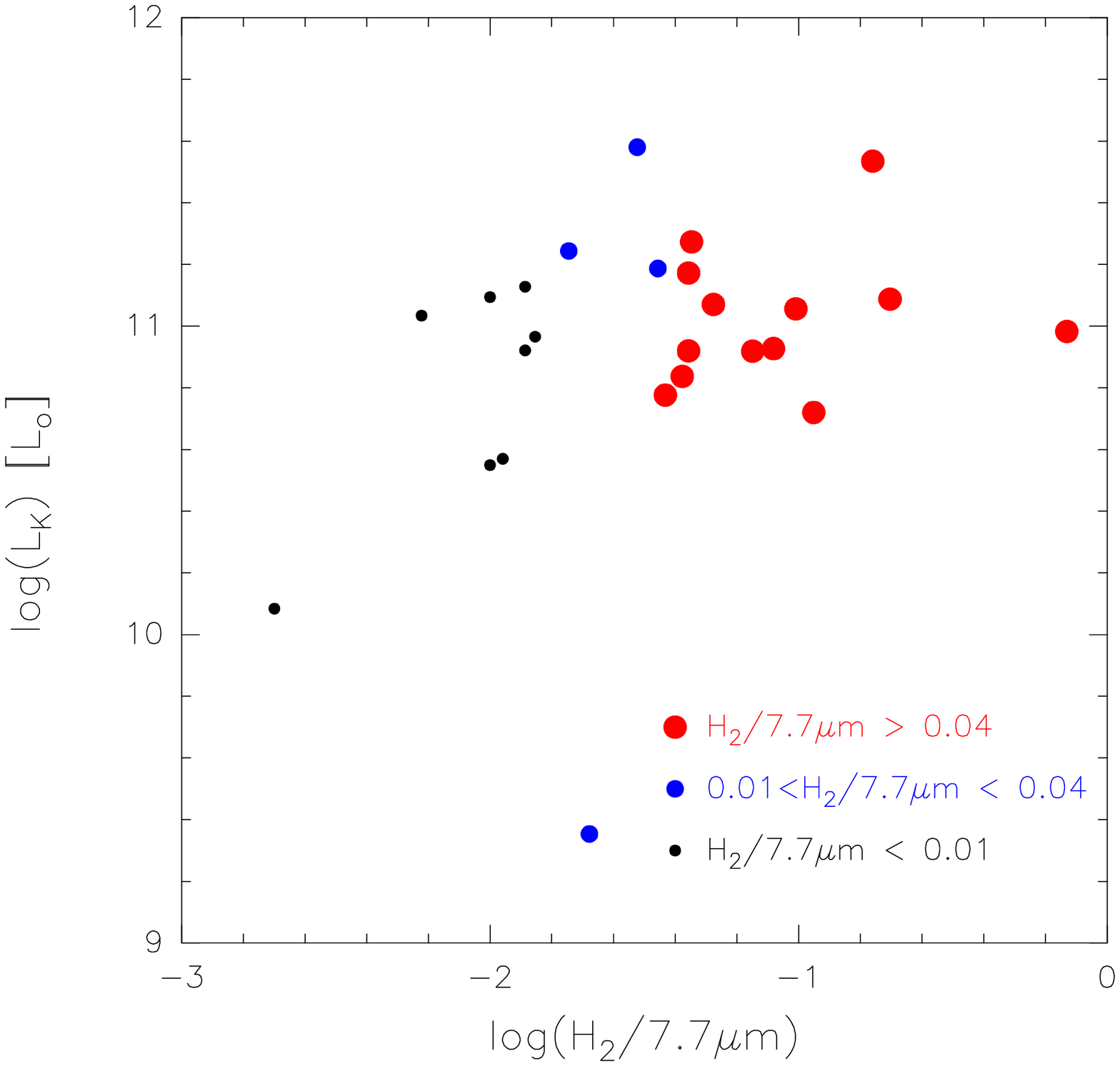}
\caption{Luminosity in the K band, \lk , as a function of \htwopah.
}
\label{fig:lk_vs_moheg}
\end{figure}

\begin{figure}
\centering
\includegraphics[width=8.cm]{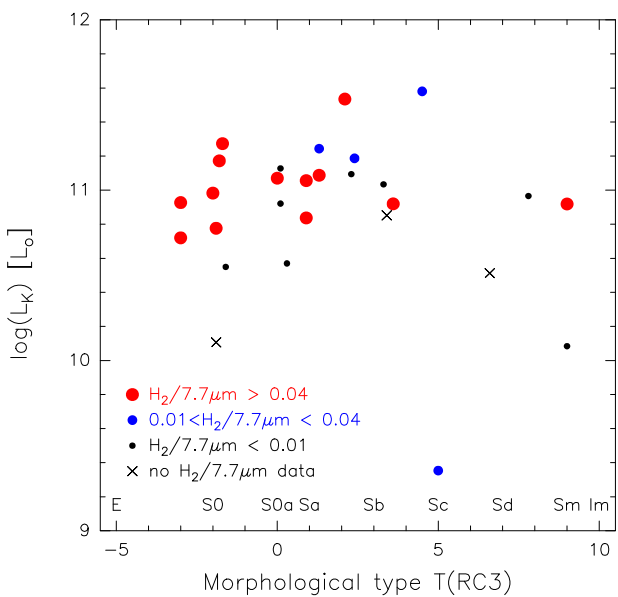}
\caption{Luminosity in the K band, \lk,  as a function of morphological type.
}
\label{fig:lk_vs_hubble}
\end{figure}

The comparatively low \mhtwotot/\lk\ ratio of MOHEGs  is {\it not} due
to a high value in \lk . In Figs.~\ref{fig:lk_vs_moheg} and
\ref{fig:lk_vs_hubble}  we show the luminosity in the K-band as a function
of \htwopah\ and morphological type, and we find no trend with
neither parameter.  Thus, MOHEG/early type galaxies have
a lower molecular gas mass, which is also evident from Fig.~\ref{fig:mh2_vs_hubble} and the results
in Sect. 4.1.

An alternative explanation for the lower \mhtwotot/\lk\ in MOHEGs could be that, instead
of having a lower molecular gas mass,  their
molecular gas is more radially extended than in non-MOHEGs. In this case our
extrapolation from the central pointing, outlined in  Sect. 4.1, would underestimate
the total molecular gas mass. {However, although we cannot exclude this possibility, and only the 
full mapping of the sample will allow us to confirm or discard this option,
there is neither observational evidence for a difference in radial extent nor did we find any indications
for such a trend when calculating the mean values of Tab.~\ref{tab:mean_mhtwo_lk}
for the restricted sample of galaxies with $f_{\rm aper} < 1.6$.}

{
\subsection{Comparison to group properties}

Apart from an analysis of the individual galaxy properties, we furthermore searched
for correlations of  \mhtwo/\lk\  and the SFE with global group properties  in order
to test whether the group dynamics or its history leaves traces on the molecular gas content
or the SF properties of the individual galaxies. Most of the group parameters that we used for our
comparison are listed in Tab.~\ref{tab:tab_group_properties}.

We looked for relations with  the median projected distance, 
the crossing time (not listed, from Hickson et al. 1992),  the HI mass and the HI deficiency,
stellar mass, baryonic mass, the dynamical mass and the compactness traced by different tracers of the
group stellar and gas content (blue luminosity, HI mass, stellar mass or the
baryonic mass)  per area covered by the group.
We approximated the baryonic mass as the sum of stellar mass and HI mass, neglecting the contribution of the molecular
gas because the latter is generally less than the HI. Furthermore, existing  observations of the molecular gas 
do not cover all members of the group so that only a lower limit to the total molecular gas content can be obtained.
In addition, we looked for trends with the evolutionary phase of the groups. We used the classification by 
\citet{2011A&A...533A.142B} who based it on the relative number of early and spiral galaxies and also
the classification by \citet{2010ApJ...710..385B} based on the HI distribution in the group. Both classifications
are in  overall agreement, but whereas the classification of \citet{2011A&A...533A.142B} is available for 16 of our
18 groups, the classification of \citet{2010ApJ...710..385B} exists only for 9 groups.

There are some trends of  group properties which are due to a relation with the morphological type of galaxies. E. g. in the late evolutionary phase,
as defined by \citet{2011A&A...533A.142B} there are more early type galaxies, which are preferably  MOHEGs and have a lower
 value of \mhtwo/\lk. In order to avoid these kind of dependencies we carried out our analysis separately for early
type ($T \le 0$) and spiral ($T \ge 1$) galaxies. For the SFE we found no trends with any of the analyzed group properties,
neither for spirals nor for early type galaxies. A similarly negative result was obtained for  \mhtwo/\lk\  in  the subsample of spiral galaxies,  but
for early type galaxies (which contained $n=12$ galaxies with 6 upper limits
in \mhtwotot) we did find some trends.
For this subsample, there might be a weak trend of  \mhtwo/\lk\  to increase with the compactness  (as traced by any of the above parameters) of the group which
will, however, need to be confirmed with a larger sample.
A stronger trend was found  for  \mhtwo/\lk\  to decrease with $M_{\rm dyn}$
 and the velocity dispersion of the group. Here, the mean \mhtwo/\lk\ for the $n=6$ galaxies with the lowest values of  $M_{\rm dyn}$ or velocity dispersion 
is 3.9$\sigma$ above the mean \mhtwo/\lk\ for the $n=6$ galaxies with the highest values.  
 This  trend could, however, be due to the fact that practically all galaxies with a 
 low $M_{\rm dyn}$ or a  low velocity dispersion are non-MOHEGs whereas practically all galaxies
with a  high $M_{\rm dyn}$ or velocity dispersion  are MOHEGs.
 It  is therefore not clear  whether the trend is really between
 the group property ($M_{\rm dyn}$ or velocity dispersion) and \mhtwo/\lk\  or rather between the group property and 
the presence of enhanced \htwo\ emission which correlates with  \mhtwo/\lk.
Thus, we conclude, that although we found some trends for the subsample of early type galaxies  studies with larger samples are needed to confirm or reject them.
}

\section{Discussion}

\subsection{Comparison to other systems}

In this paper, we have studied the properties of the cold molecular gas and compared
them to those of the warm molecular gas.  In order to draw further conclusions
about their relation and in particular the excitation mechanism of \htwo, it
is instructive to compare our results to  two well-studied systems
where  the observational situation allows a good assessment of the processes that
are at play.

HCG~92 (Stephan's Quintet, hereafter SQ) contains a kpc-size ridge 
which exhibits a very high
\htwo\ luminosity caused by a galaxy colliding with intragroup gas.
The presence of shocks can be clearly inferred from various indicators
as X-ray emission  from a hot post-shock plasma  \citep{2003A&A...401..173T, 2005A&A...444..697T},
 radio synchrotron emission \citep[e.g.,][]{2001AJ....122.2993S,2002AJ....123.2417W}
 from cosmic ray electrons 
accelerated in the shocks  or excitation diagnostics from the 
optical \citep{2003ApJ...595..665X} and mid-IR \citep{2010ApJ...710..248C} emission.
Apart from warm molecular gas, abundant cold (i.e.  CO traced) molecular gas is
present \citep{2012ApJ...747...95G}.
\citet{2009A&A...502..515G}  modelled the warm
\htwo\  emission and showed that it is most likely powered by the dissipation of kinetic 
turbulent energy produced by shocks. In order for this process to work
the energy injected on large scales by the collision must cascade on smaller scales
which is possible in a multi-phase ISM.

Another system with strong \htwo\ emission is the bridge between the Taffy galaxies \citep{1993AJ....105.1730C,2002AJ....123.1881C},
a system of two galaxies  which have just experienced a face-on collision.
The presence of strong synchrotron emission in the bridge region strongly suggests
the presence of shocks accelerating cosmic rays \citep{2010A&A...524A..27L}.
\citet{2012ApJ...751...11P}  carried out a study of the warm and cold molecular gas in the 
bridge and concluded that, similar to SQ, the most likely 
excitation mechanism are shocks powered by kinetic energy injected in the collision.

In both systems the observations clearly show that collisions have taken place and
have produced shocks. These processes 
provide the power for the warm \htwo\ emission. 
Is it possible that in our systems similar processes are taking place?

Our sample lacks the spatial resolution in CO and the large amount of ancillary
data necessary to directly see  whether a gas dynamic interaction and
shocks are present. However, we can at least test whether such a scenario would be 
possible  from an
energetical point of view.
If the dominant energy input is given by shocks and turbulent energy, then
these energy sources must be able to balance the cooling rate which is given by
the \htwo\ luminosity. The turbulent heating rate can be written as
$\Gamma_{\it turb} = \frac{3}{2} M_{\it H_2, warm} \sigma_{\it T}^3/l$
\citep{1999ApJ...524..169M}
where $\sigma_{\it T}$ is the \htwo\ velocity dispersion and   $l$ the characteristic length over
which the velocity dispersion is measured. 
%
Following \citet{2009A&A...502..515G}, we can then make a rough estimate  whether such a process is energetically
possible by calculating the turbulent heating rate and requiring

\begin{equation}
\frac{3}{2} \frac{\sigma_{\it T}^3}{l}  \ge \frac{L(H_2)}{M_{\it H_2,warm}}.
\label{eq:cooling-heating}
\end{equation}
We cannot determine $\sigma_{\it T}$ from the observations {because disentangling the
turbulent velocity field from the galaxy rotation and tidal effects requires high spatial resolution observations},
but we can compare our values of
$L(H_2)/M_{\it H_{2,warm}}$ {(where we compiled $L(H_2)$ from Tab. 2 in  \citet{2013ApJ...765...93C})} to those of SQ and the Taffy bridge.
The ratio $L(H_2)/M_{\it H_{2,warm}}$  for the galaxies in our sample
ranges from 0.03 to 0.3  \lsun \msun$^{-1}$. For the  ridge  in SQ the ratio is
 $L(H_2)/M_{\it H_{2,warm}} = 0.1 \pm 0.02 $ \lsun \msun$^{-1}$  \citep{2012ApJ...747...95G} 
 and for the Taffy bridge  $L(H_2)/M_{\it {H_2,warm}} = 0.08 $ \lsun \msun$^{-1}$ \citep{2012ApJ...751...11P} ,
 i.e. in both systems the values are in the range of our sample.
 
 In a further crude test to examine the energy that could be available in shocks
 we compare the kinetic energy in the cold molecular gas \mhtwocenter\ 
to   the cooling rate  $L(H_2)$ by adopting a corresponding relation as
Eq.~\ref{eq:cooling-heating}  for the cold molecular gas. 
The spatial resolution of our data does not allow
us to derive the {turbulent} velocity dispersion (the line width is mainly driven by rotation) or dissipation scale $l$ {directly} from
the CO spectra and therefore we have to rely on plausibility arguments and a comparison
to SQ and the Taffy bridge. 
The ratios $L(H_2)$/\mhtwocenter\ for our galaxies range between 0.003 and 0.04 \lsun \msun$^{-1}$ which 
is somewhat lower than
the corresponding values for SQ (0.03  \lsun\msun$^{-1}$) and the Taffy bridge (0.06  \lsun \msun$^{-1}$).
The higher values for the two latter objects is due to their high warm-to-cold mass ratio
(\mhtwowarm/\mhtwo = 0.3 for SQ and 0.7 for the Taffy bridge).

In order to quantify our estimate further, we calculate from  Eq.~\ref{eq:cooling-heating},
assuming that the kinetic energy dissipates in our 
galaxies over a  scales of 1 kpc (a size plausible for a collision with an
intragroup  gas cloud),  the
velocity dispersion necessary to power the \htwo\ luminosity.
We derive a range of $\sim$ 20-55 \kms , 
which is much smaller than our CO line widths and thus easily
 possible to be present.

Therefore, even though spatial resolution of the CO data is not sufficient 
to firmly decide where the perturbations in the CO and the \htwo\ emission come
from,  we can conclude that shocks produced by a collision with intragroup gas are
an energetically feasible option.

\subsection{{The role of CO  for the evolution of galaxies in HCGs}}

The results of \citet{2013ApJ...765...93C} showed  that MOHEG galaxies in HCGs are in a short
transition phase between star forming, blue (spiral)  galaxies and quiescent, red
(early-type) galaxies. 
This might be a coincidence or it might indicate that the process responsible for the
elevated \htwo\ emission  is somehow related to  the morphological transition.
\citet{2013ApJ...765...93C}
discuss different possible processes that could produce the shocks  responsible for the elevated H$_2$ emission,
and conclude that the most likely is an interaction with the intergalactic medium, possibly
in the form of viscous stripping. 
How do our results fit into this picture?

Apart from the energetical argument discussed in the previous subsection,
our CO measurements show several pieces of evidence that are consistent with this  general scenario.
The CO distribution in the  mapped objects  is in three cases clearly  asymmetric (HCG~91a, HCC~57a, HGC~25b)
with strong, narrow lines towards a position where warm \htwo\ was detected (HCG~91a).
There is evidence for  CO emission with a high scale-height above the disk in HCG~40b, {also
towards a position with enhanced \htwo .}
The low SFE in some objects and positions
could be an indication for the presence of diffuse {kinematically perturbed} molecular gas, not
actively forming stars.
As far as the kinematics of the spectra is concerned, we found that several of the
MOHEG galaxies had unusually wide lines (up to $\sim$ 1000 \kms\ at 0-level) with 
indications of several components. 
Even though the low spatial resolution makes it difficult to 
draw firm conclusions about what causes all these features, 
it is clear that the molecular gas shows abnormalities which might be
related to the processes causing the enhanced H$_2$ emission and the
 intermediate IRAC colours.
Indeed,  higher resolution ($\sim 5$ \arcsec ) CARMA observations of HCG~57a have 
confirmed the peculiarities of the molecular gas distribution and
kinematics and have shown that the motion of part of the molecular gas does
not follow the general rotation {but is rather due to and inflow or outflow }\citep{alatalo_sub}.

Not only the molecular gas distribution and its kinematics, but also the total
molecular gas mass shows differences between MOHEGs and non-MOHEGs.
We found that both the  molecular gas mass as well as 
\mhtwotot/\lk\ in 
MOHEGs is low compared to non-MOHEG galaxies (both early-types and
spirals).  
This could indicate a real decrease in the molecular gas mass or -- as we
only measured the centre and then extrapolated to the total molecular gas mass assuming
a radial scale-length -- a more radially extended emission of the molecular gas.
Models suggest that  S0 galaxies in groups are transformed spiral galaxies
\citep[e.g.][]{2000Sci...288.1617Q}. The fact that the stellar masses in both morphological groups are 
the same, is consistent with this picture, because the stellar mass does not
change on short time-scales.  In this picture, 
{the lower  molecular gas mass of  MOHEGs compared to
non-MOHEGs}   would indicate that a 
decrease in the molecular gas mass might be an important step in
the transition from spirals to early-type galaxies.
The low sSFR found in early-type HCGs galaxies \citep{2013ApJ...765...93C,2011A&A...533A.142B}
would therefore be caused by a decrease in neutral gas, then molecular gas, followed by
a decrease in the SF activity. Repeated
harassment and tidal stripping could transform late-type disks to early-type disks rapidly.



\section{Summary and conclusions}

We studied the warm and cold molecular gas mass in a sample of 31 galaxies in 
18 HCGs.  The cold molecular gas mass, \mhtwo, is traced by the CO(1-0) emission which was mostly
observed by us  at the IRAM 30m telescope (for 20 galaxies) and partly taken from the
literature (for 11 galaxies). Most galaxies were observed only at their central
position,  and we extrapolated this measurement to the total  molecular gas mass 
in the disk, \mhtwotot, following the prescription of \citet{2011A&A...534A.102L}.
Four objects were mapped at several positions over their disks.
27 galaxies have spectroscopic data from the \spitzer\ which allowed us to derive
the luminosity and mass,  \mhtwowarm, of the warm molecular gas mass.
About half of the sample (14 objects)  has a high \htwo\ luminosity compared to the PAH emission
at 7.7~\mi, well above the level
expected from SF and most likely excited by shocks, and are classified as
MOHEGs. Nine galaxies have levels of their \htwo\ emission expected from SF 
and the remaining four galaxies have elevated  \htwo\ emission at an intermediate level.

The goal of this study was to characterized the properties of the cold molecular gas in these
objects, compare them to those of the warm molecular gas and to look
for differences between MOHEG and non-MOHEG galaxies that could shed light on the
mechanisms that promote the morphological evolution of galaxies in HCG.

The main results of our analysis are:

\begin{itemize}

\item 
The ratio between the warm and cold molecular gas mass is between 
a few percent and $\sim$ 50 \%, covering the same range as found for SINGS galaxies
\citep{2007ApJ...669..959R}.
No trend with neither the  morphological type  nor the
MOHEG parameter, \htwopah , has been found. 

\item  Some galaxies, most of them MOHEGs, show very broad spectra (up to
1000 \kms ) with irregular shapes and indications of several components,
which cannot be explained by pure rotation. This might be 
indicative of either a projection effect, adding several components within
the same beam, or a perturbation of the molecular gas motions in the galaxies.

\item In the galaxies where the CO emission has been mapped, we found evidence for an asymmetric
   distribution (in HCG~25b, HCG~91c, HCG~57a) and for off-planar emission (in HCG~40c). 
   Both the asymmetry and the off-planar emission
   are in the direction of a position where off-centre \htwo\ emission was found. 
   
\item The SFE 
in the central positions lies for most cases in the range 
found for spiral galaxies \citep{2011ApJ...730L..13B}.  In a few objects and at some off-centre positions,
the SFE is  considerably lower which might indicate the
presence of diffuse {or kinematically perturbed} gas not participating in SF.
We found no trend of the SFE with morphological type, \htwopah\ {or any tested group parameters (HI mass and deficiency, velocity
dispersion, dynamical and stellar masses or compactness}). Early-type
galaxies in HCGs seems to be forming stars with the same efficiency as spiral galaxies.

\item We found evidence for a lower cold molecular gas mass in MOHEG  galaxies, {compared to non-MOHEG galaxies}.
Both the molecular gas mass, \mhtwotot, as well as the ratio \mhtwotot/\lk\ 
was lower than the corresponding value in non-MOHEG galaxies (both for the entire sample and 
separately for early-types and for
spirals). 
{We found a trend  for galaxies in groups with a low dynamical mass or velocity dispersion to have a higher \mhtwotot/\lk\  and to be
more likely a non-MOHEG.}

\end{itemize}

These results indicate pecularities in the kinematics, distribution and total content of the
cold molecular gas in MOHEGs, compared to non-MOHEG galaxies.
The decrease in the molecular gas content is expected to cause a  decrease in the SFR,
{which could produce} an evolution of galaxies from blue to the red sequence.
{Thus, our CO data are
consistent with a picture of morphological evolution of galaxies in HCGs in which MOHEGs
play a central role as galaxies being in the process of evolution.
Furthermore, our data  hint to a possible link of these signs of evolution to
kinematical perturbations of the molecular gas which appears energized by
the dissipation of turbulent energy. However,  much higher spatial 
resolution is needed to prove this picture by locating those perturbations and separating
the contributions from tidal effects, in/outflows, and turbulence.}


\bibliographystyle{aa}
\bibliography{Biblio_hcg} 

\begin{acknowledgements}
This work has been supported by the research projects   AYA2011-24728 from the Spanish Ministerio de Ciencia y Educaci\'on and the Junta de Andaluc\'\i a (Spain) grants FQM108.
It is based on observations with the Instituto de Radioastronom\'ia Milim\'etrica IRAM 30\,m.

This research has made use of the NASA/IPAC Extragalactic Database (NED) which is operated by the Jet Propulsion Laboratory, California Institute of Technology, under contract with the National Aeronautics and Space Administration. We also acknowledge the use of the HyperLeda database (http://leda.univ-lyon1.fr).
\end{acknowledgements}

\appendix


\section{Spectra of detected positions}


\begin{figure*}
\centerline{
\includegraphics[width=3.5cm,angle=-0]{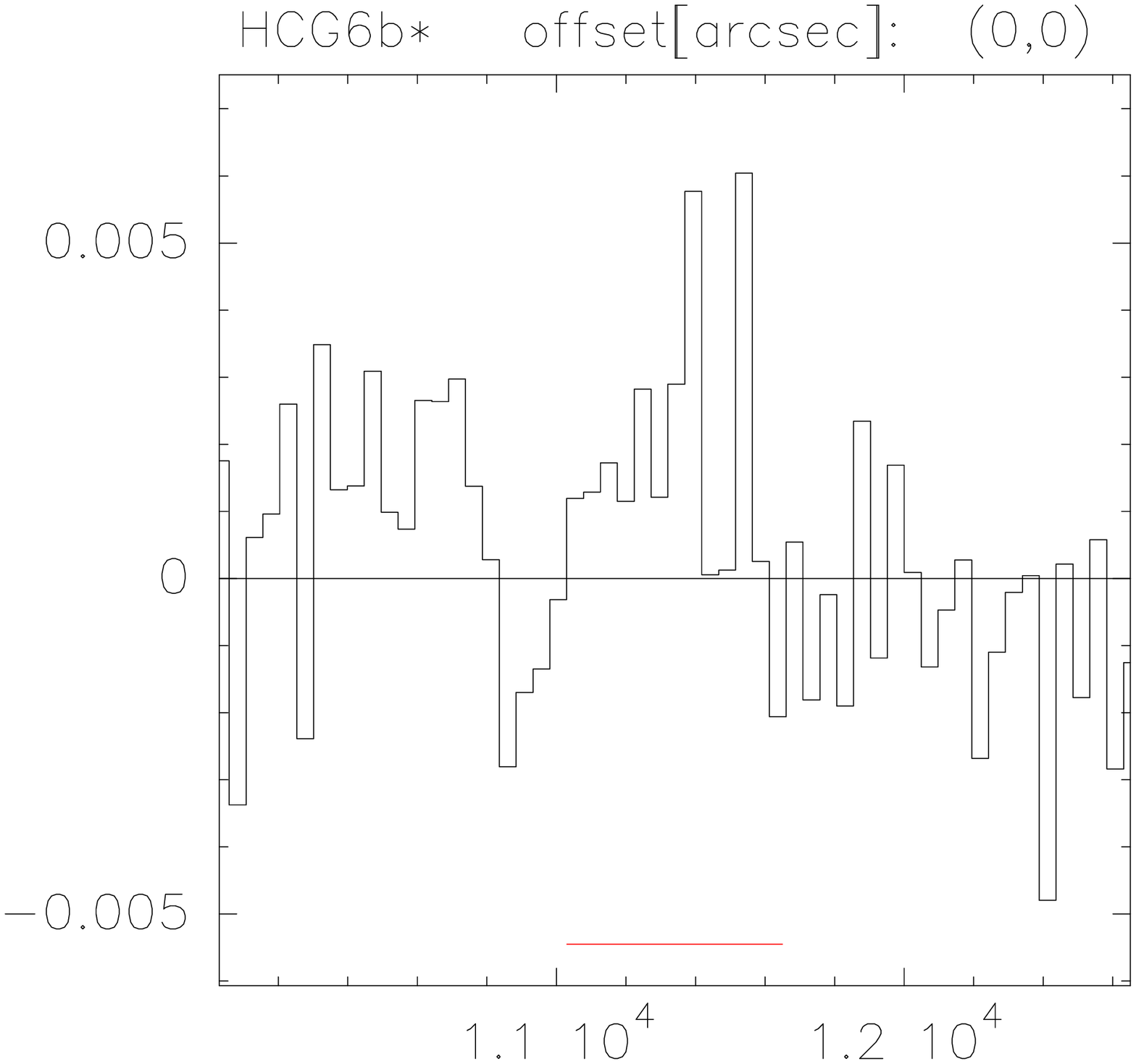}
\includegraphics[width=3.5cm,angle=-0]{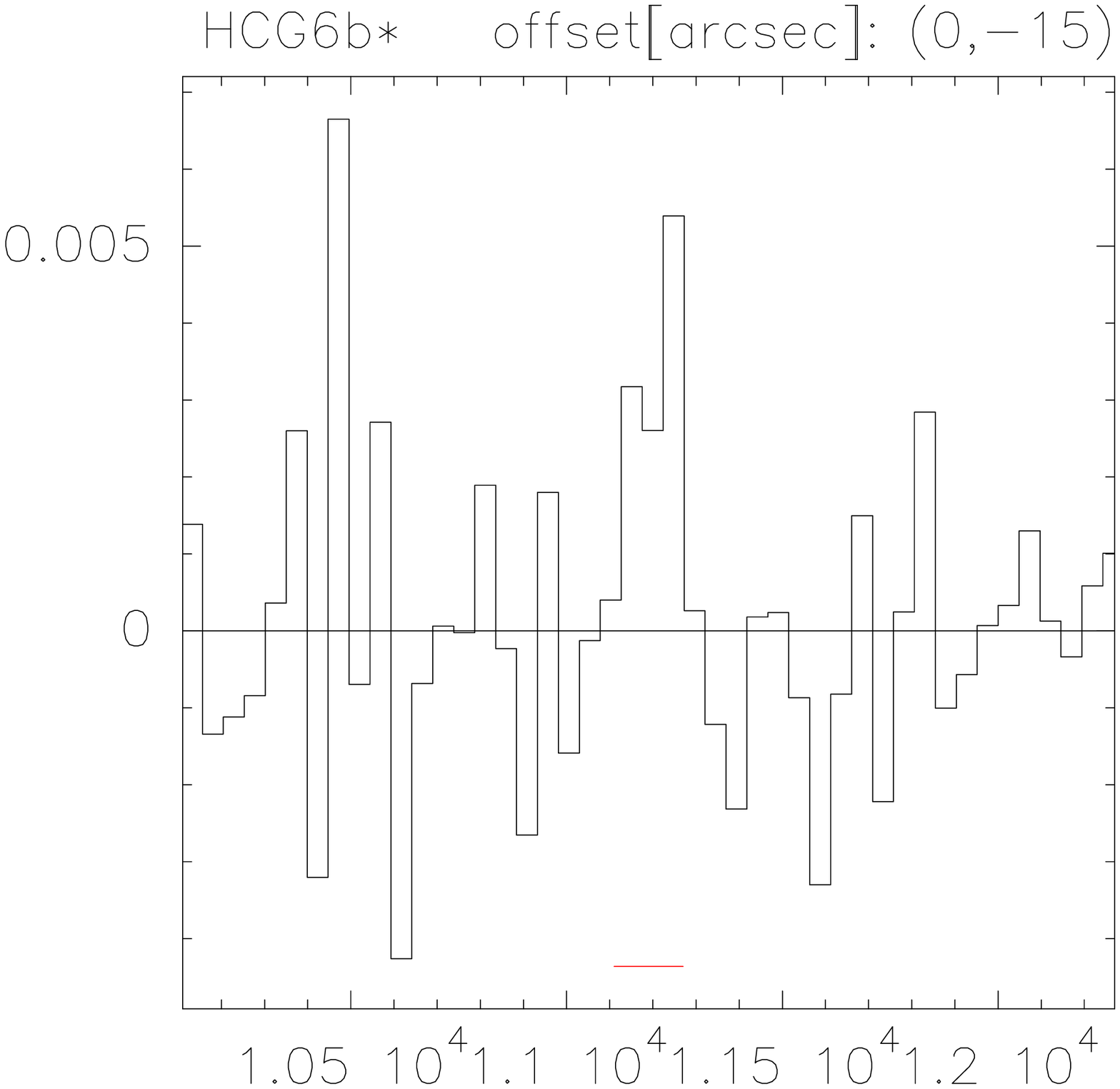}
\includegraphics[width=3.5cm,angle=-0]{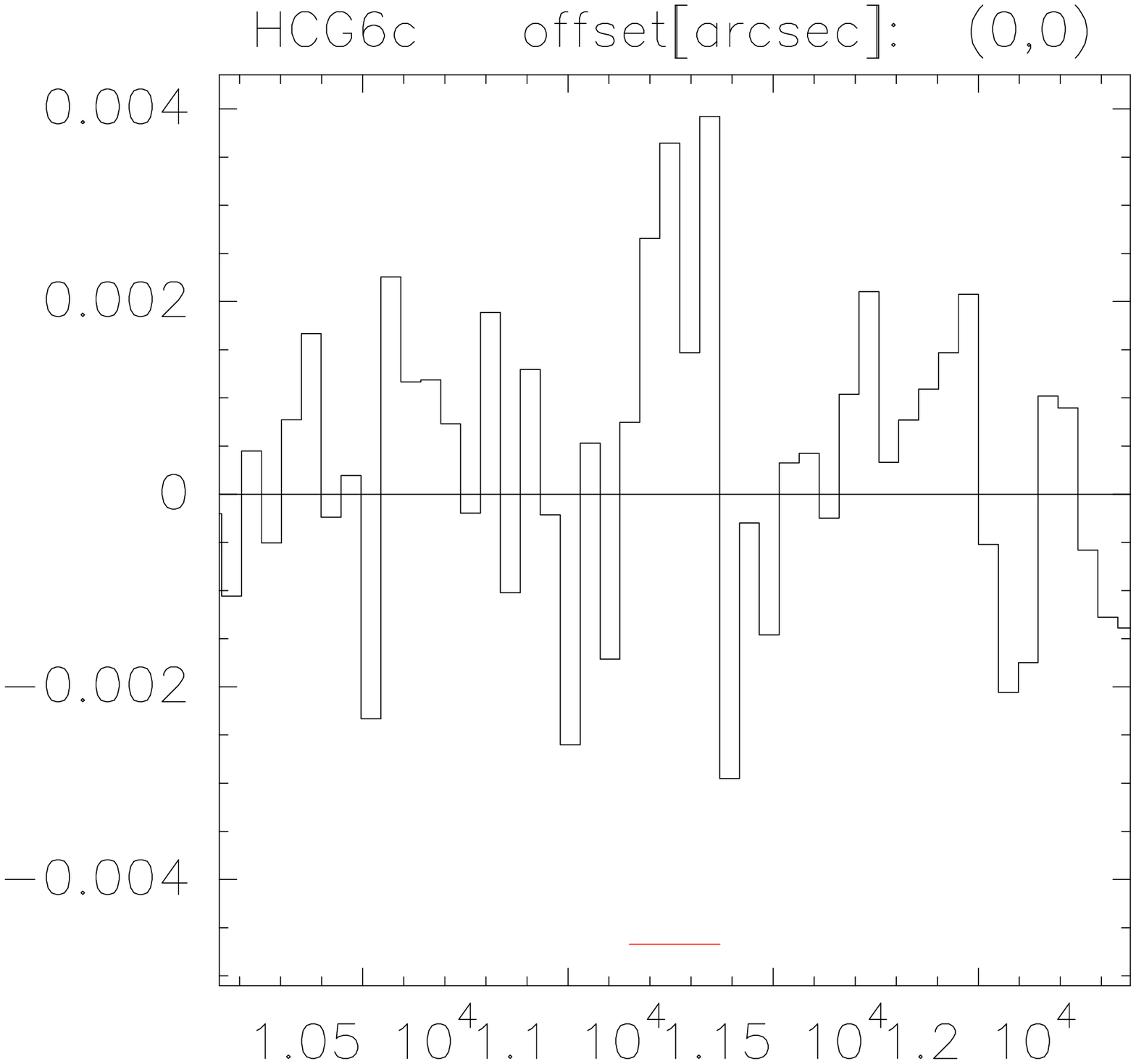}
\includegraphics[width=3.5cm,angle=-0]{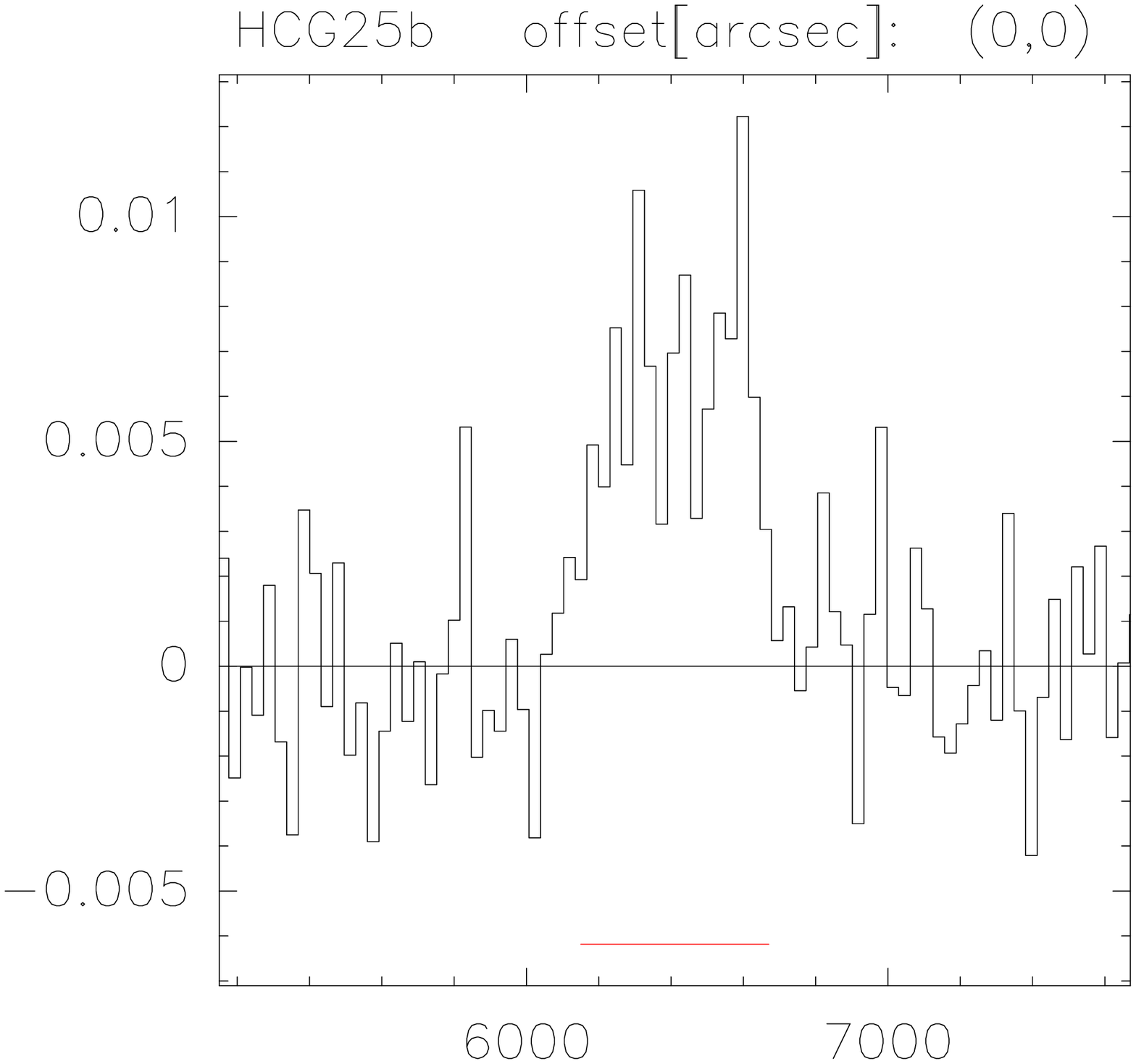}
\includegraphics[width=3.5cm,angle=-0]{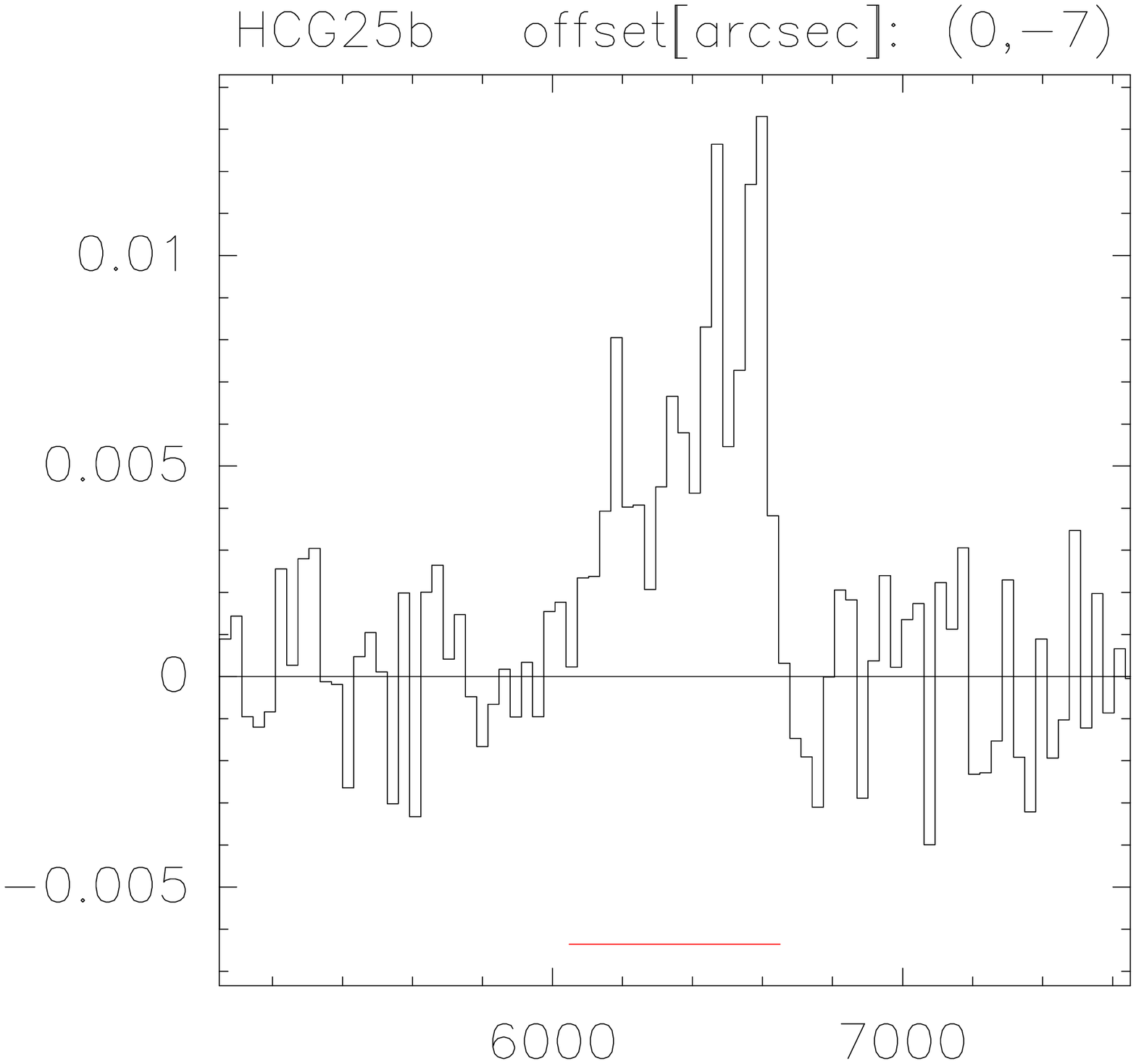}
}
\centerline{
\includegraphics[width=3.5cm,angle=-0]{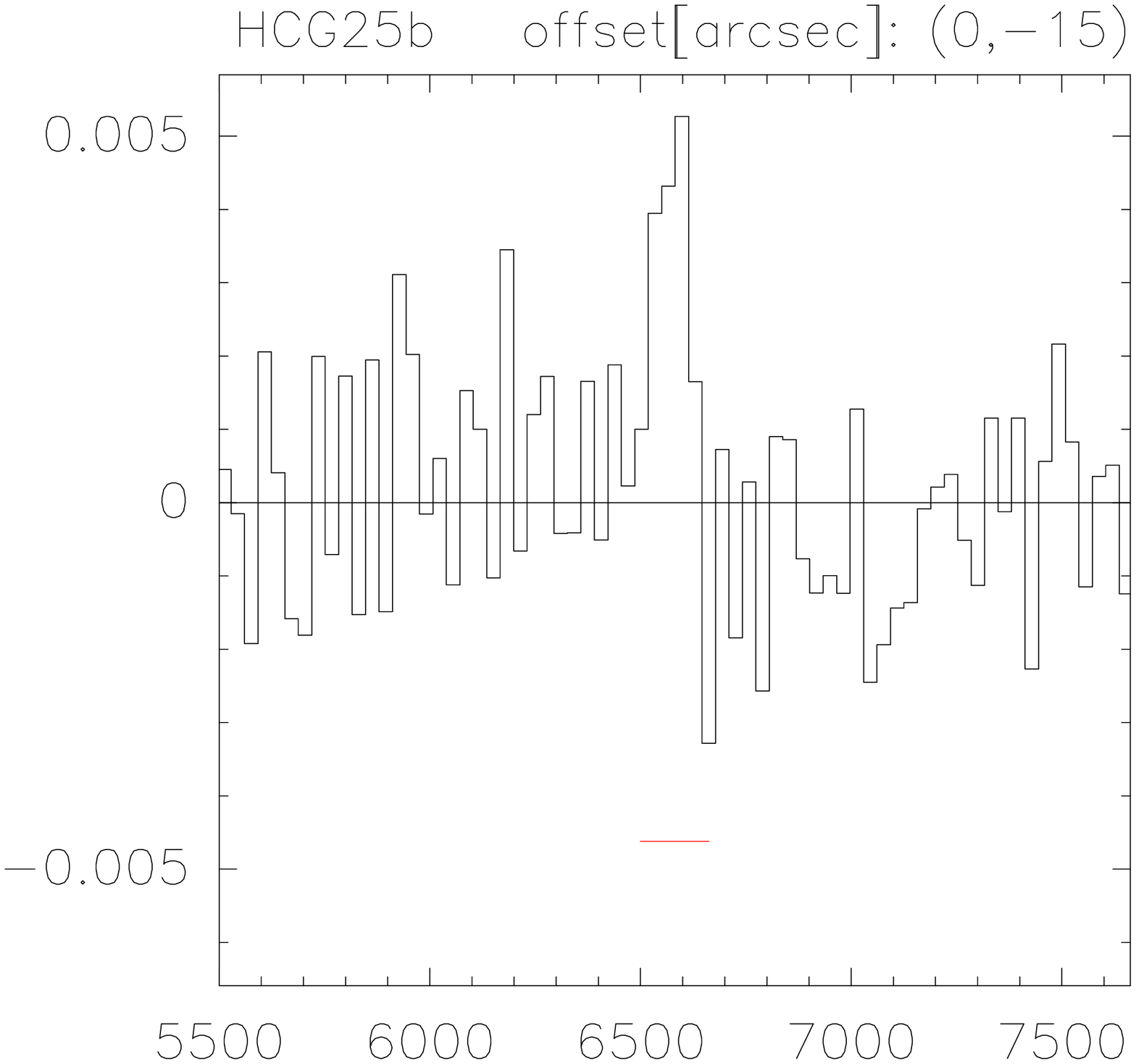}
\includegraphics[width=3.5cm,angle=-0]{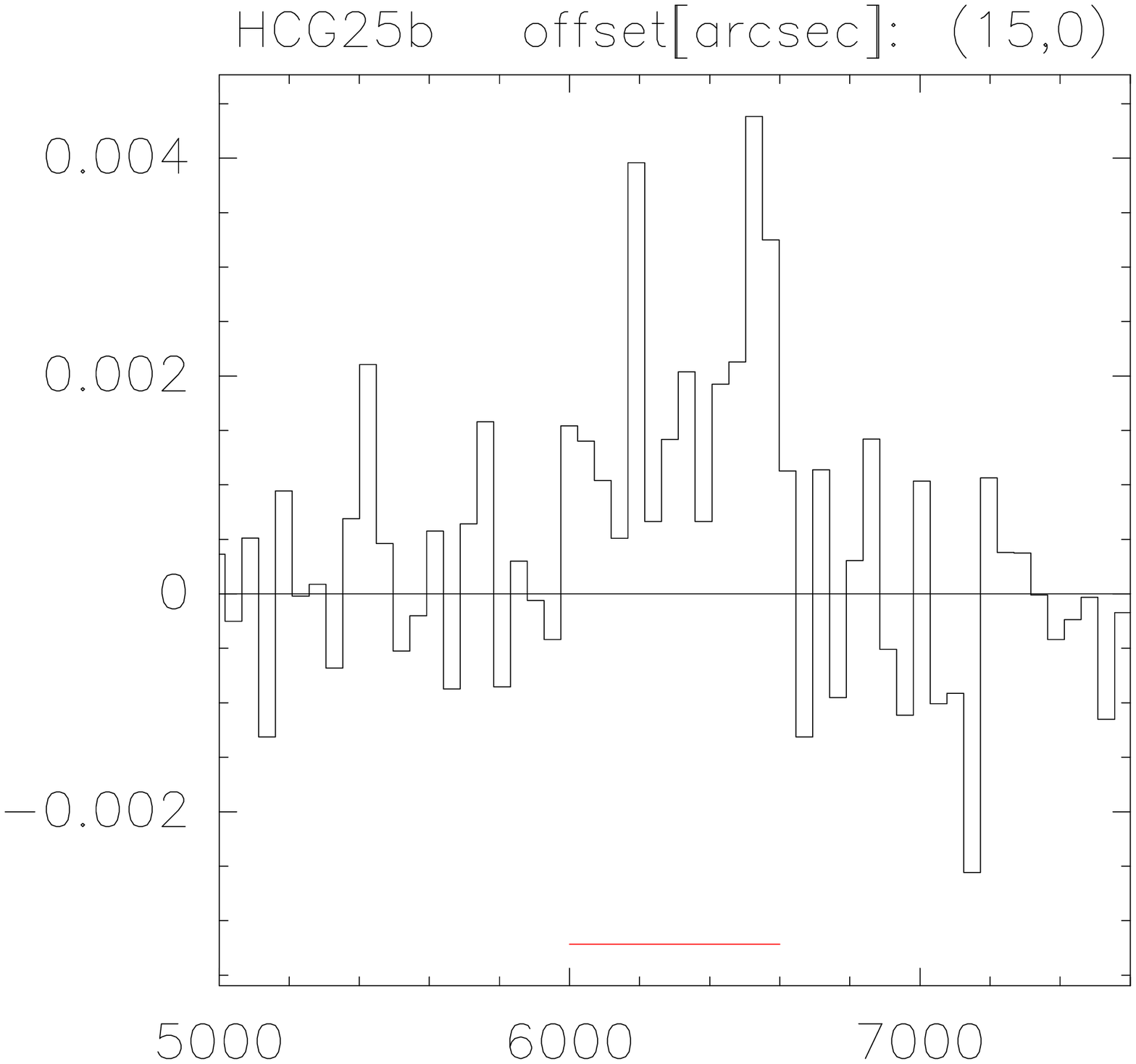}
\includegraphics[width=3.5cm,angle=-0]{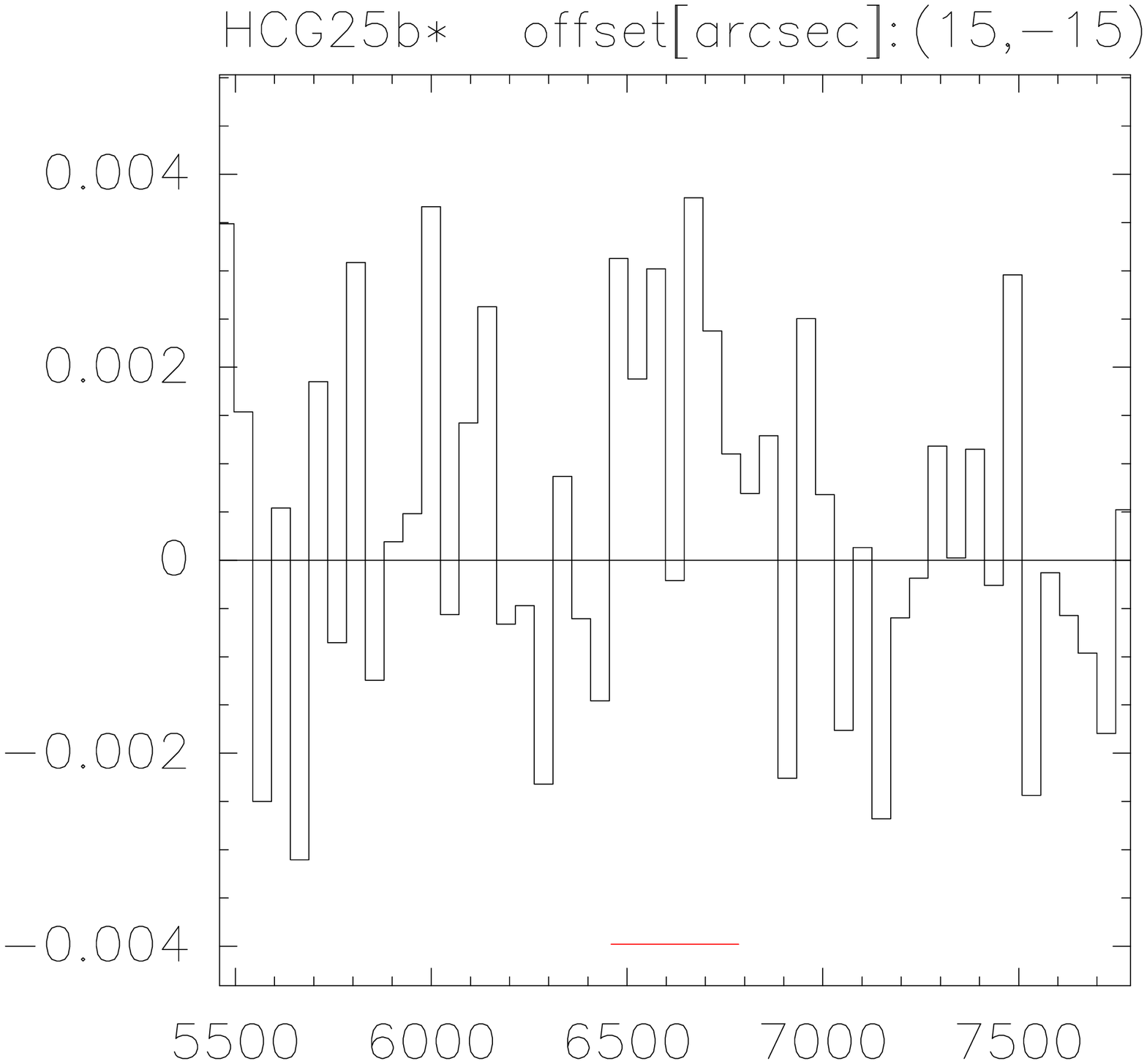}
\includegraphics[width=3.5cm,angle=-0]{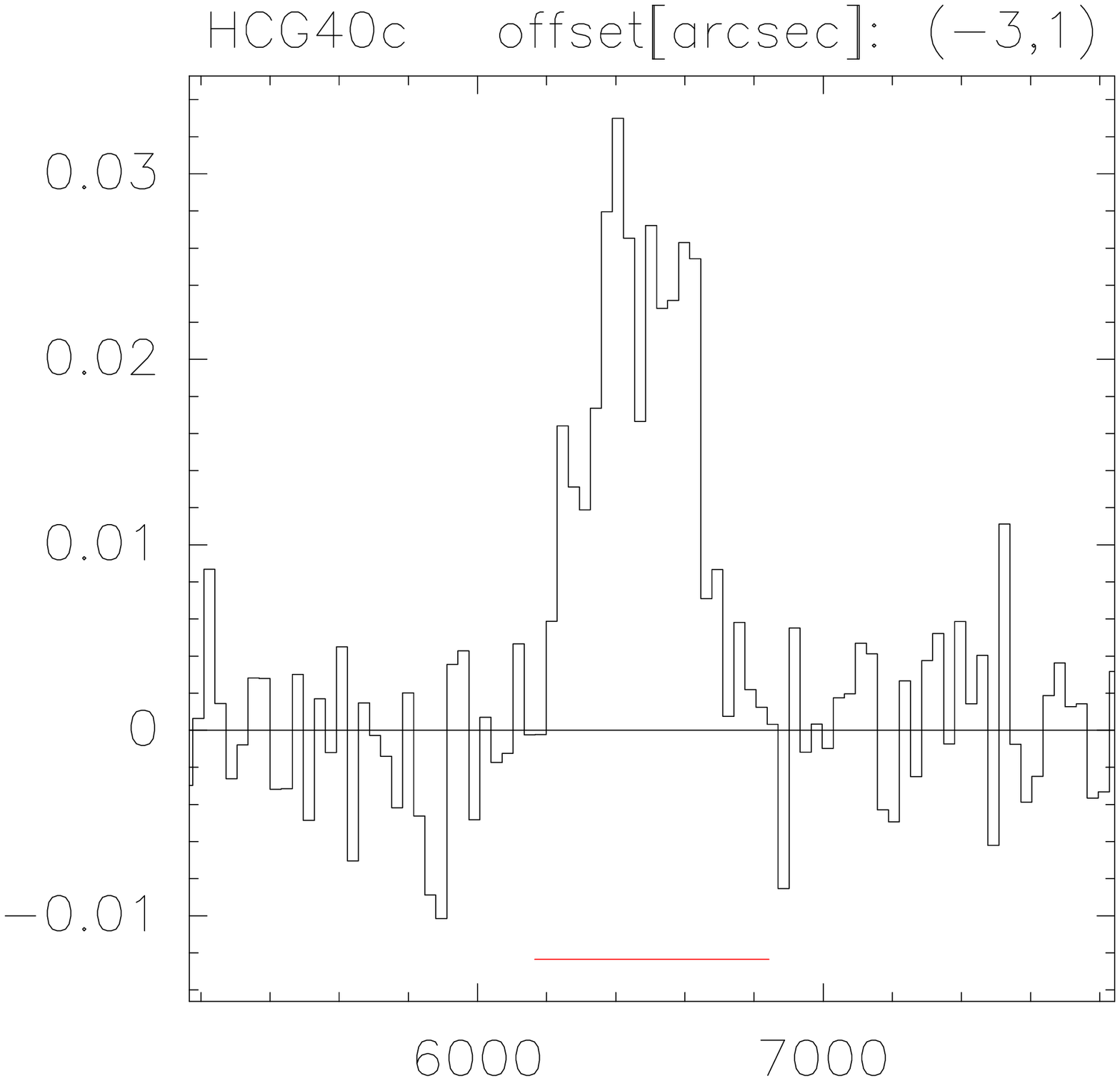}
\includegraphics[width=3.5cm,angle=-0]{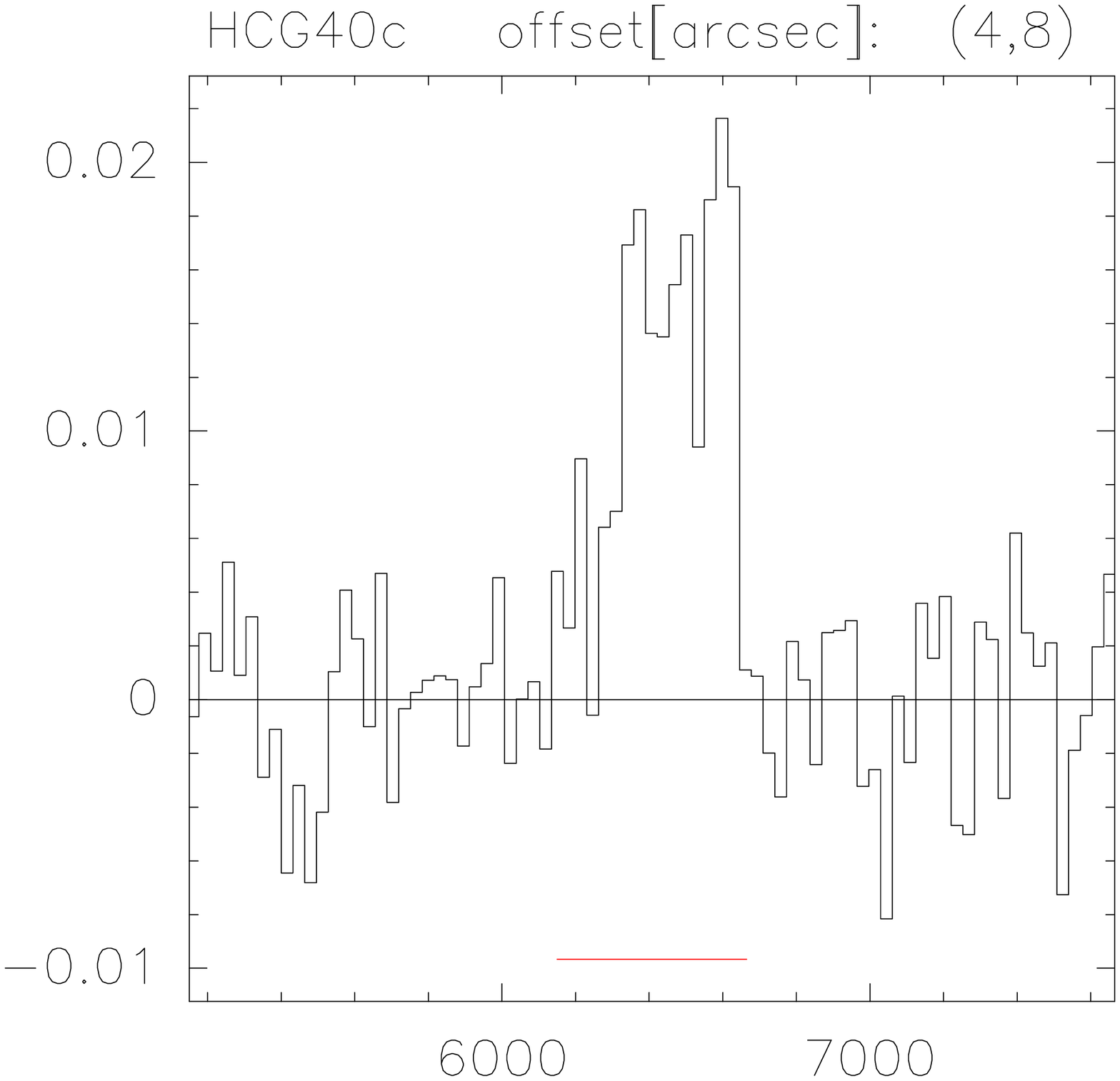}
}

\centerline{
\includegraphics[width=3.5cm,angle=-0]{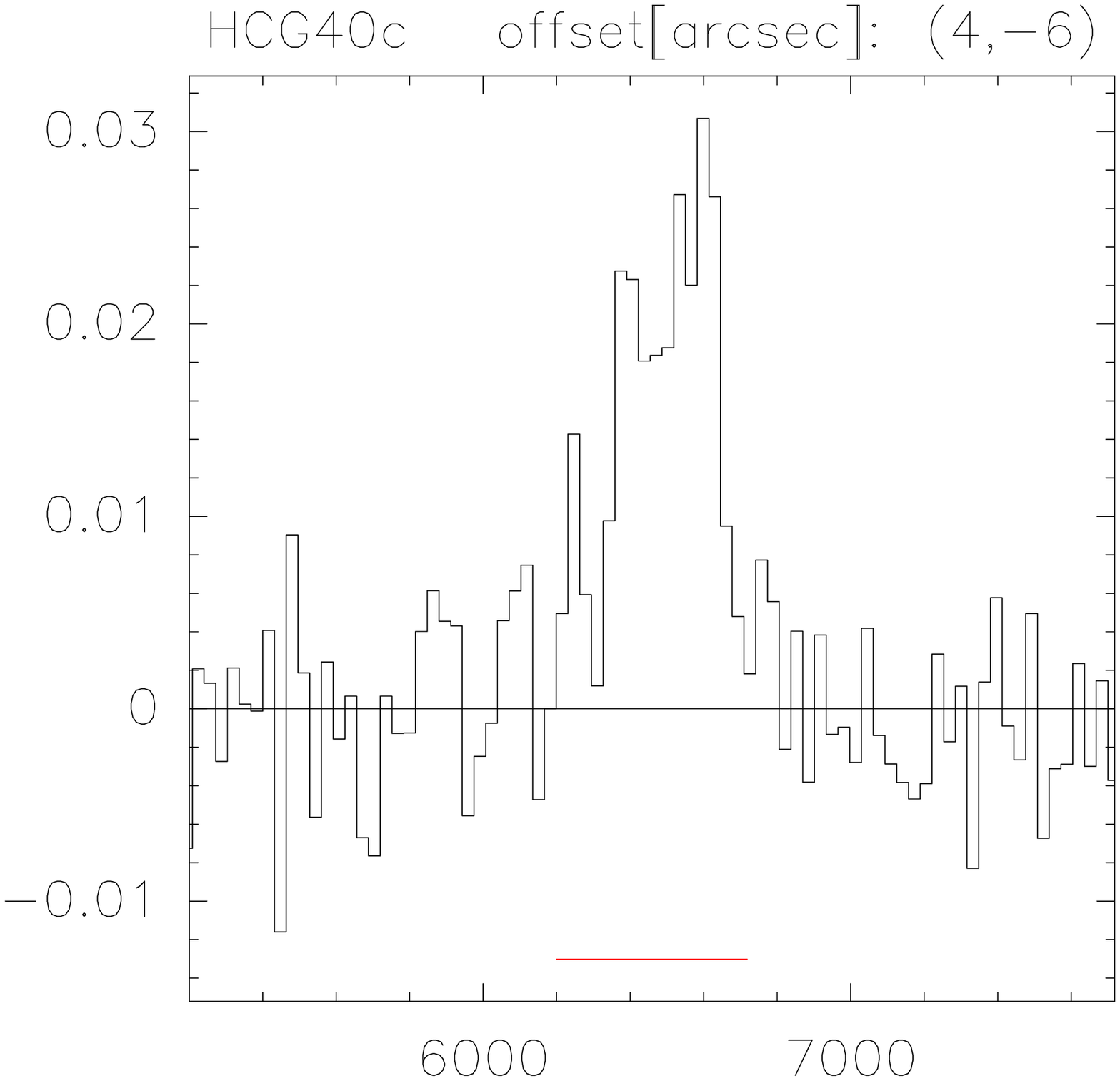}
\includegraphics[width=3.5cm,angle=-0]{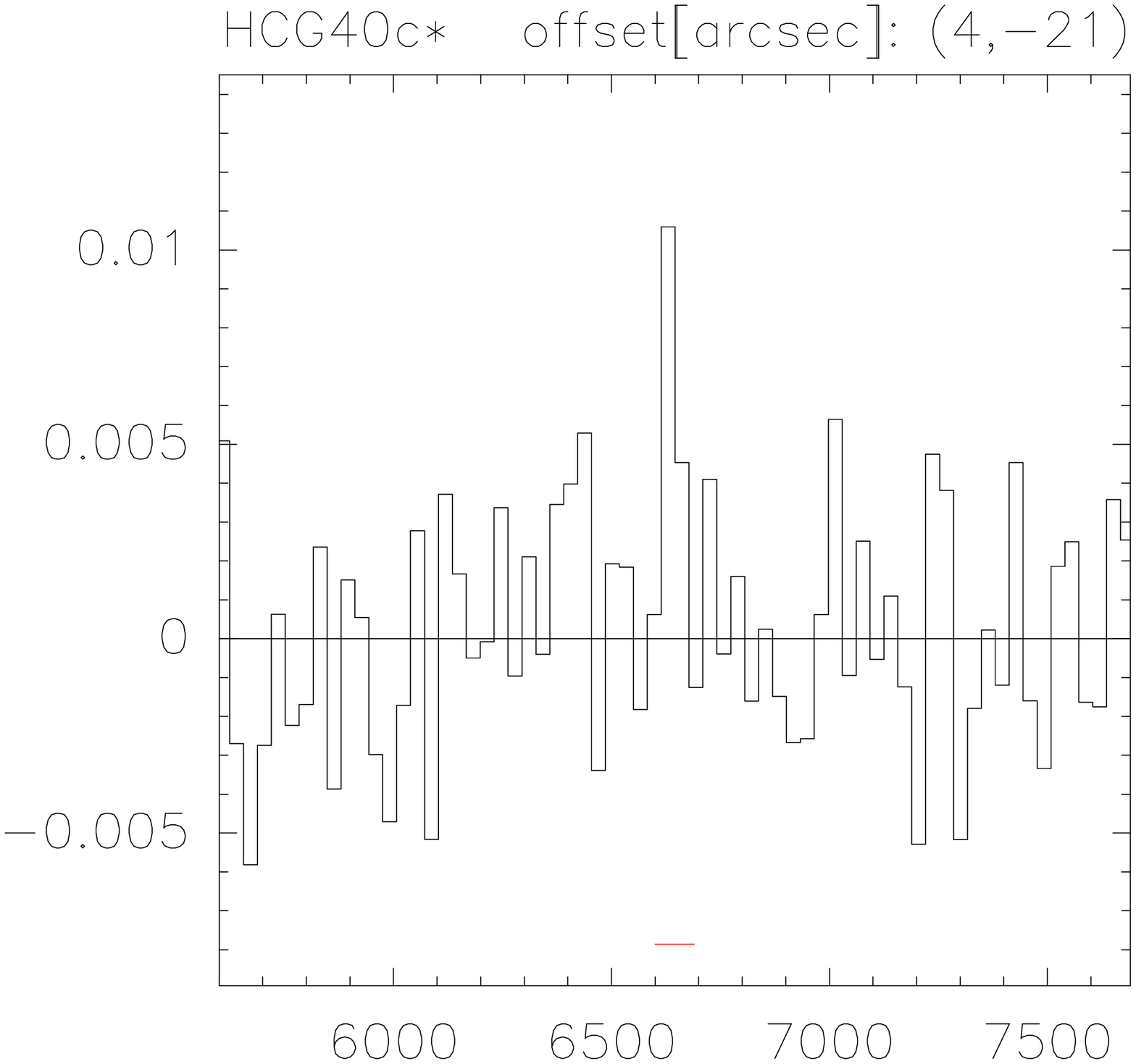}
\includegraphics[width=3.5cm,angle=-0]{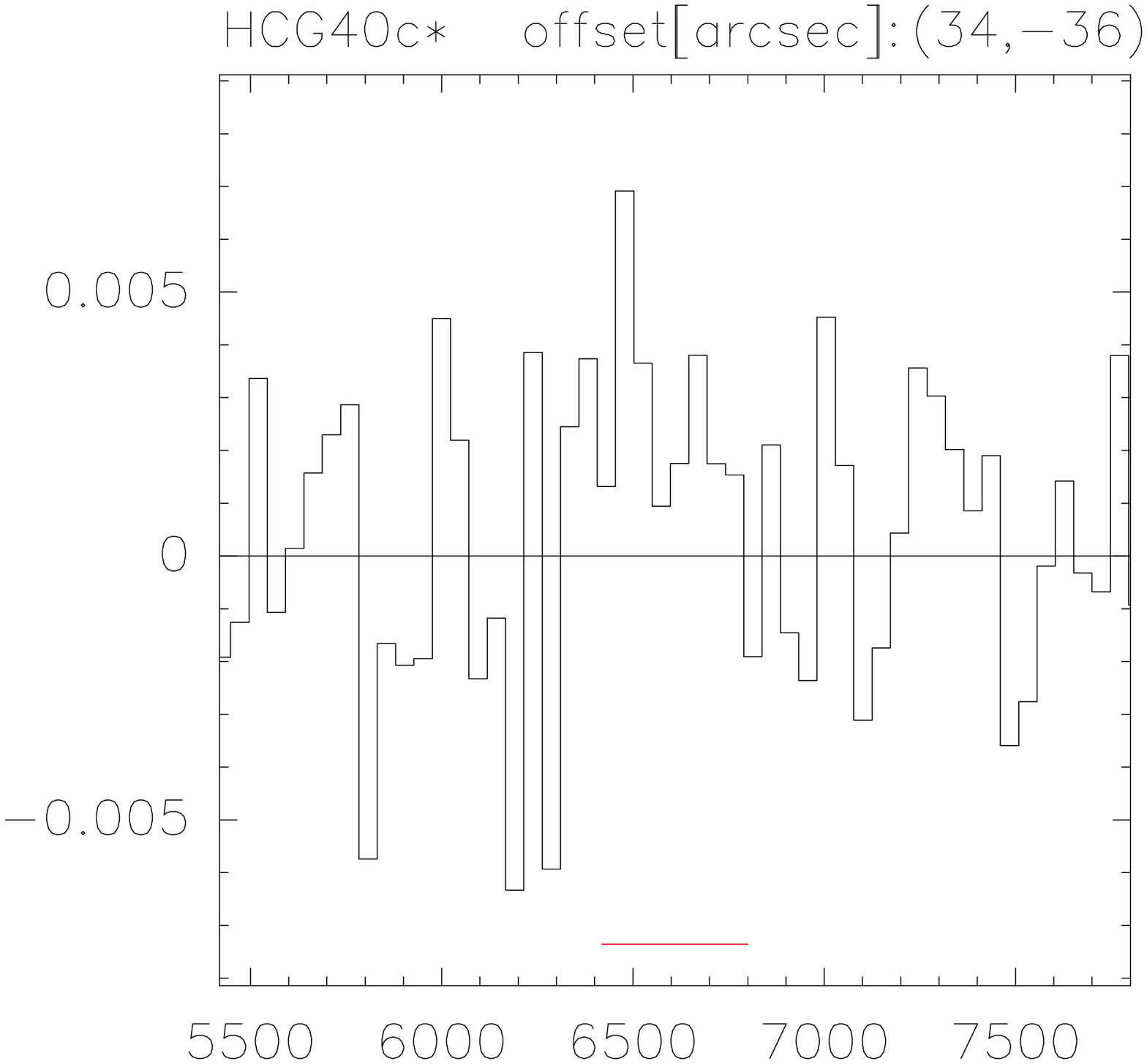}
\includegraphics[width=3.5cm,angle=-0]{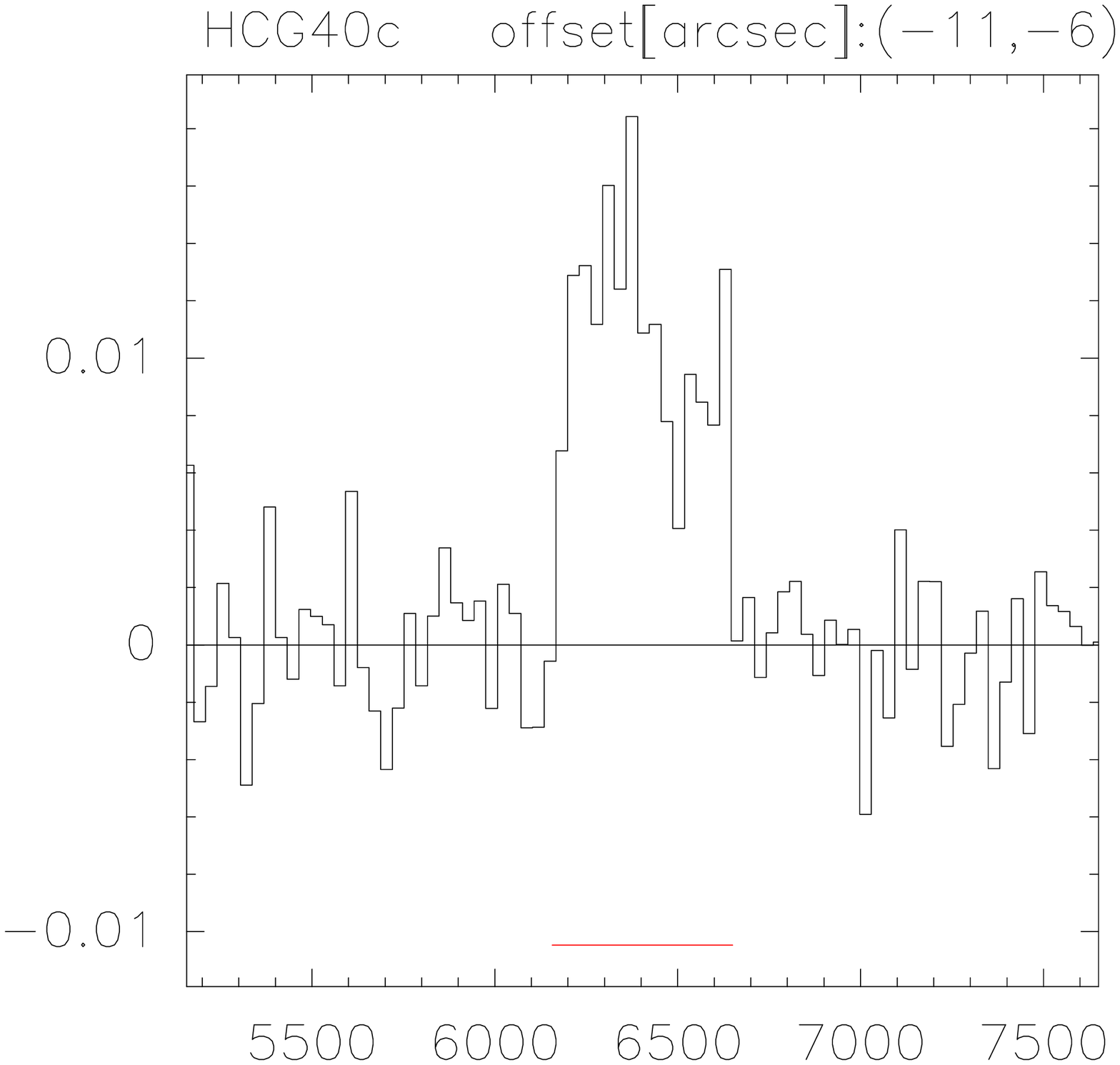}
\includegraphics[width=3.5cm,angle=-0]{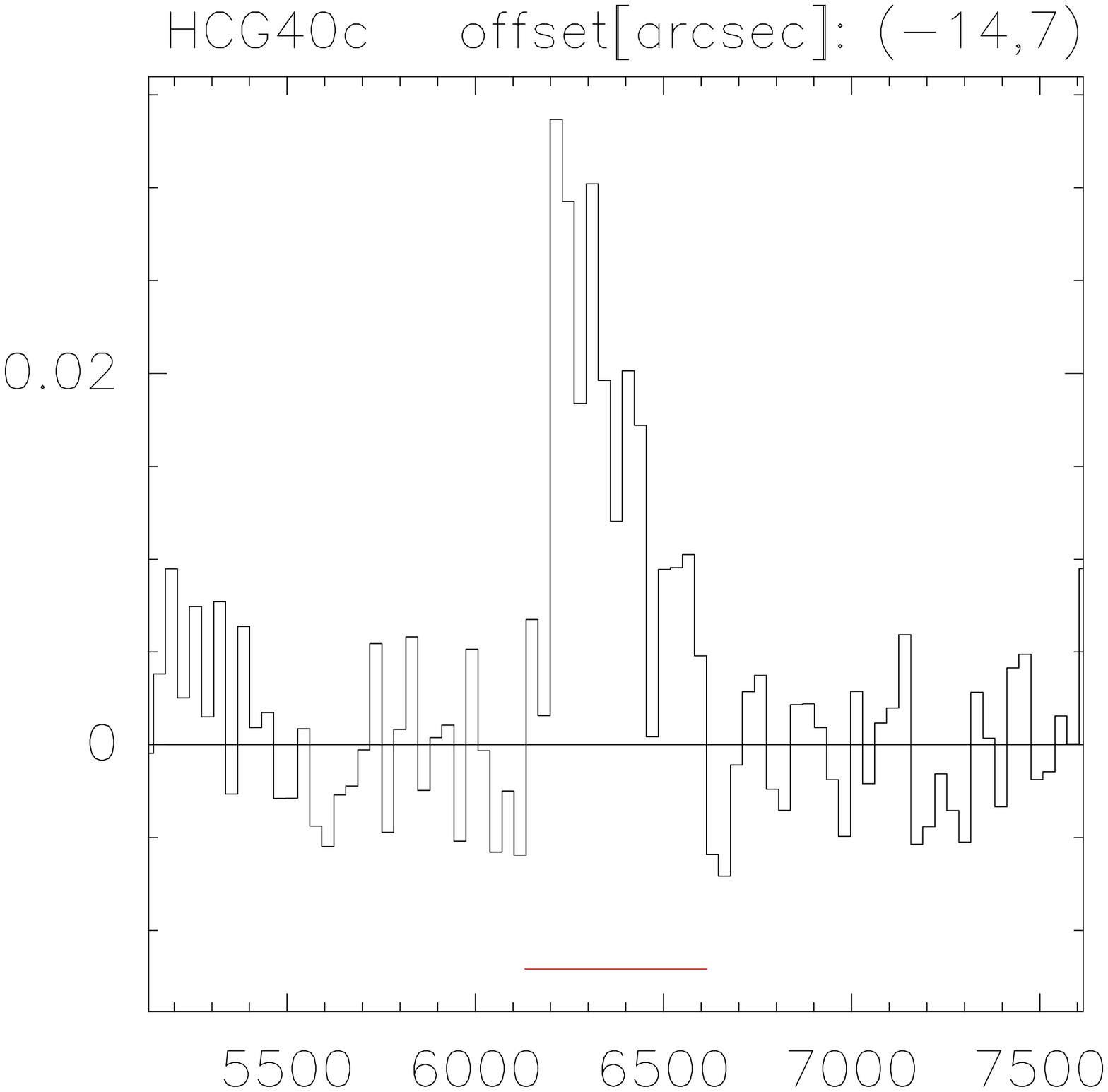}
}
\centerline{
\includegraphics[width=3.7cm,angle=-0]{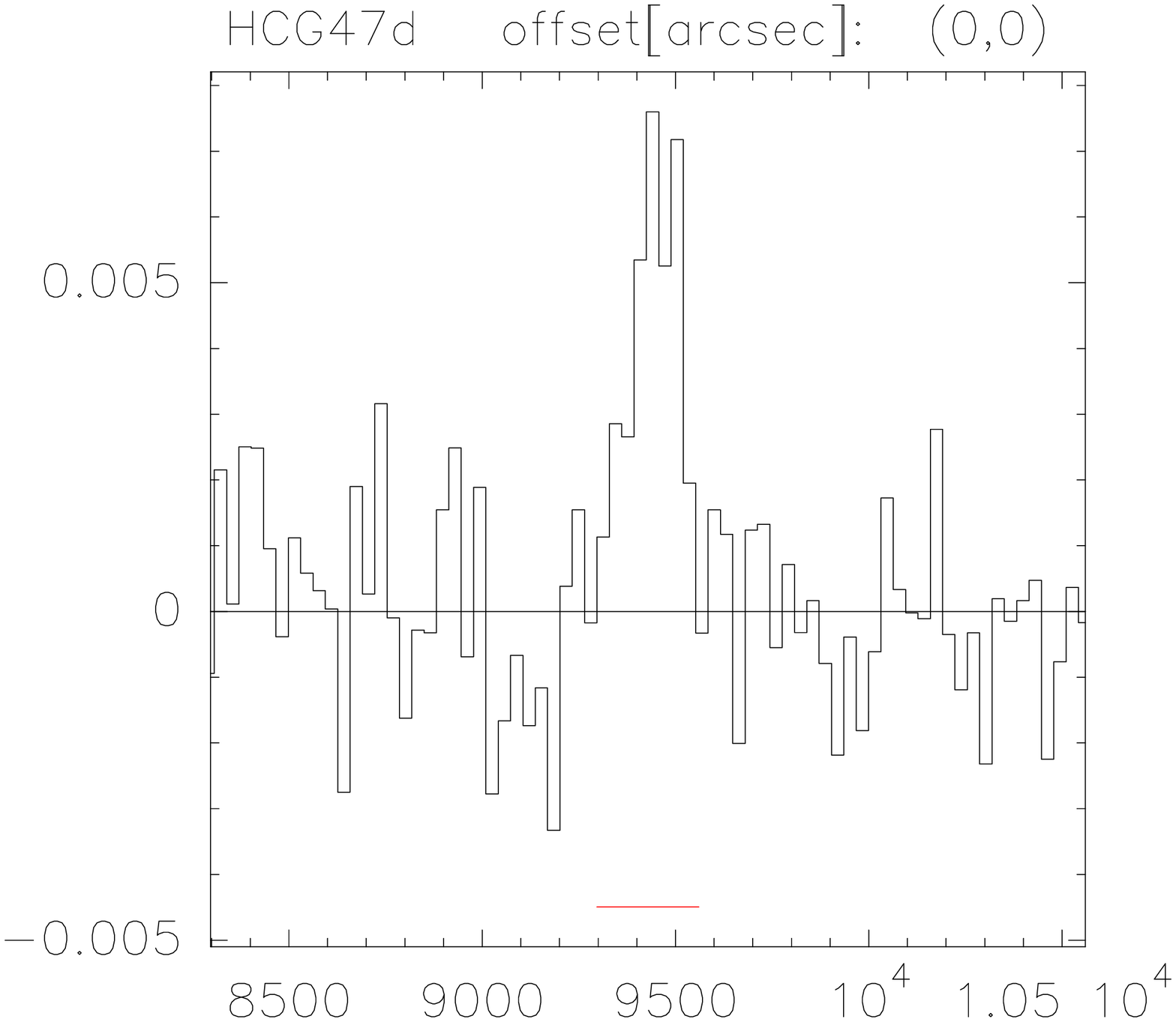}
\includegraphics[width=3.5cm,angle=-0]{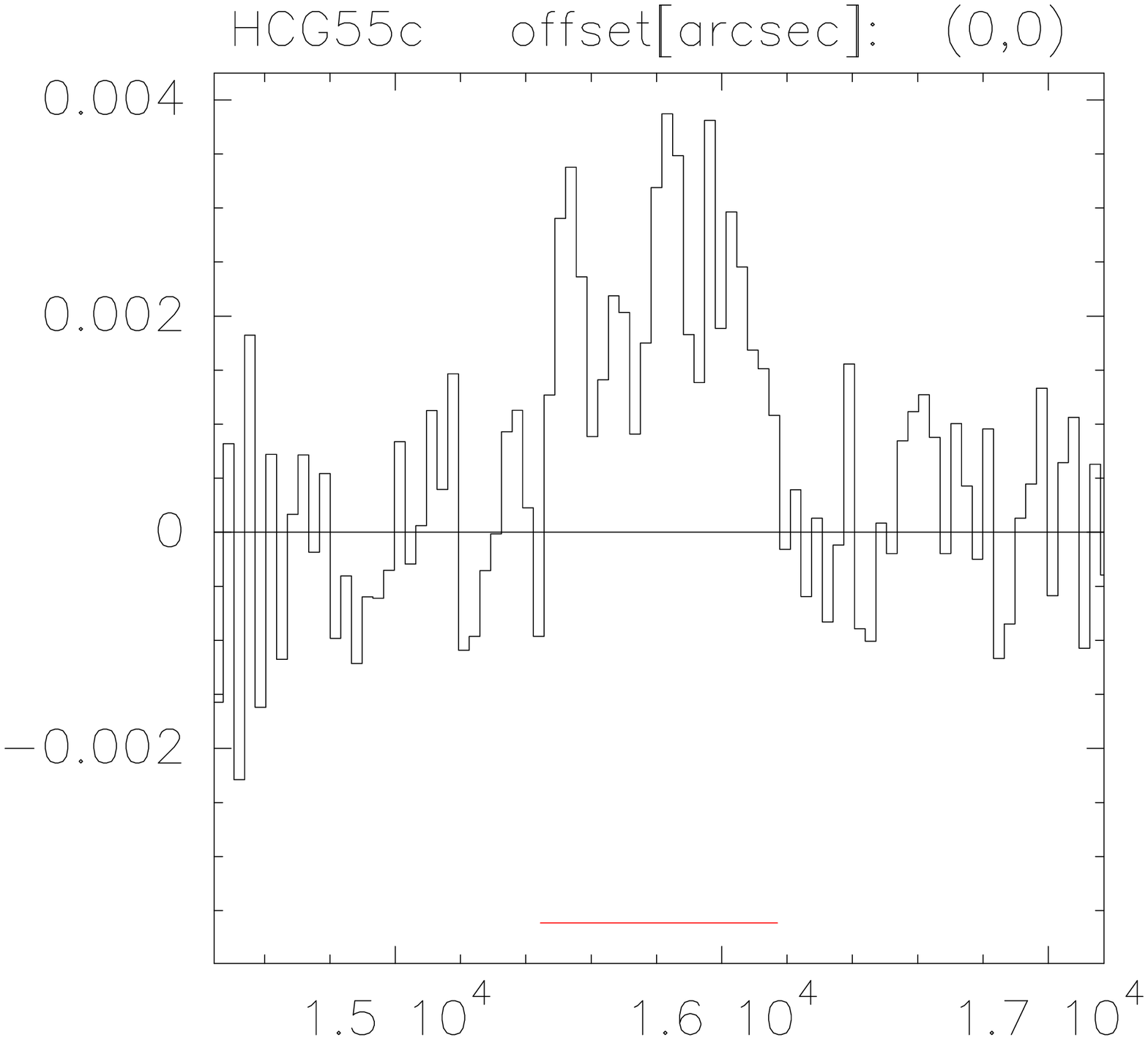}
\includegraphics[width=3.5cm,angle=-0]{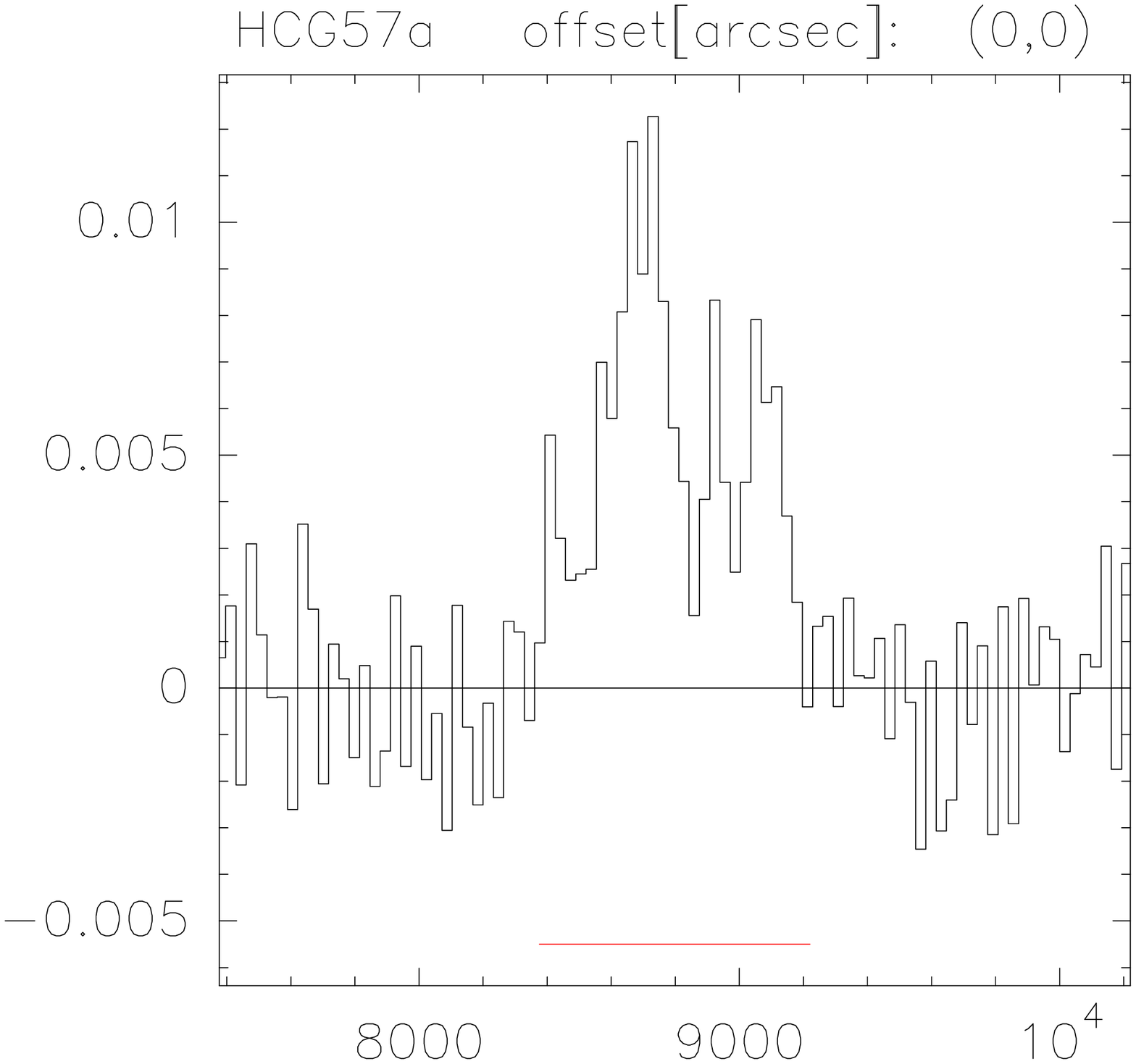}
\includegraphics[width=3.5cm,angle=-0]{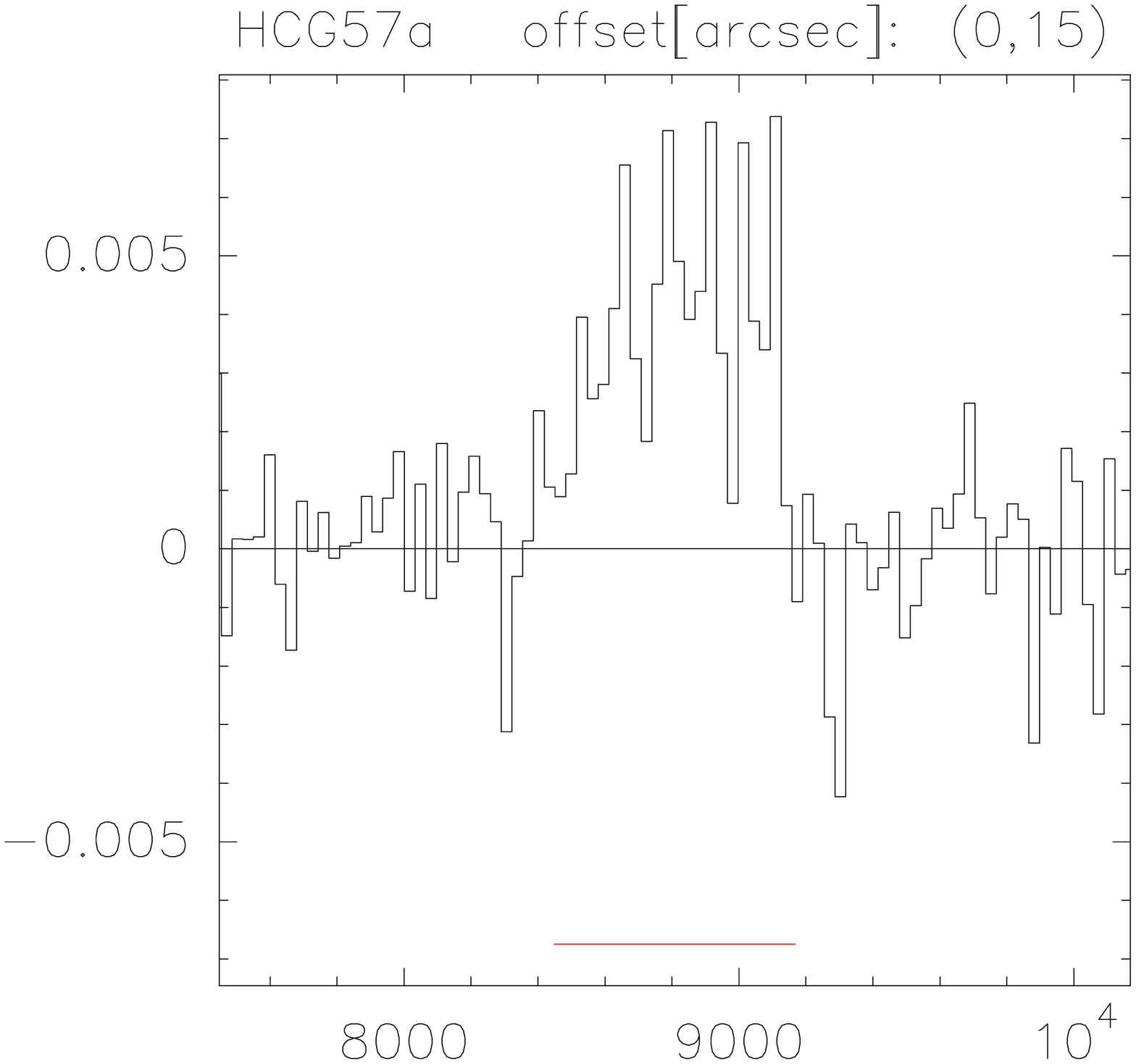}
\includegraphics[width=3.5cm,angle=-0]{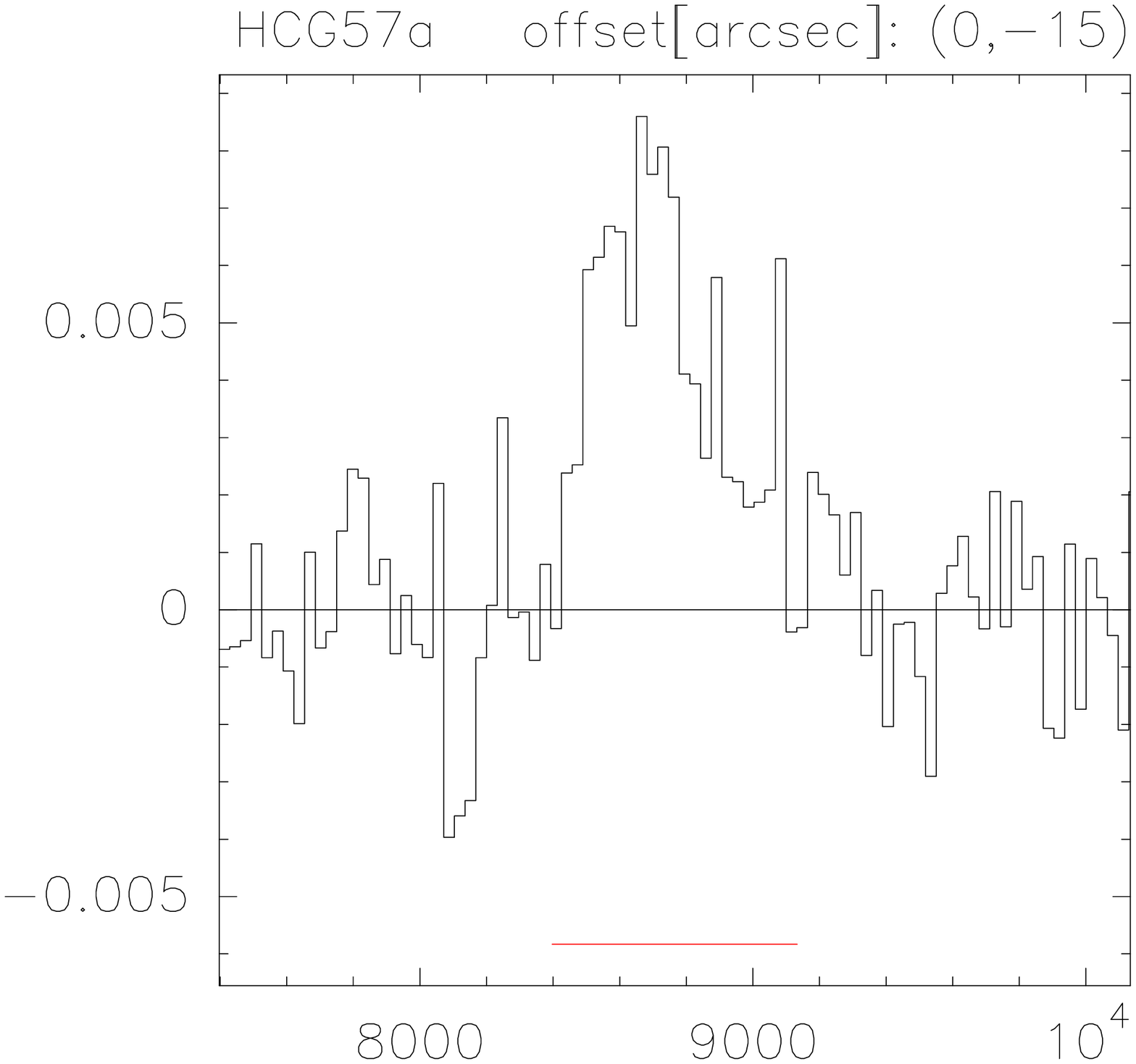}
}

\centerline{
\includegraphics[width=3.5cm,angle=-0]{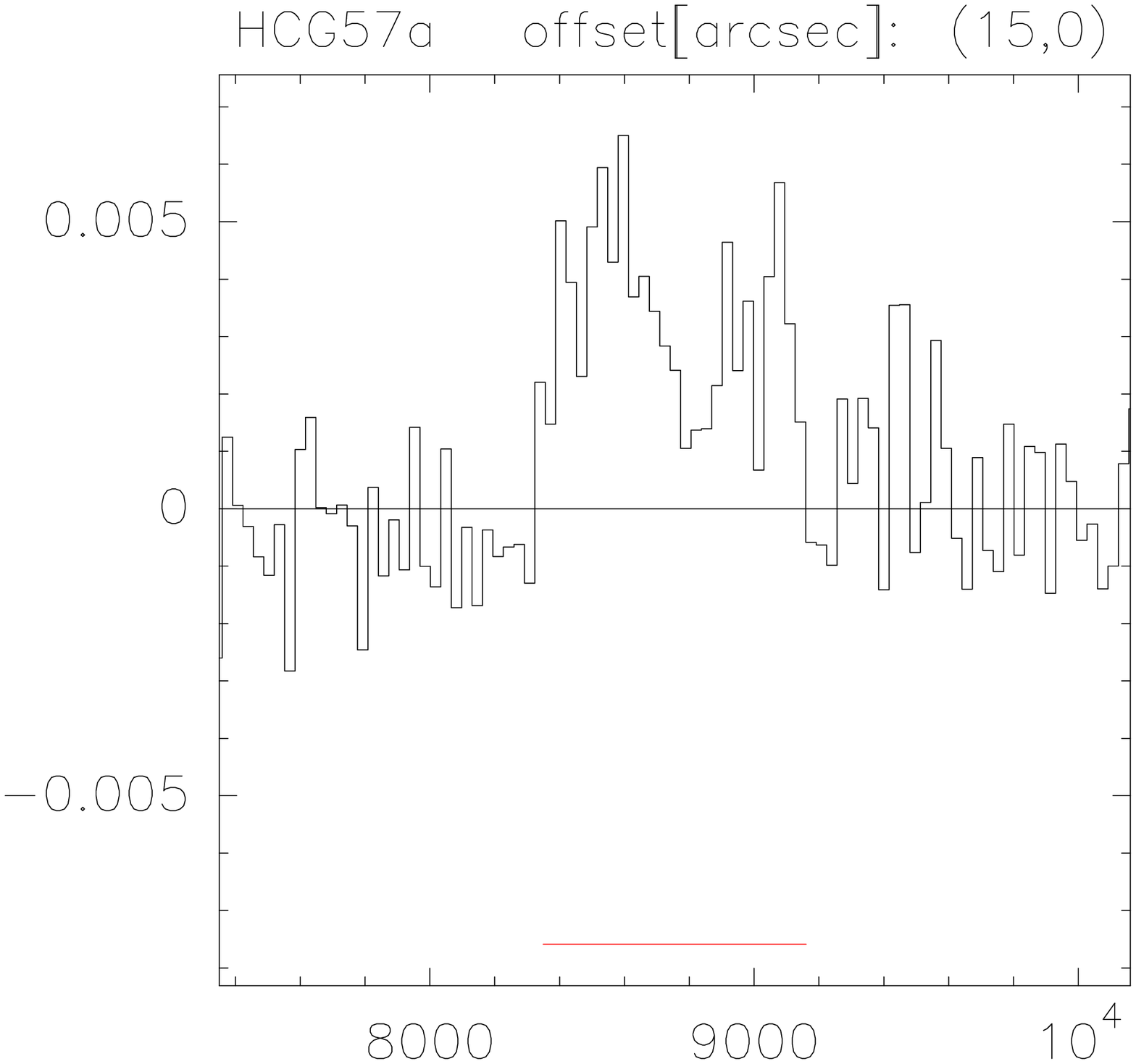}
\includegraphics[width=3.5cm,angle=-0]{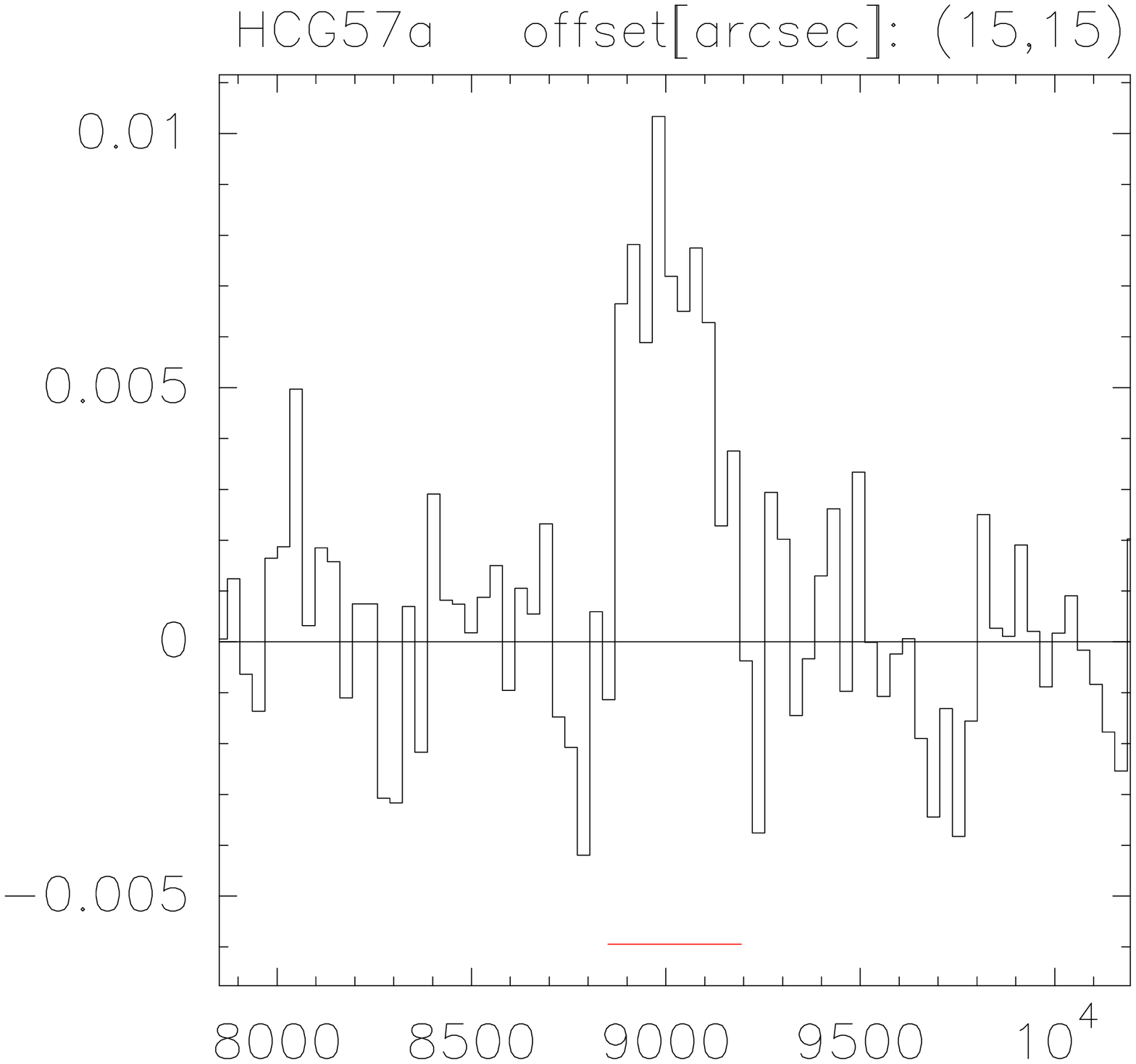}
\includegraphics[width=3.5cm,angle=-0]{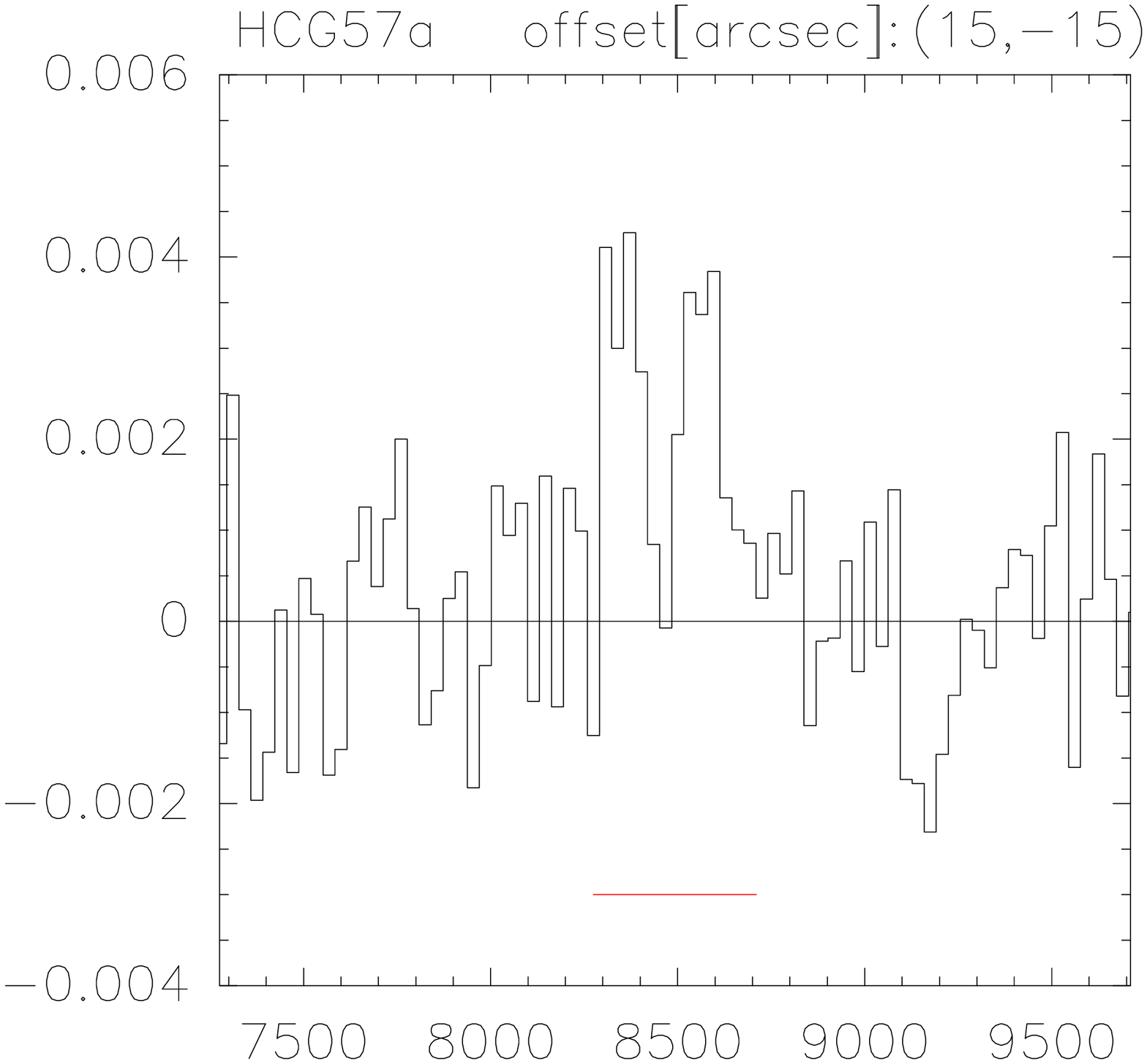}
\includegraphics[width=3.5cm,angle=-0]{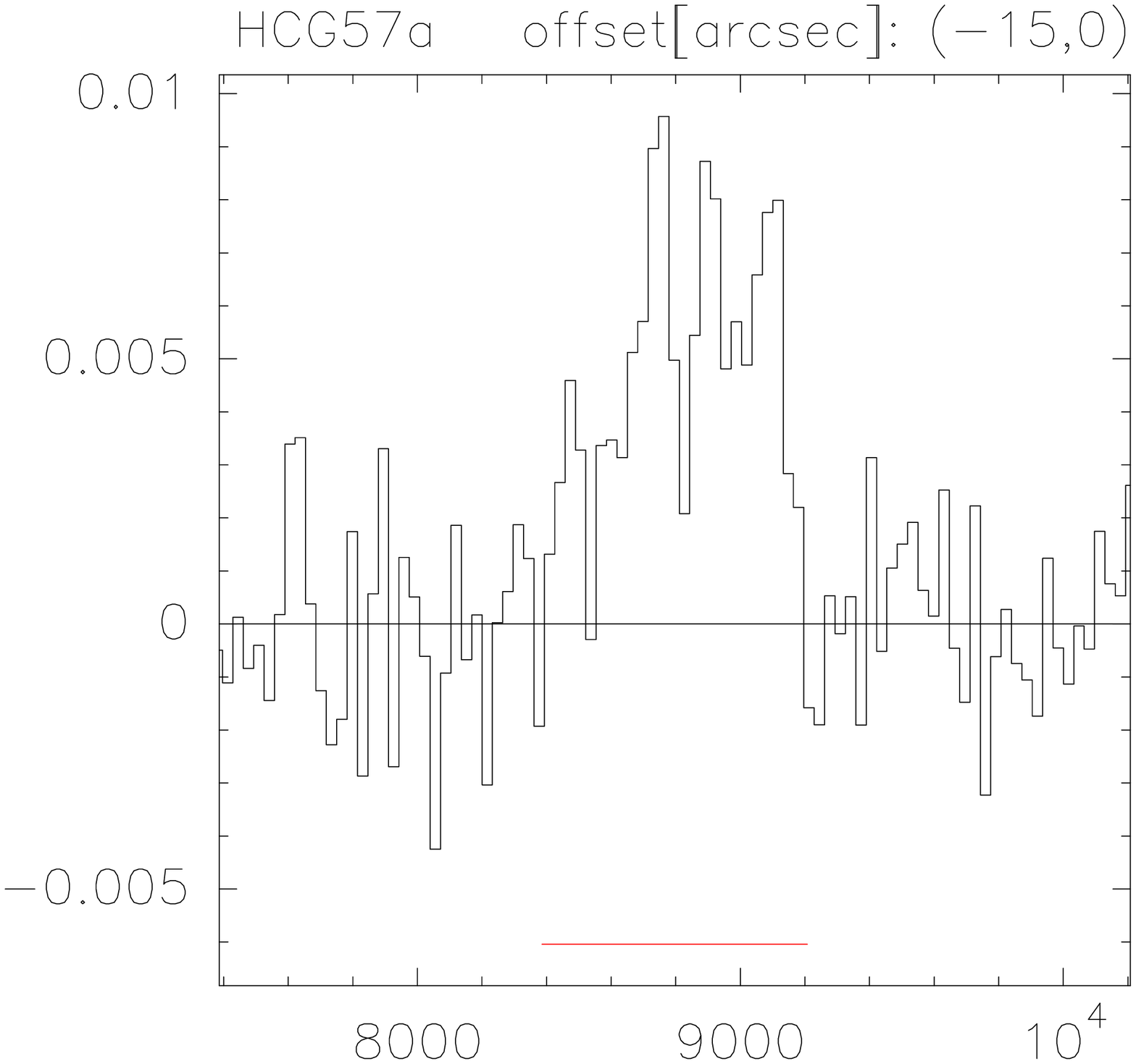}
\includegraphics[width=3.5cm,angle=-0]{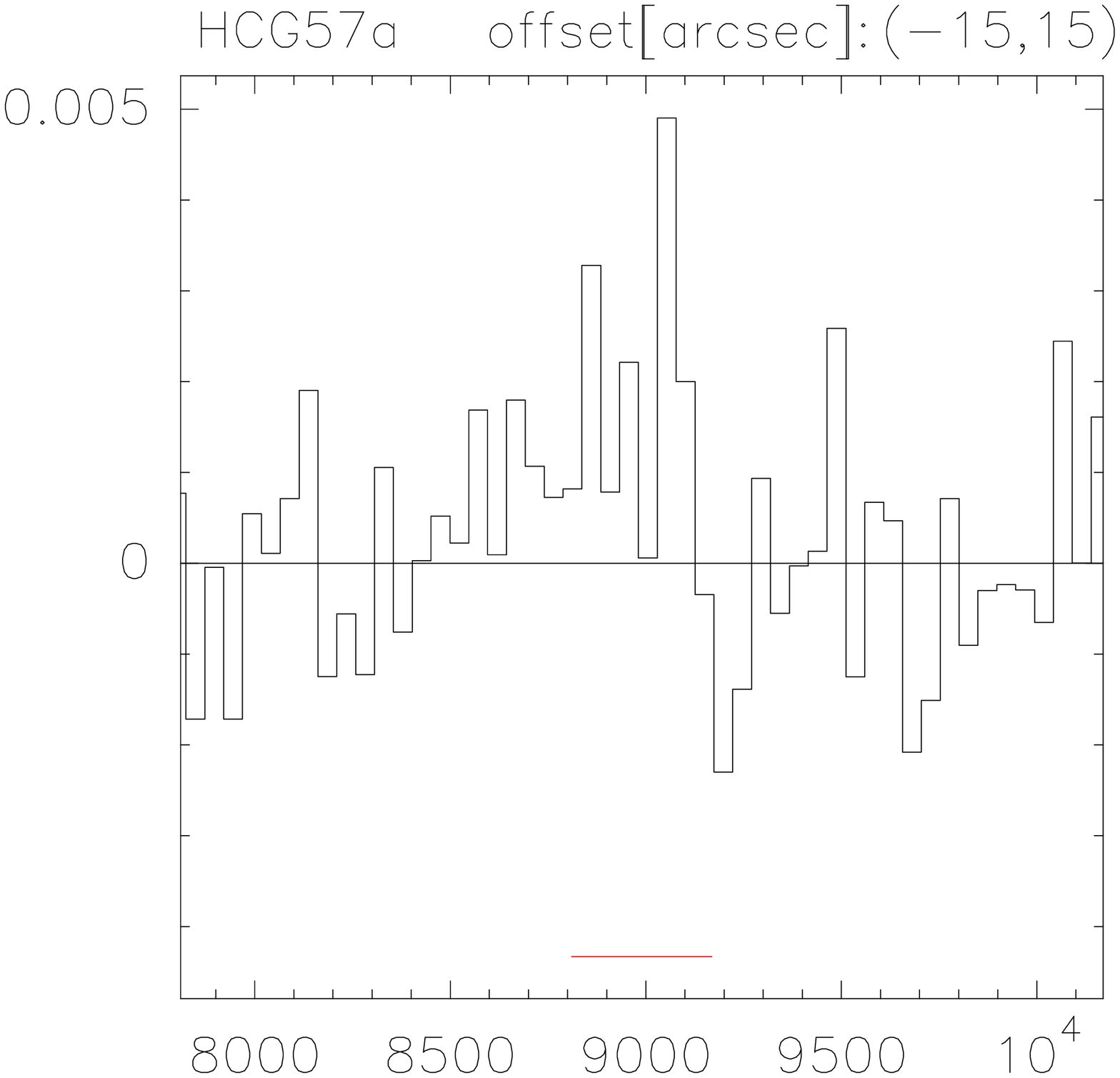}
}

\centerline{
\includegraphics[width=3.6cm,angle=-0]{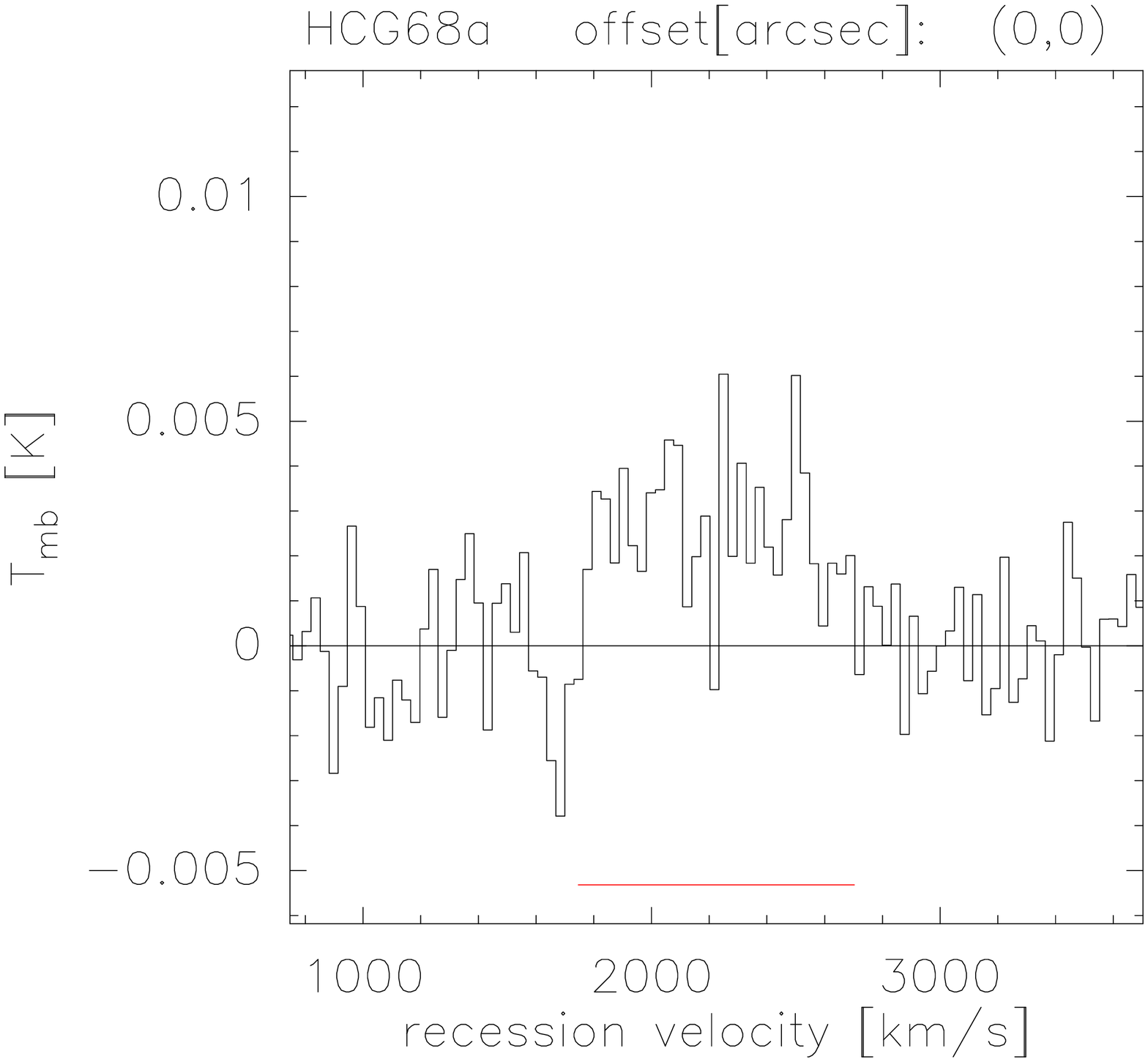}
\includegraphics[width=3.5cm,angle=-0]{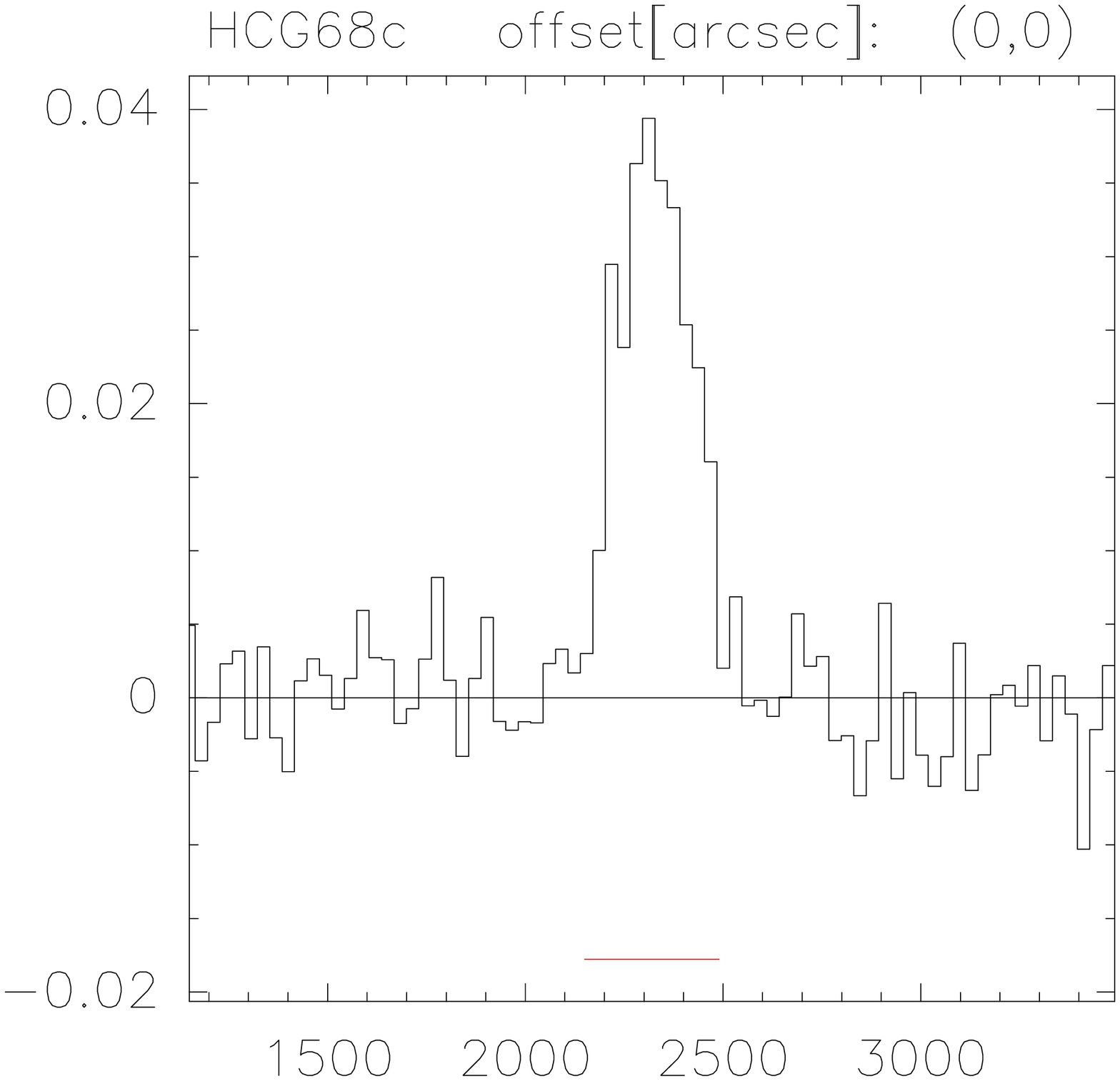}
\includegraphics[width=3.5cm,angle=-0]{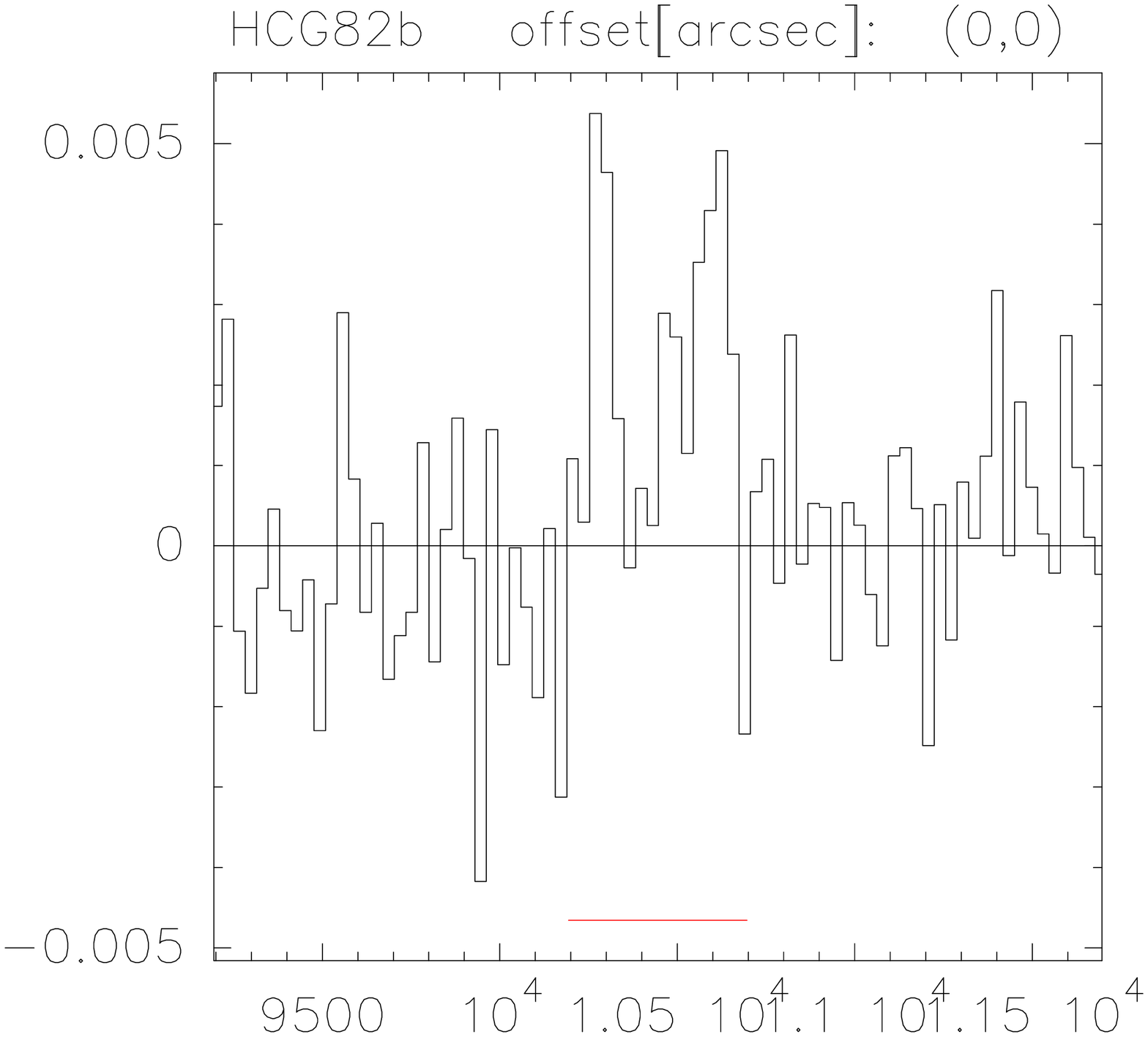}
\includegraphics[width=3.5cm,angle=-0]{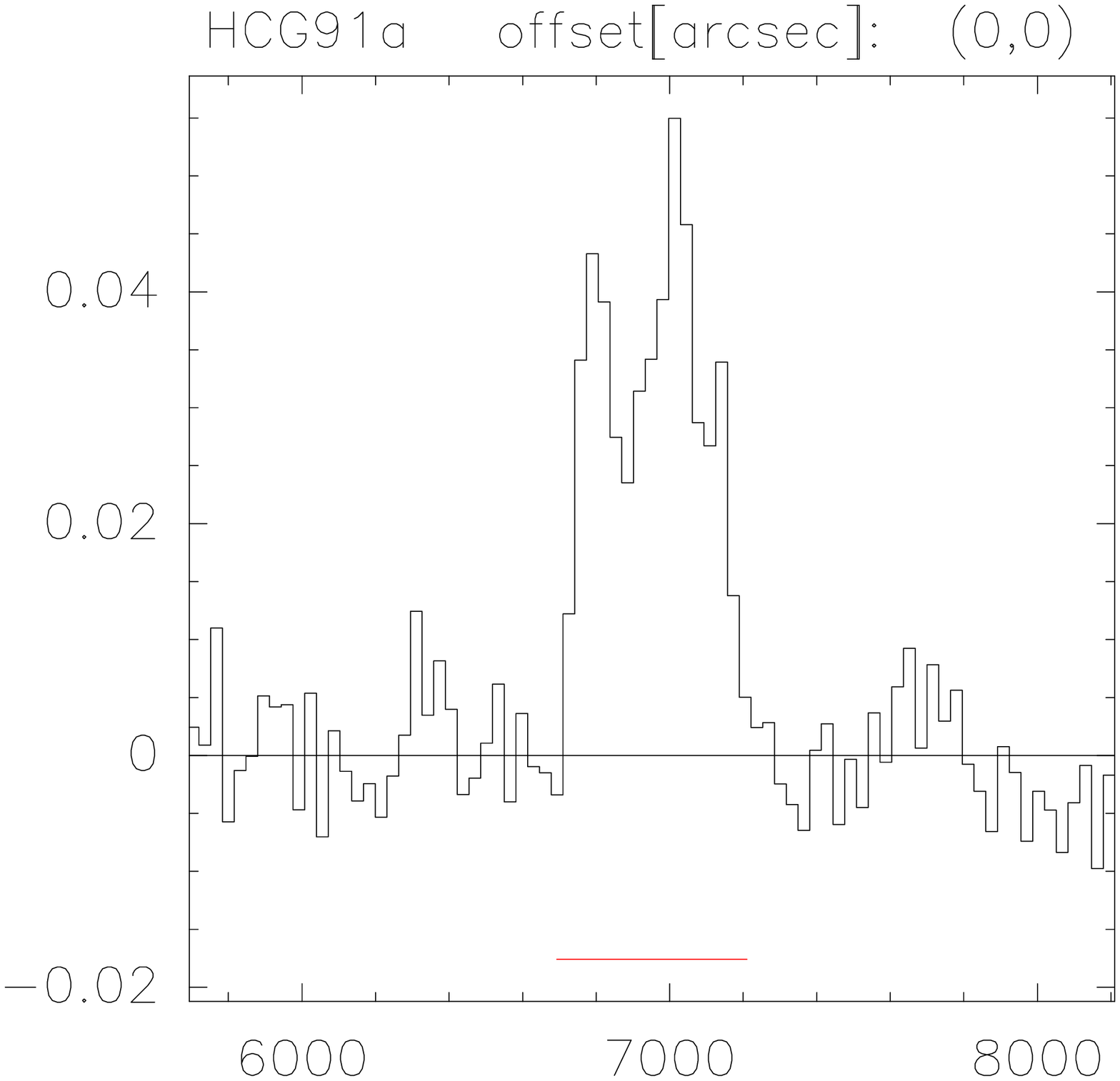}
\includegraphics[width=3.5cm,angle=-0]{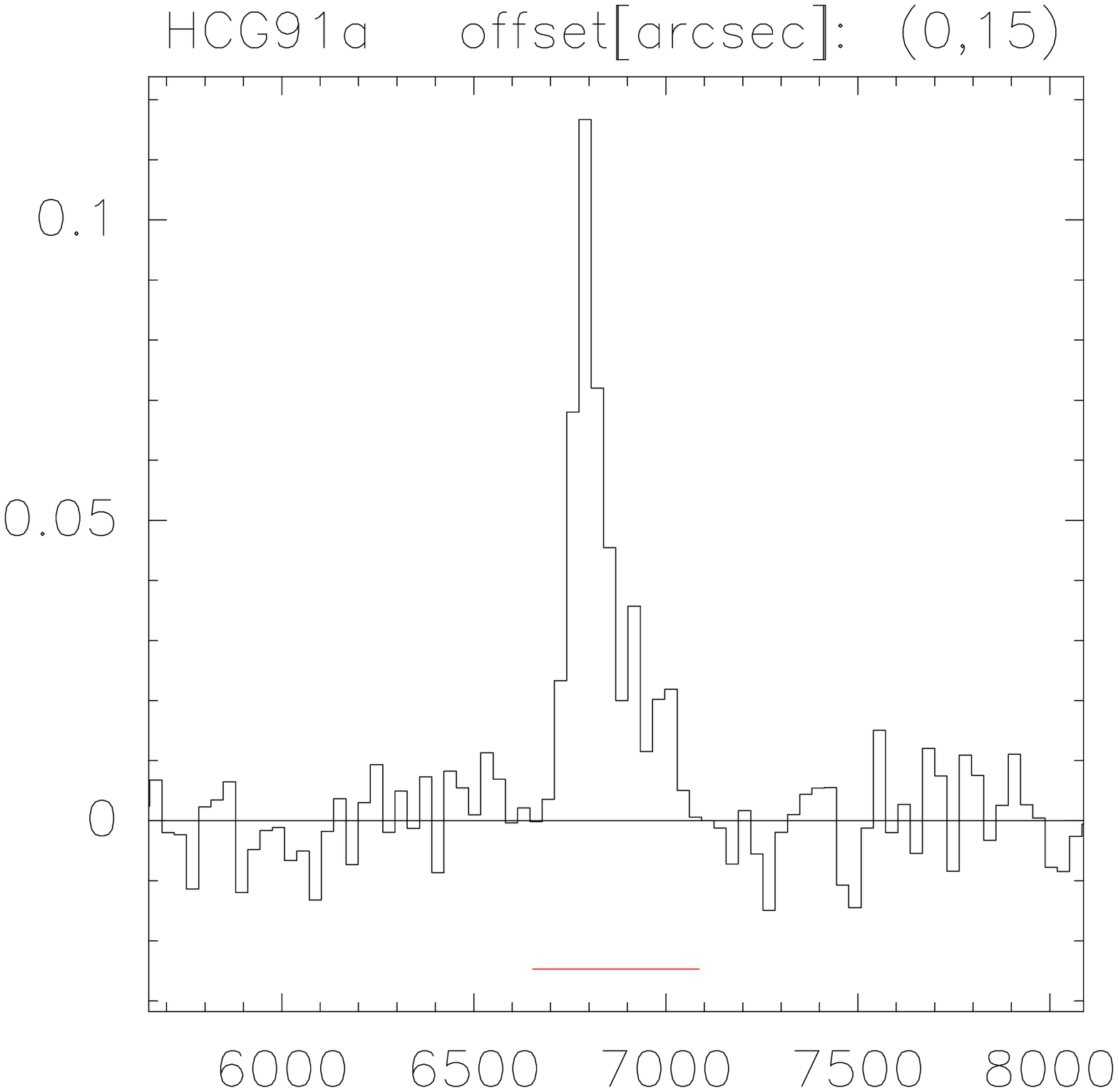}
}

\caption{CO(1-0) spectra of the detected (including tentative detections) spectra. The velocity resolution
is $\sim$ 32 \kms\ for most spectra and $\sim$  48 \kms\ for some cases where a lower resolution was required
to clearly see the line. The red line segment shows the zero-level line width of the 
CO line adopted for the determination of the velocity integrated intensity. 
An asterisk next to the name indicates a tentative detection.
}
\label{fig:spectra_co10}
\end{figure*}

\begin{figure*}
\centerline{
\includegraphics[width=3.5cm,angle=-0]{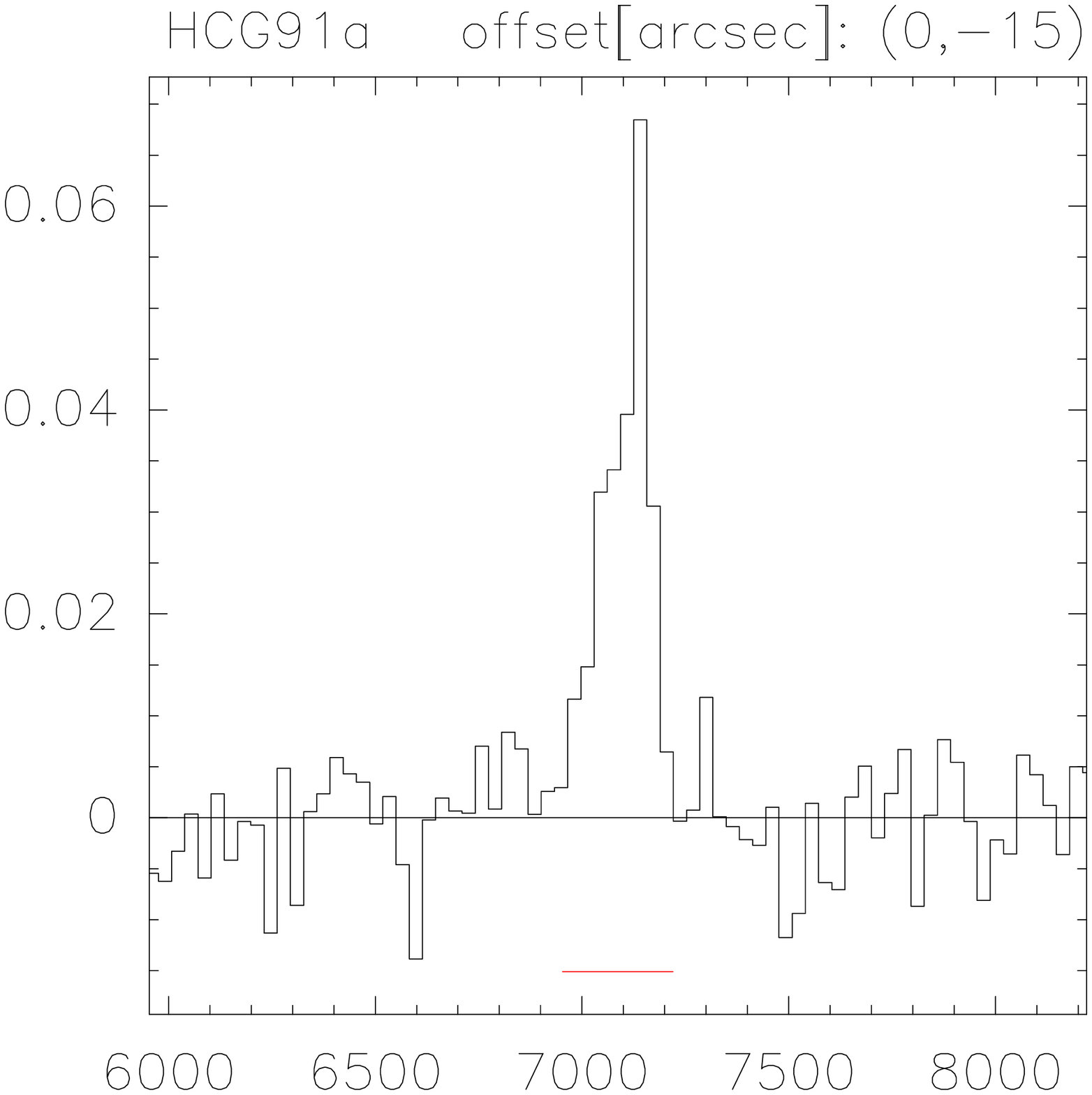}
\includegraphics[width=3.5cm,angle=-0]{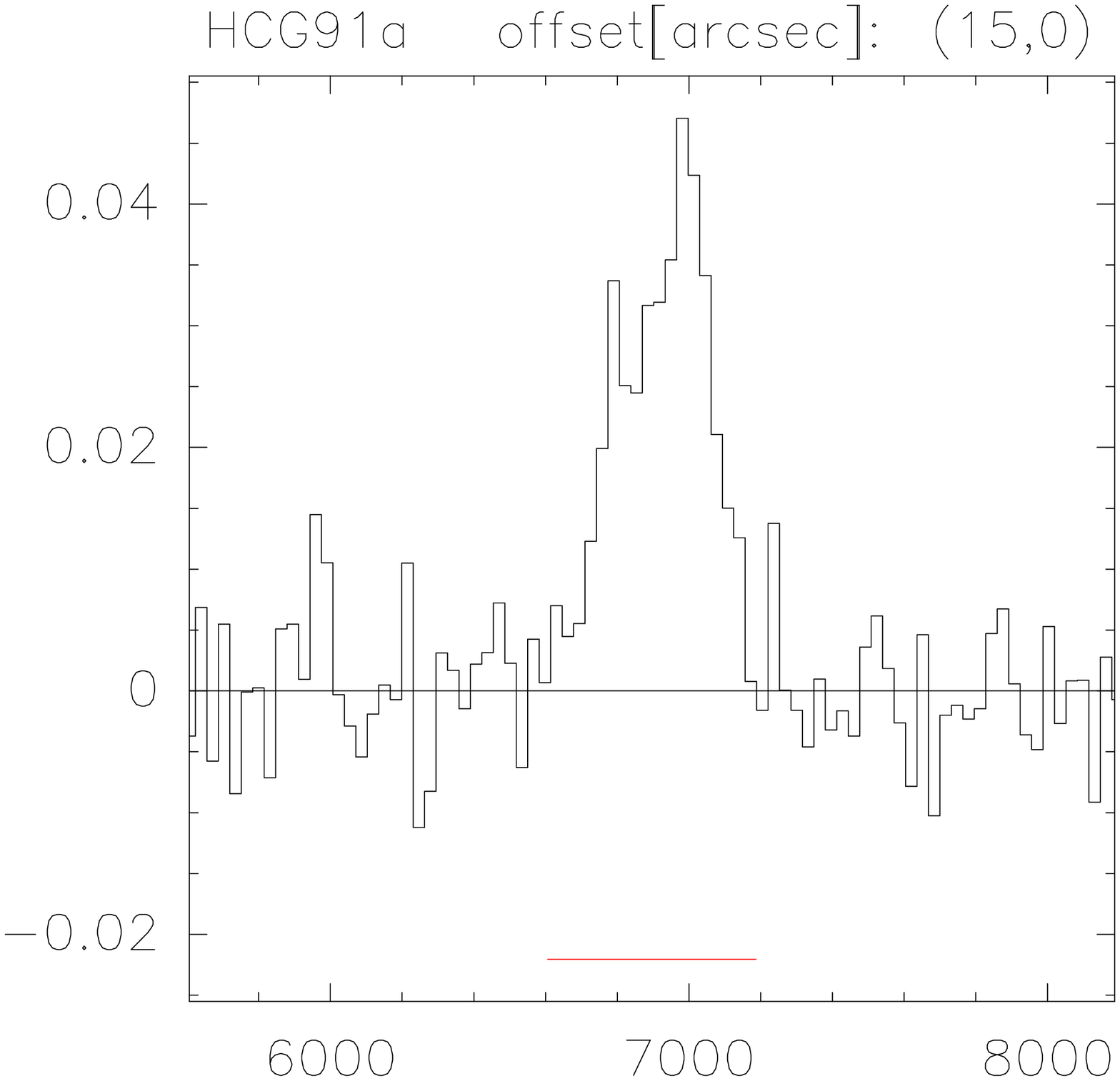}
\includegraphics[width=3.5cm,angle=-0]{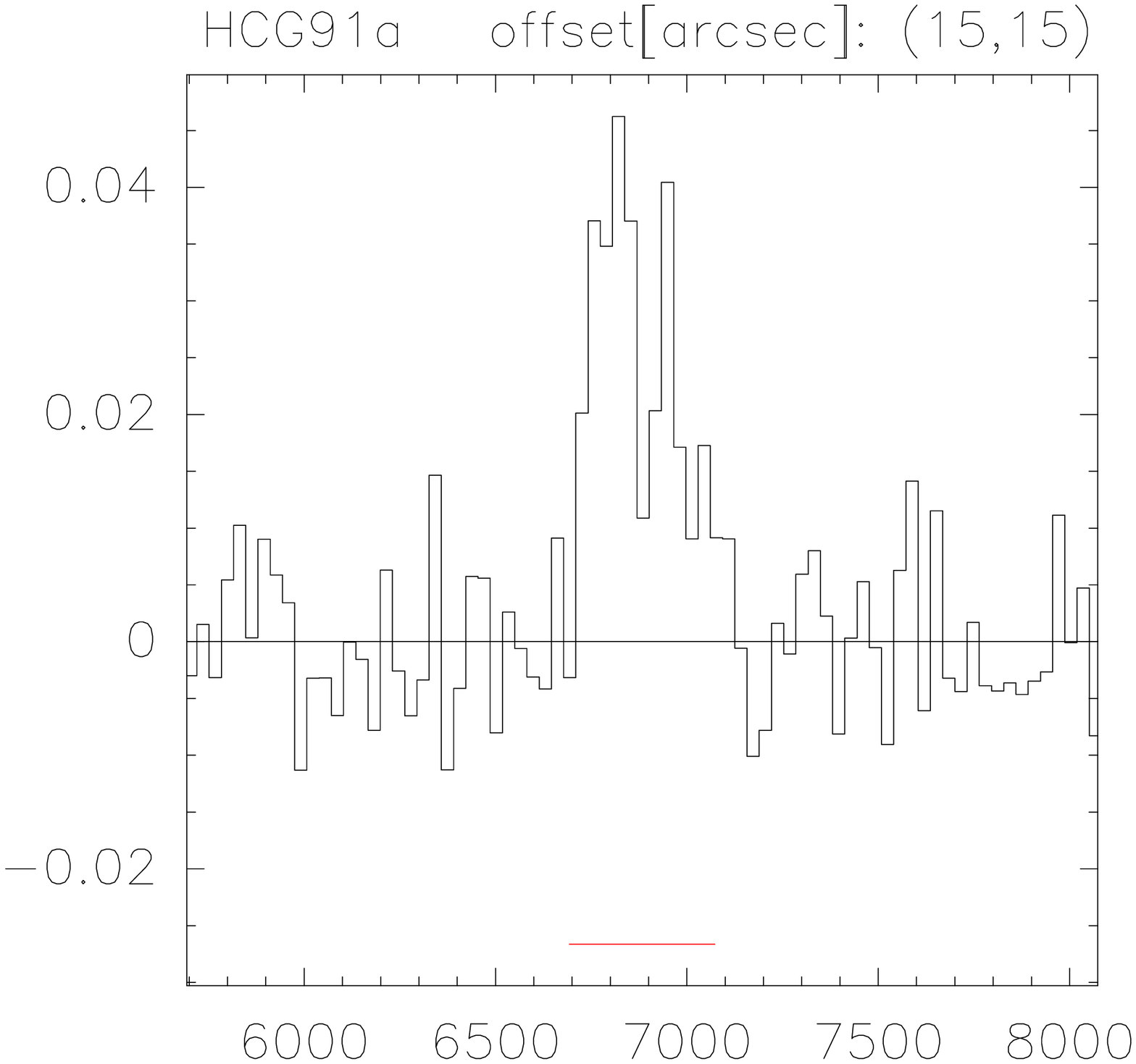}
\includegraphics[width=3.5cm,angle=-0]{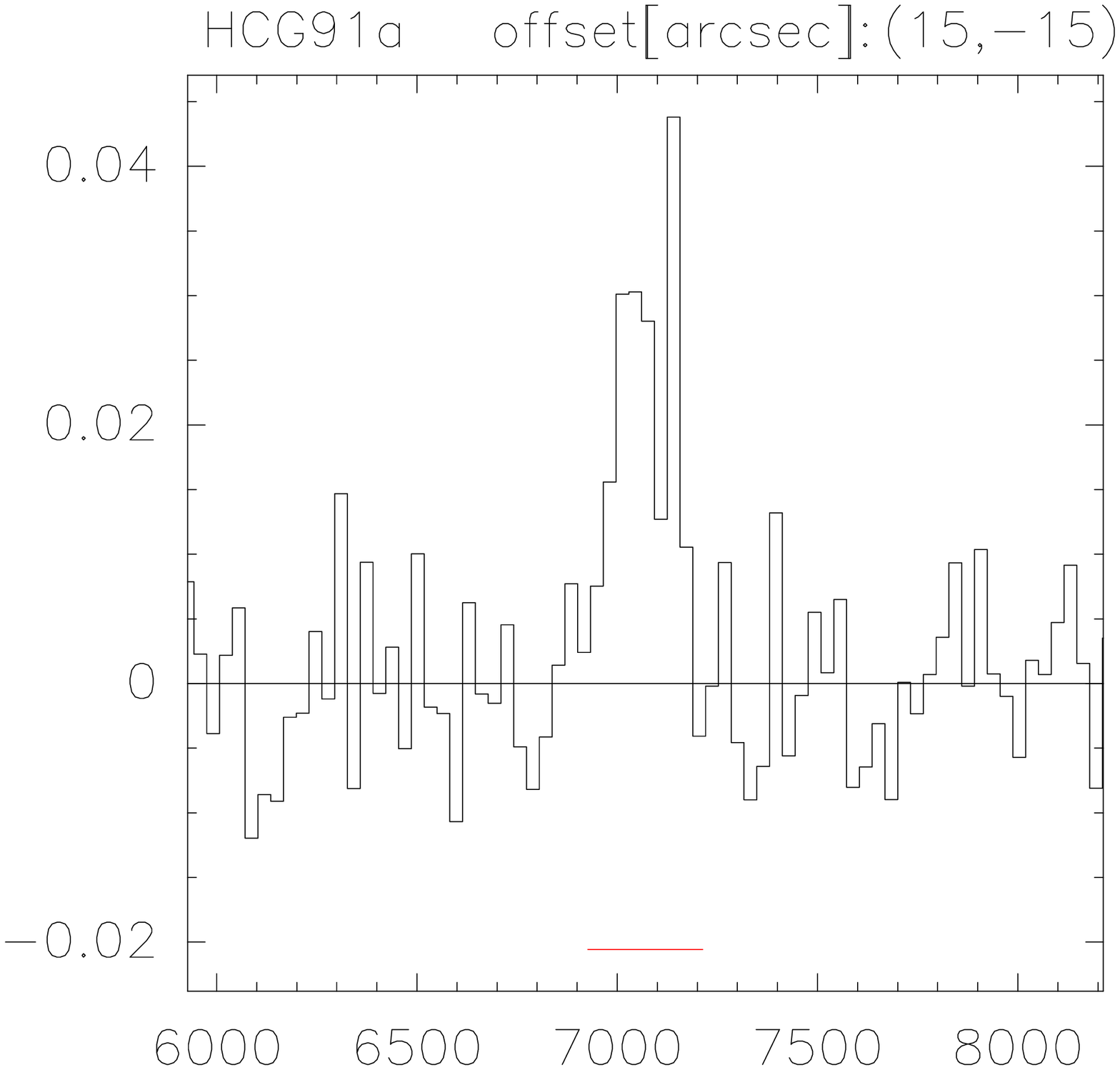}
\includegraphics[width=3.5cm,angle=-0]{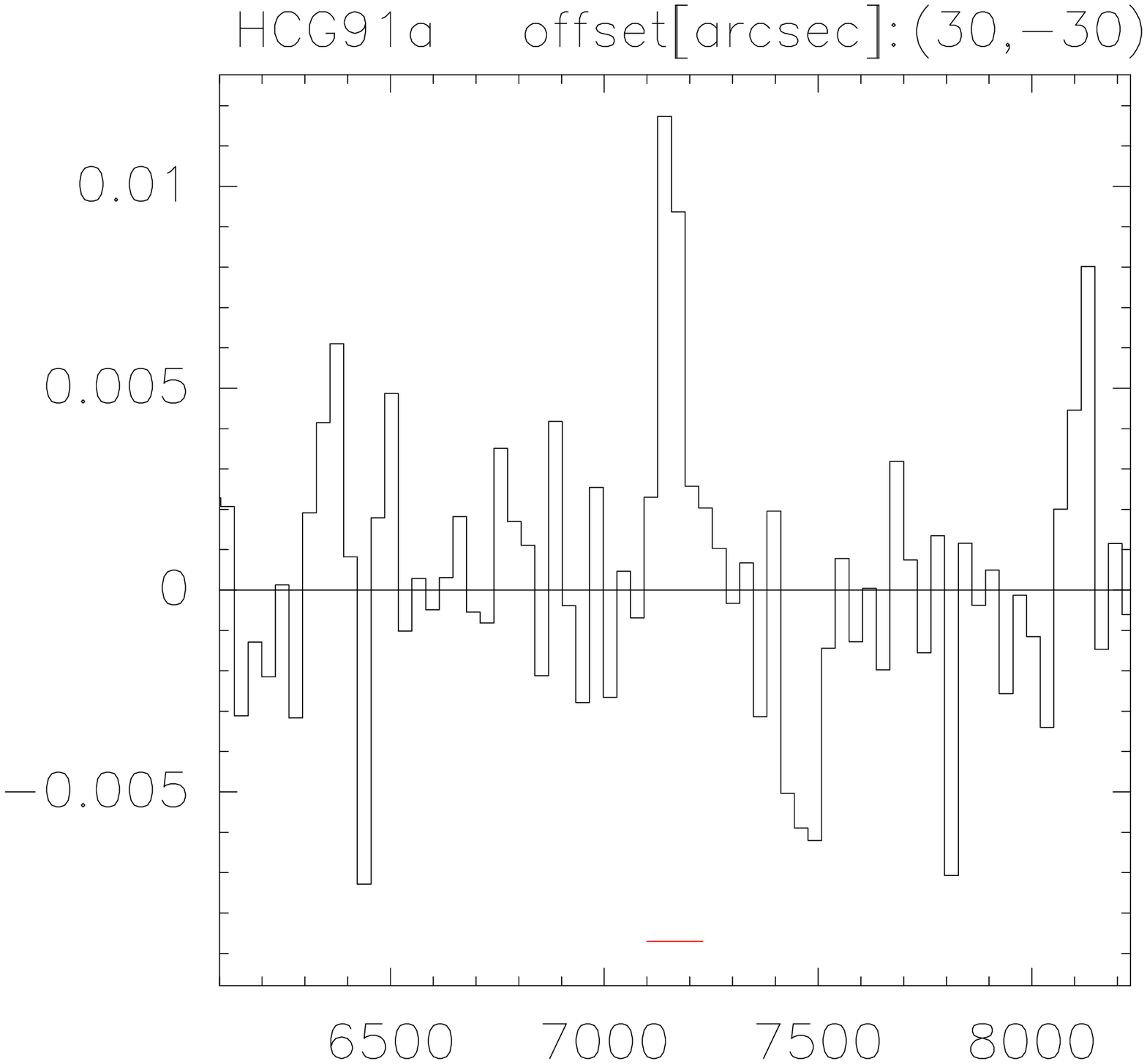}
}
\centerline{
\includegraphics[width=3.5cm,angle=-0]{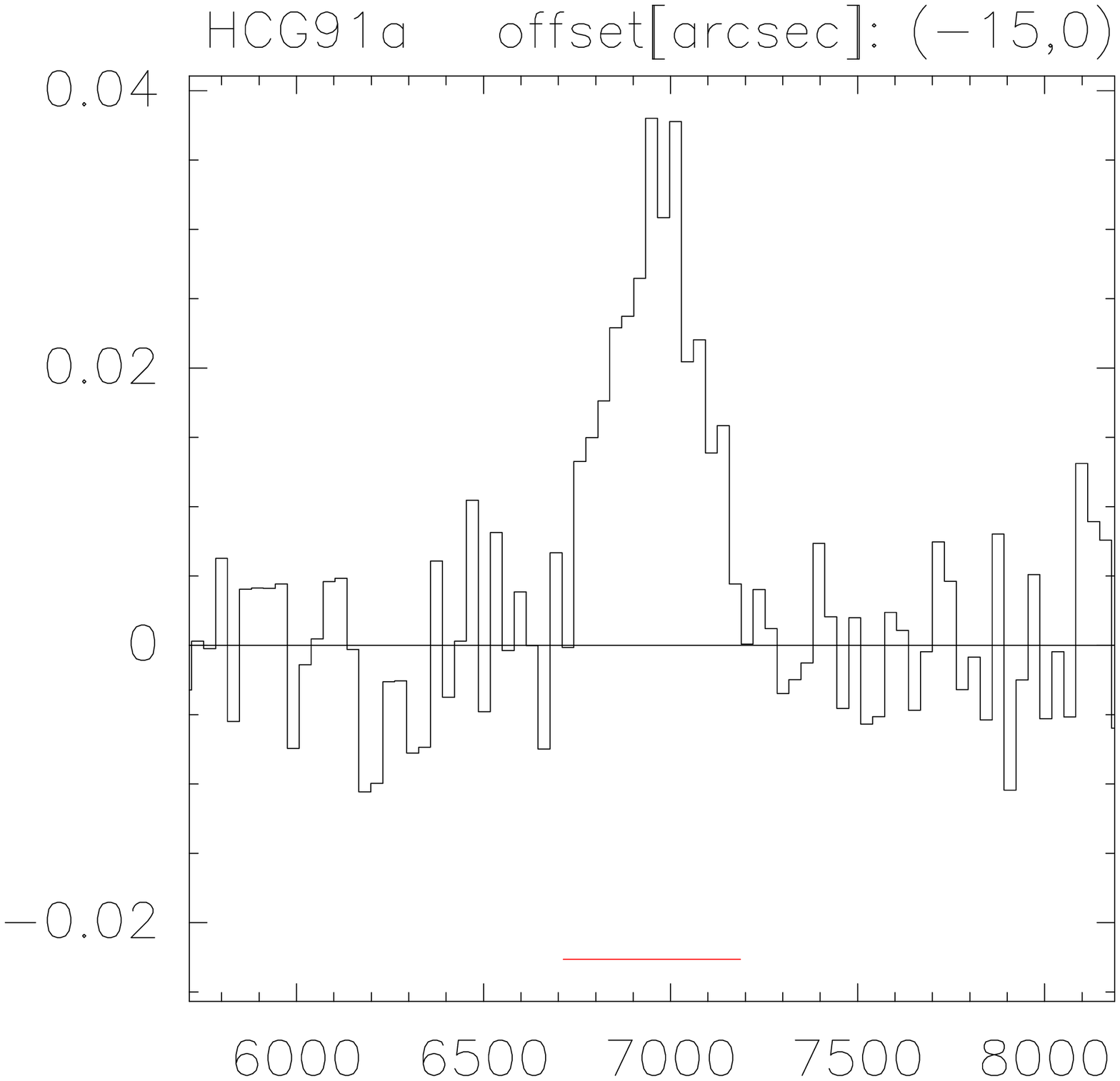}
\includegraphics[width=3.5cm,angle=-0]{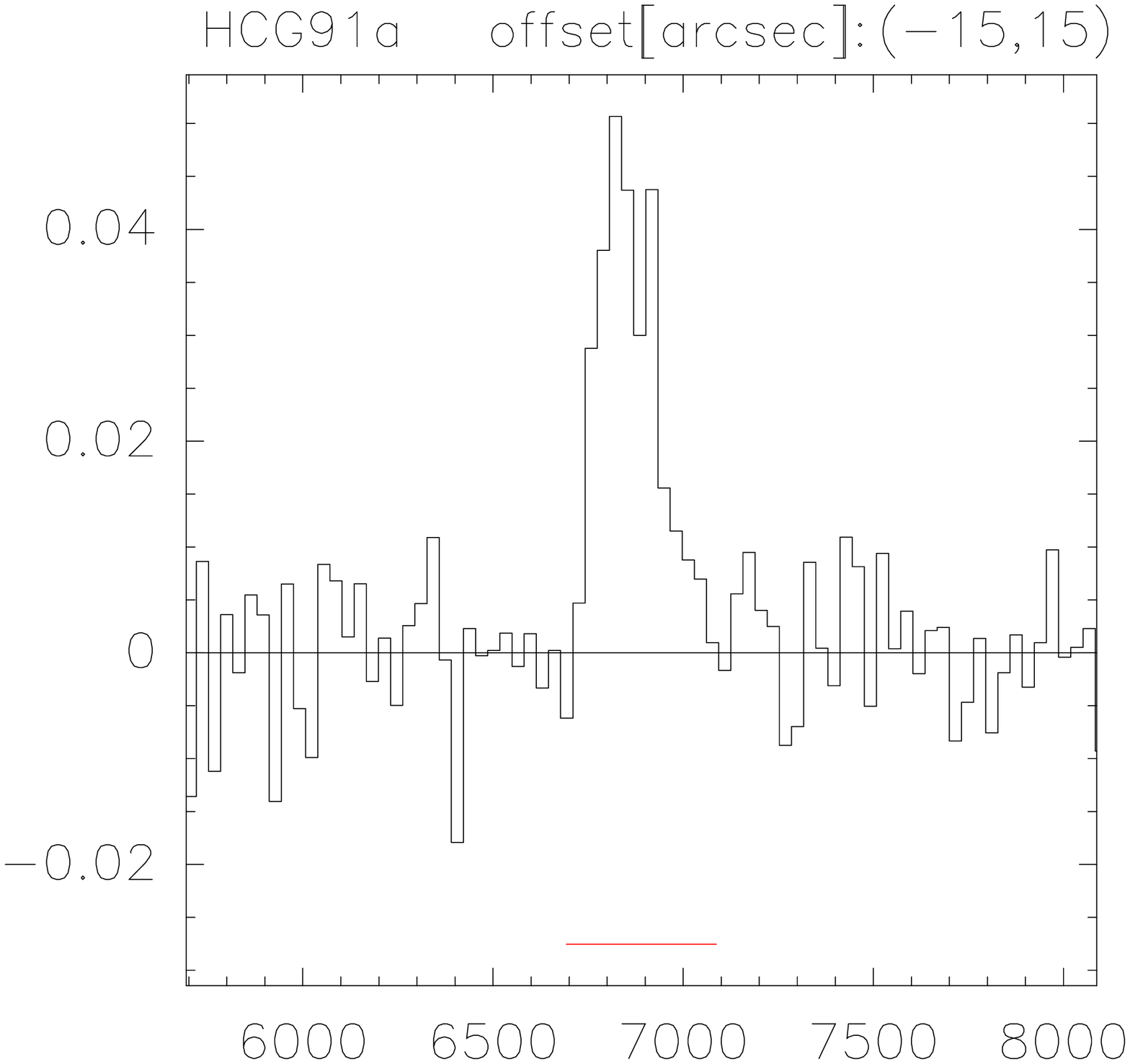}
\includegraphics[width=3.5cm,angle=-0]{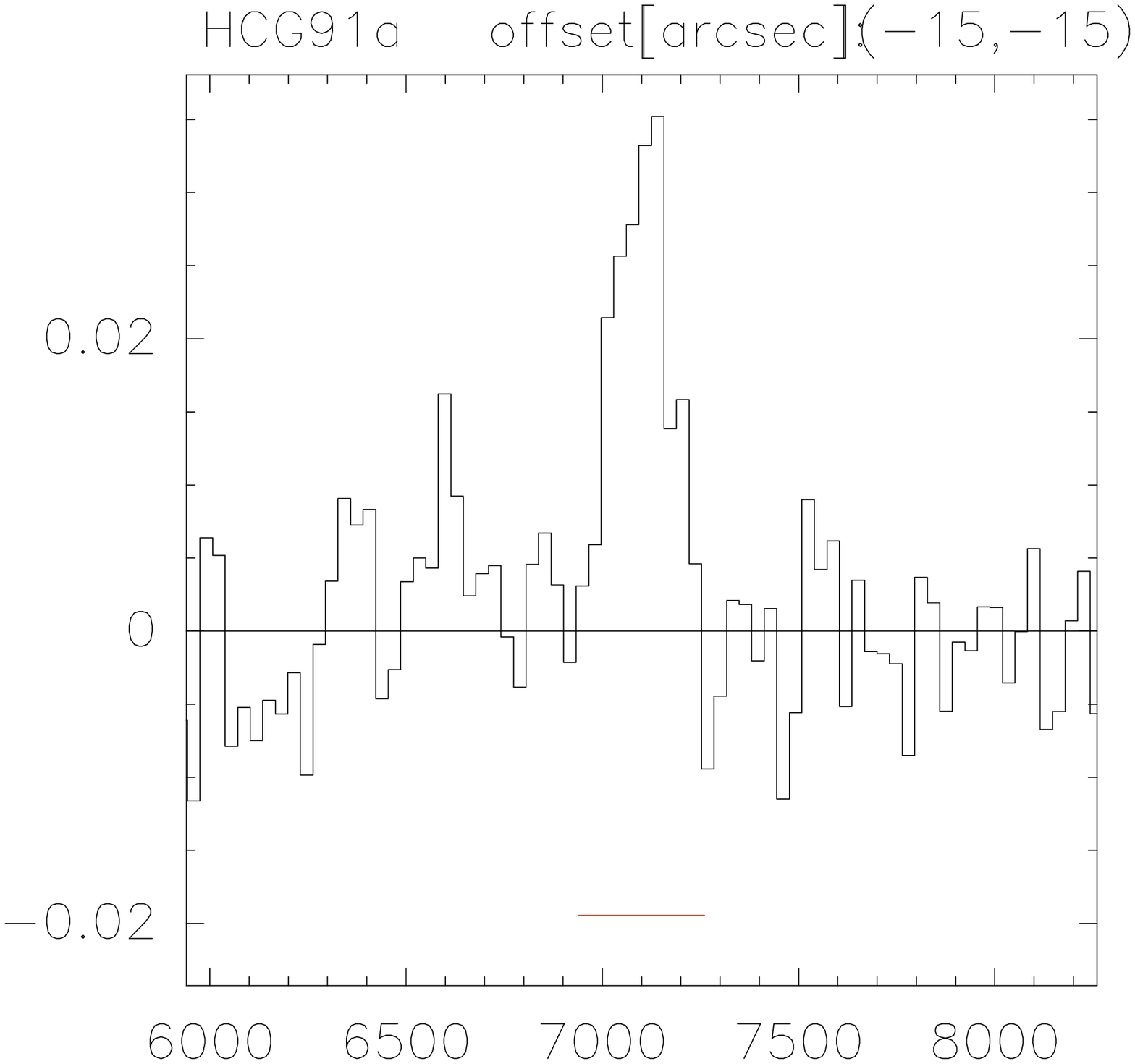}
\includegraphics[width=3.5cm,angle=-0]{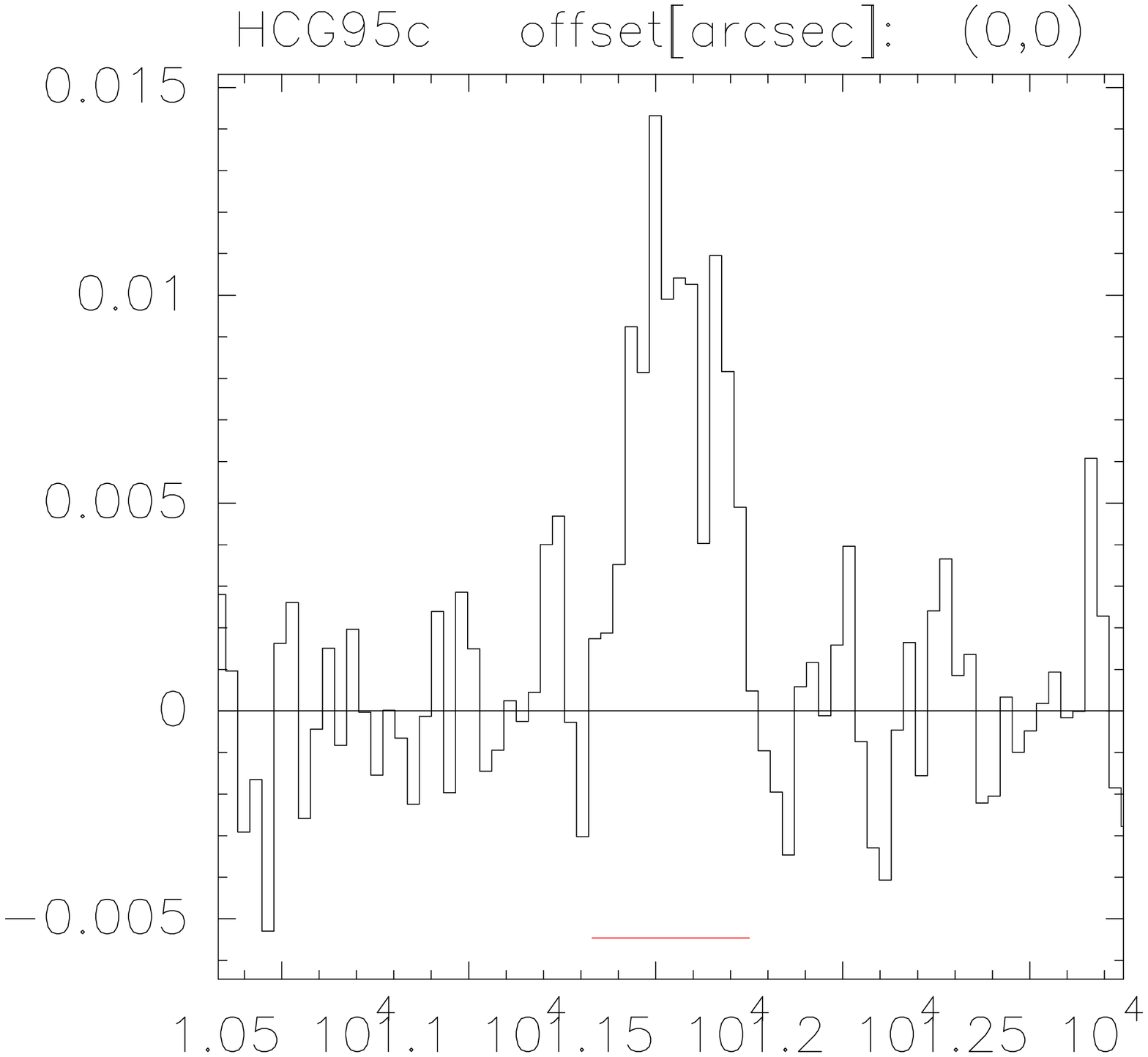}
\includegraphics[width=3.5cm,angle=-0]{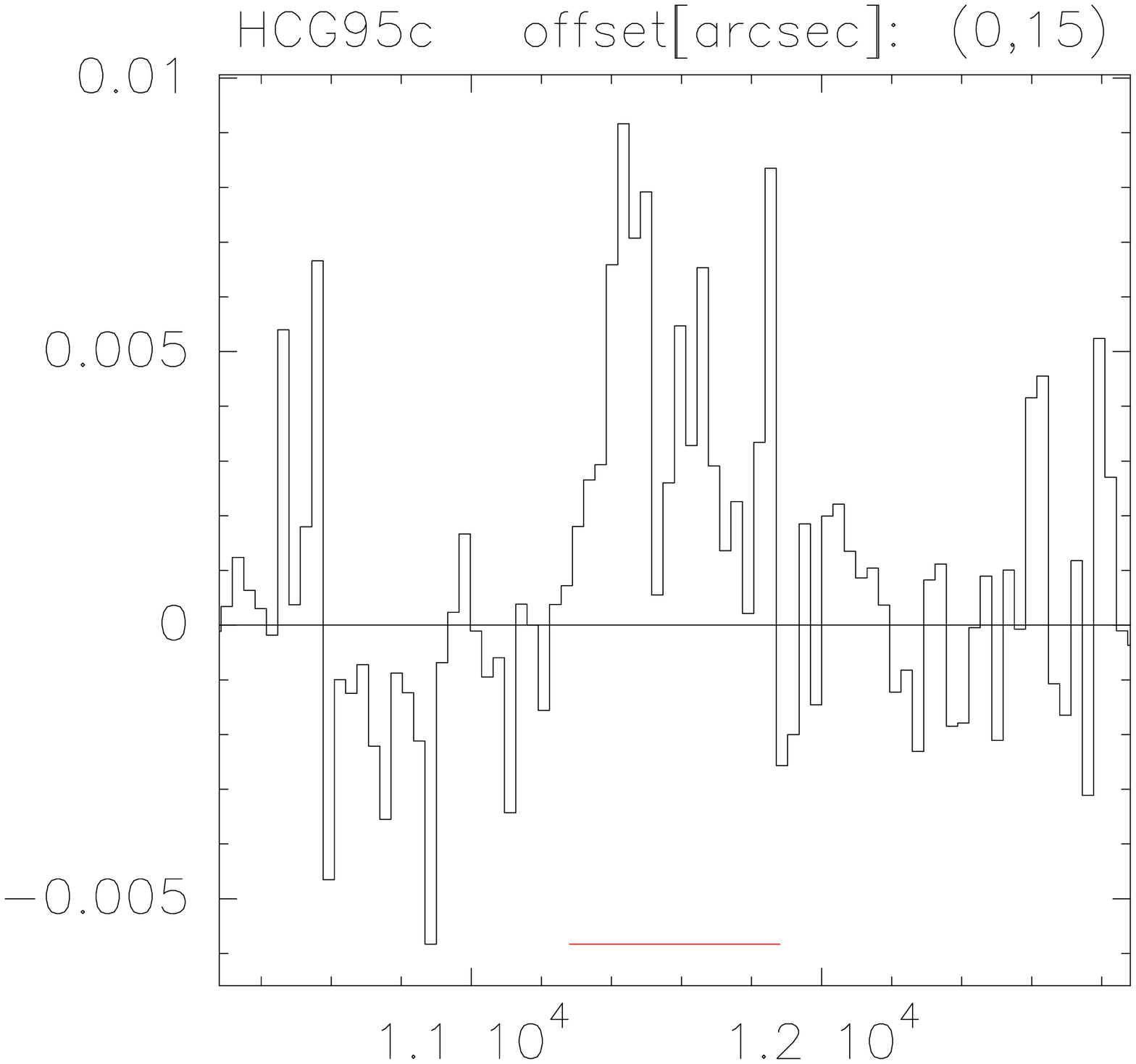}
}

\centerline{
\includegraphics[width=3.5cm,angle=-0]{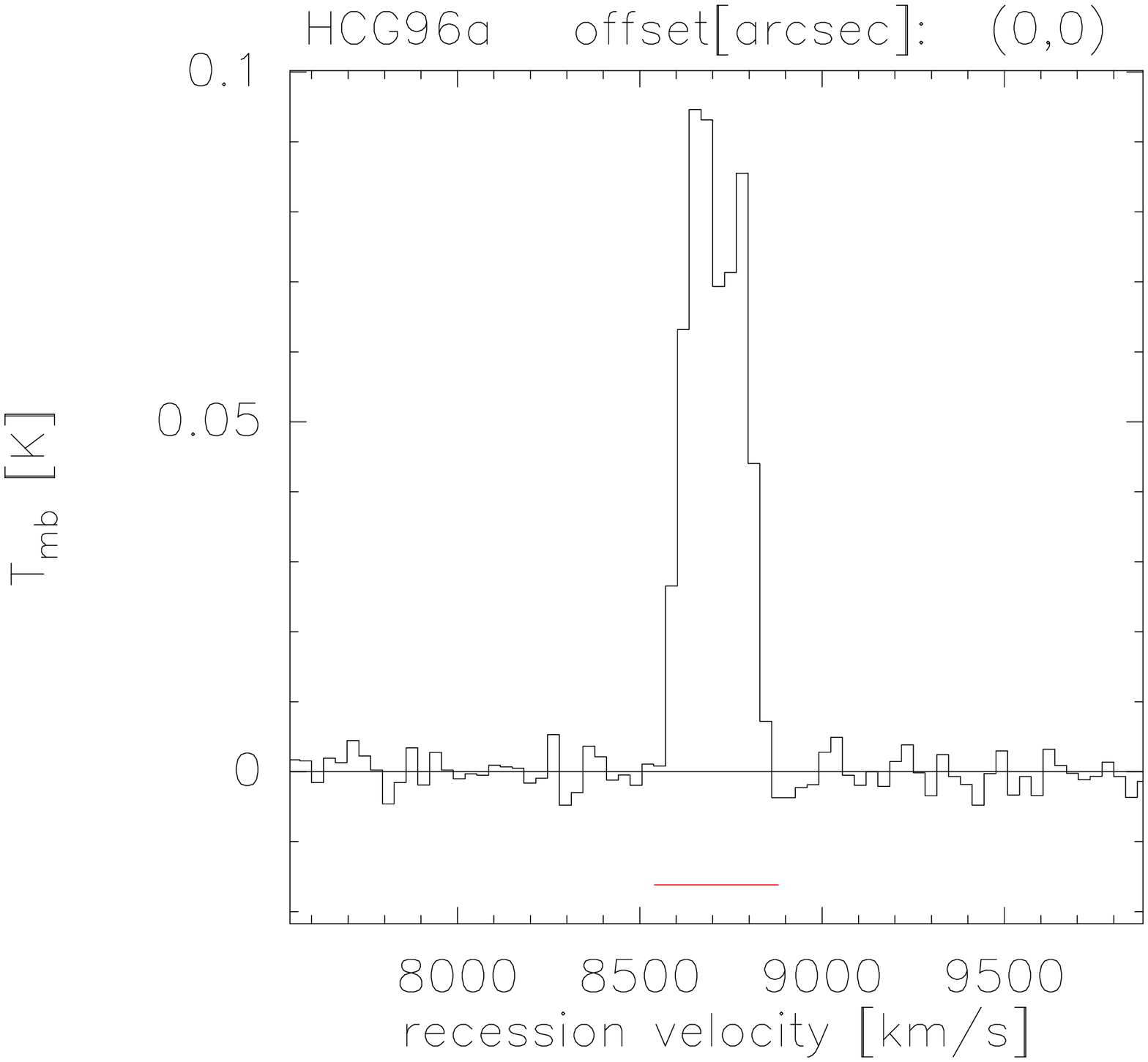}
}
\addtocounter{figure}{-1} 
\caption{(continued)} 

\end{figure*}


\begin{figure*}
\centerline{
\includegraphics[width=3.5cm,angle=-0]{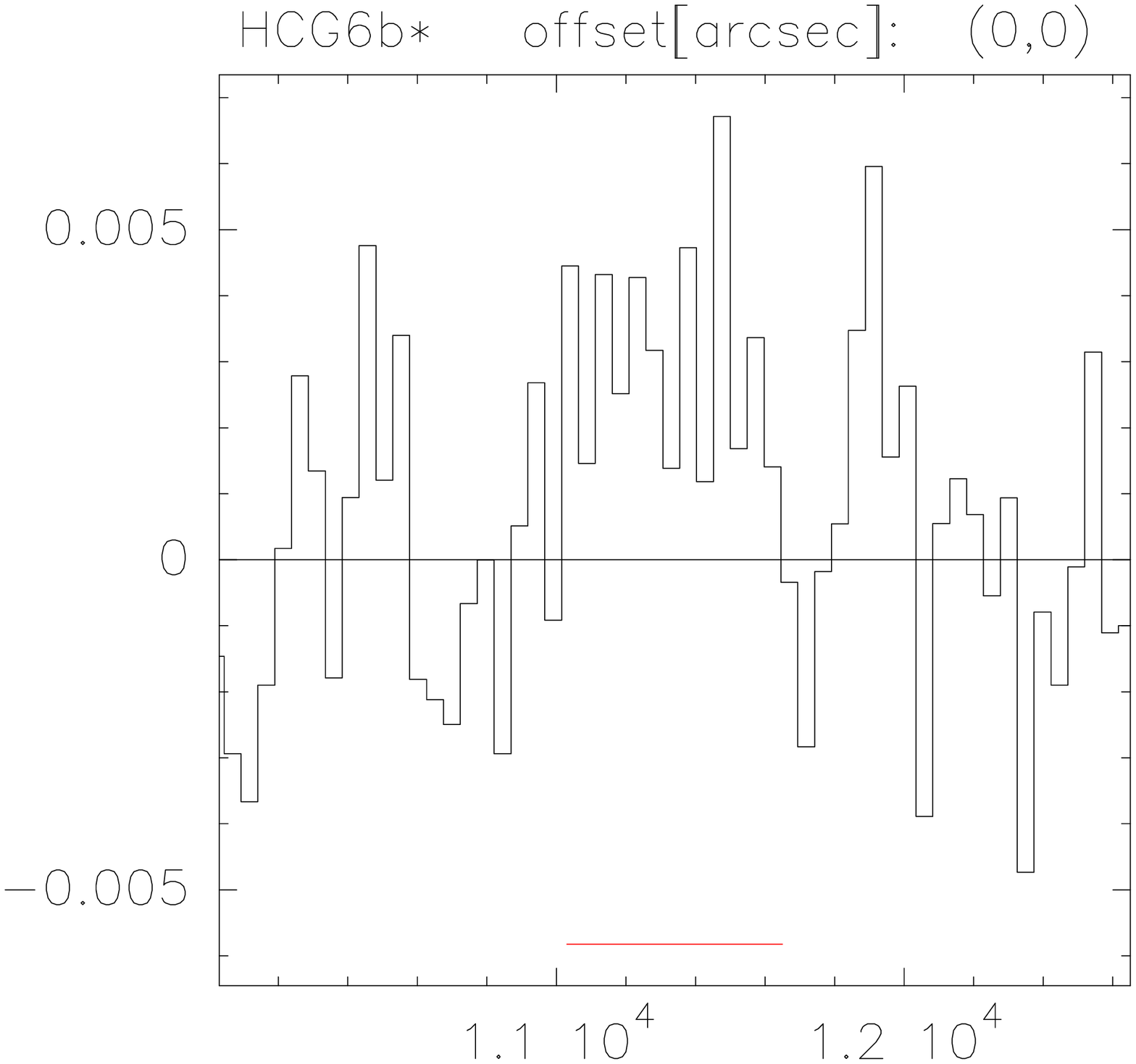}
\includegraphics[width=3.5cm,angle=-0]{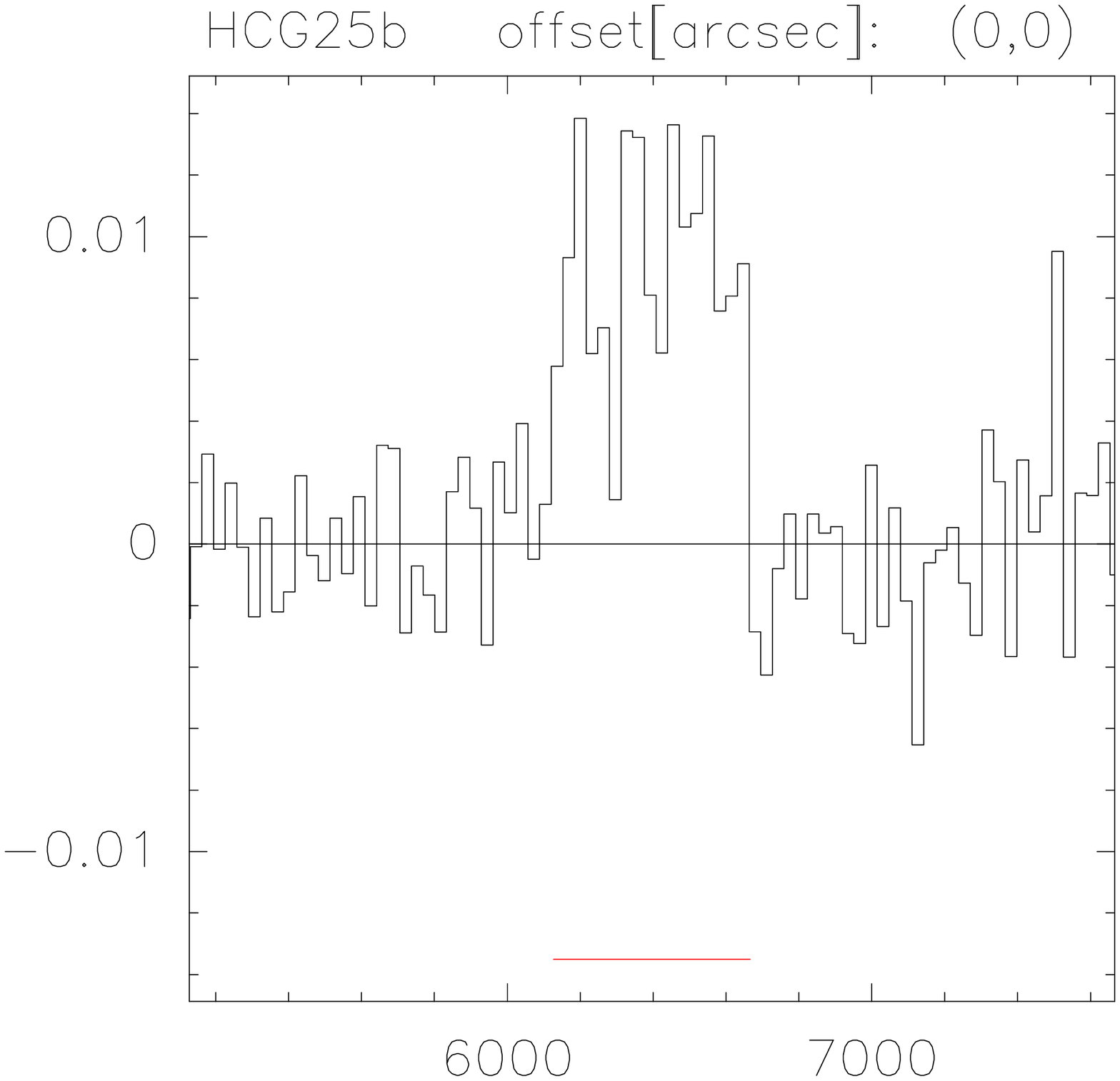}
\includegraphics[width=3.5cm,angle=-0]{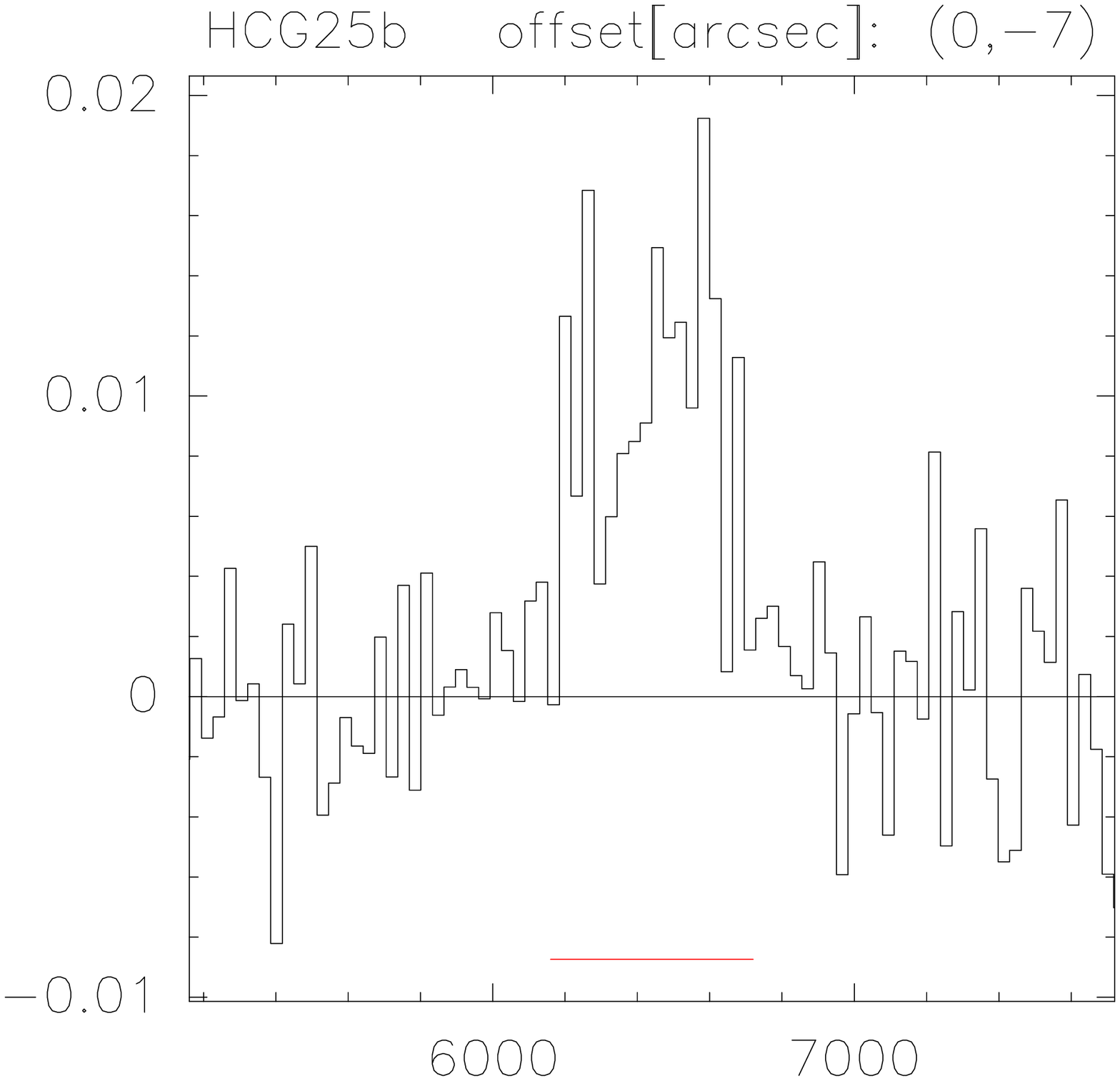}
\includegraphics[width=3.5cm,angle=-0]{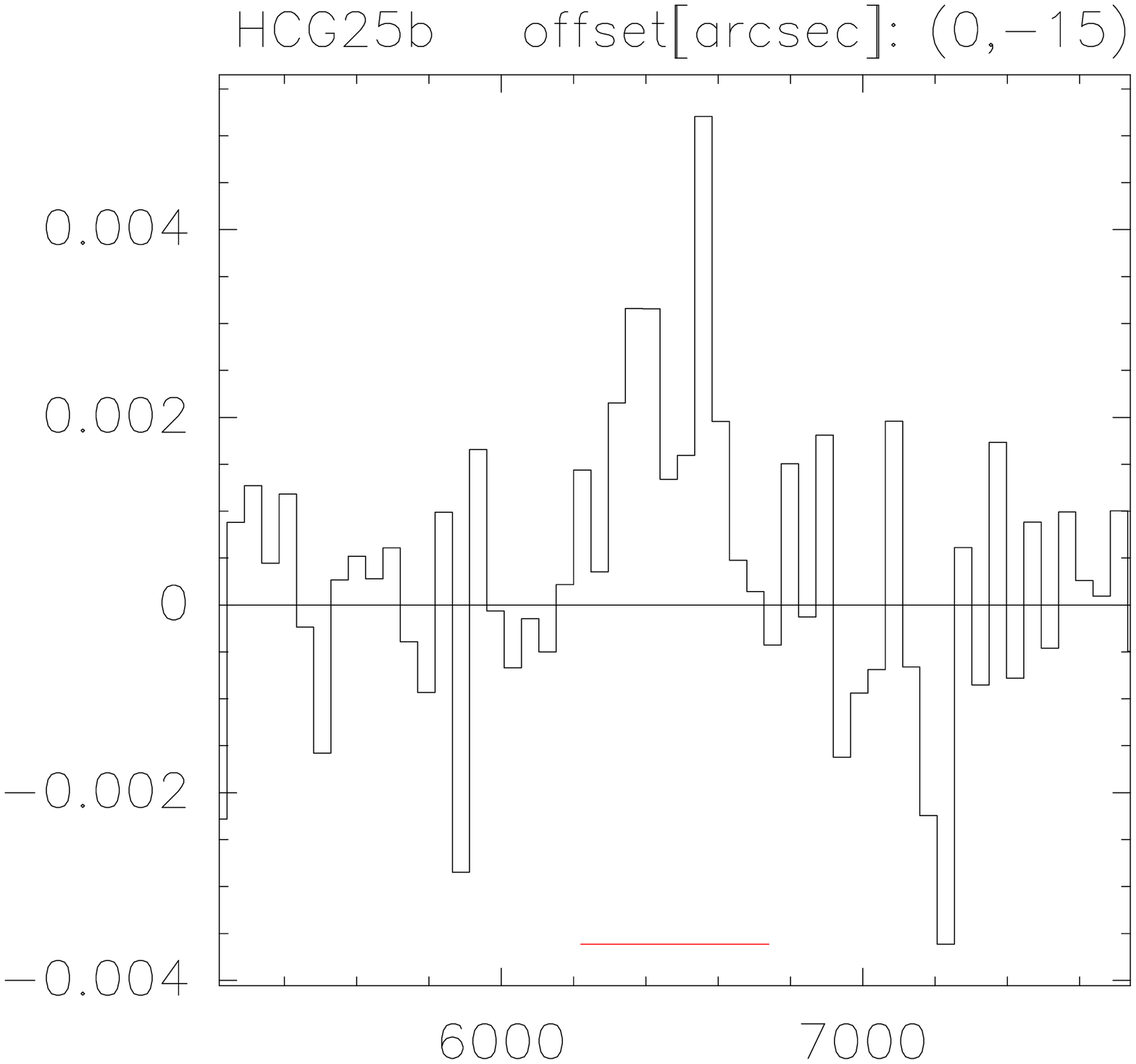}
\includegraphics[width=3.5cm,angle=-0]{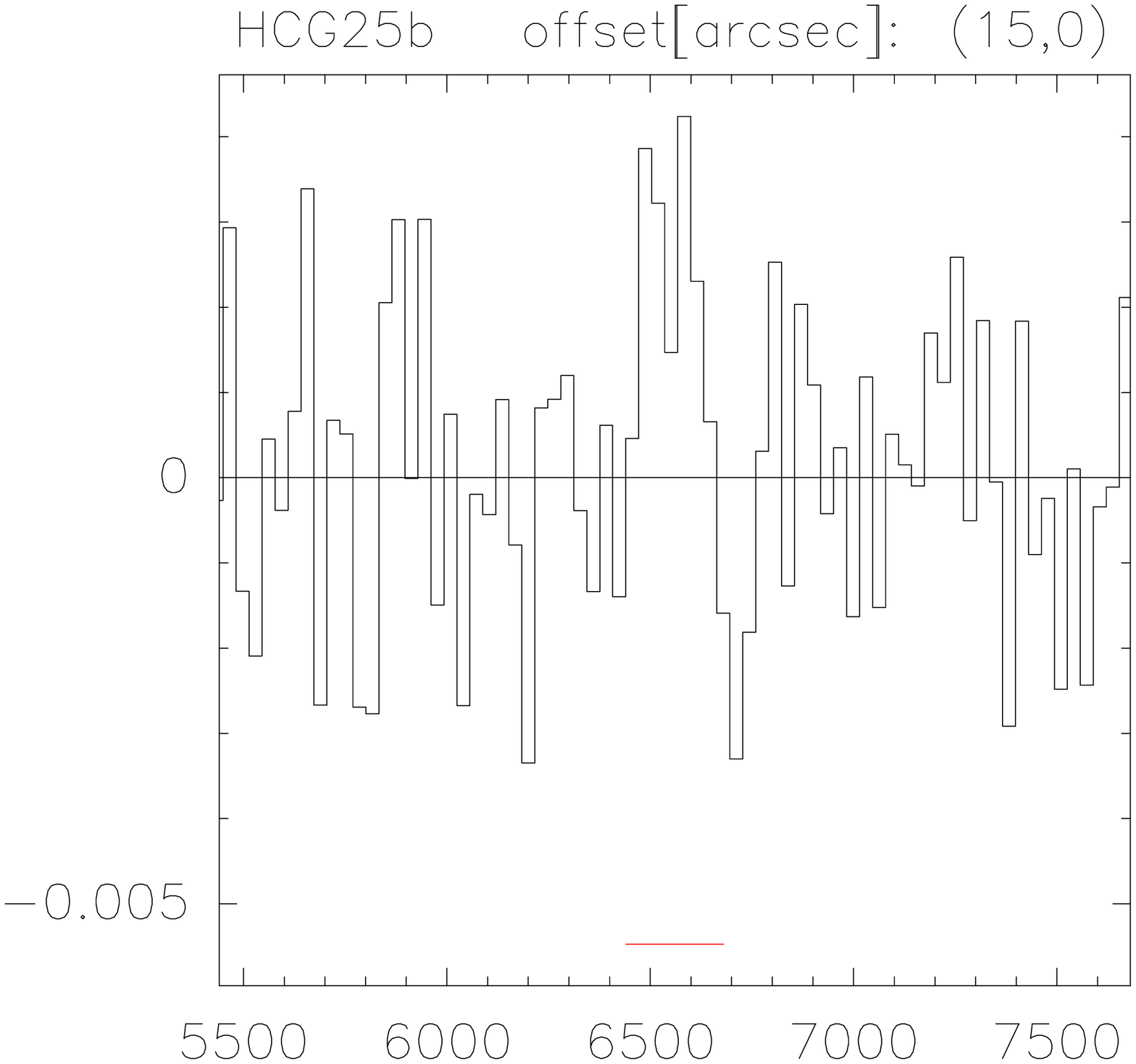}
}
\centerline{
\includegraphics[width=3.5cm,angle=-0]{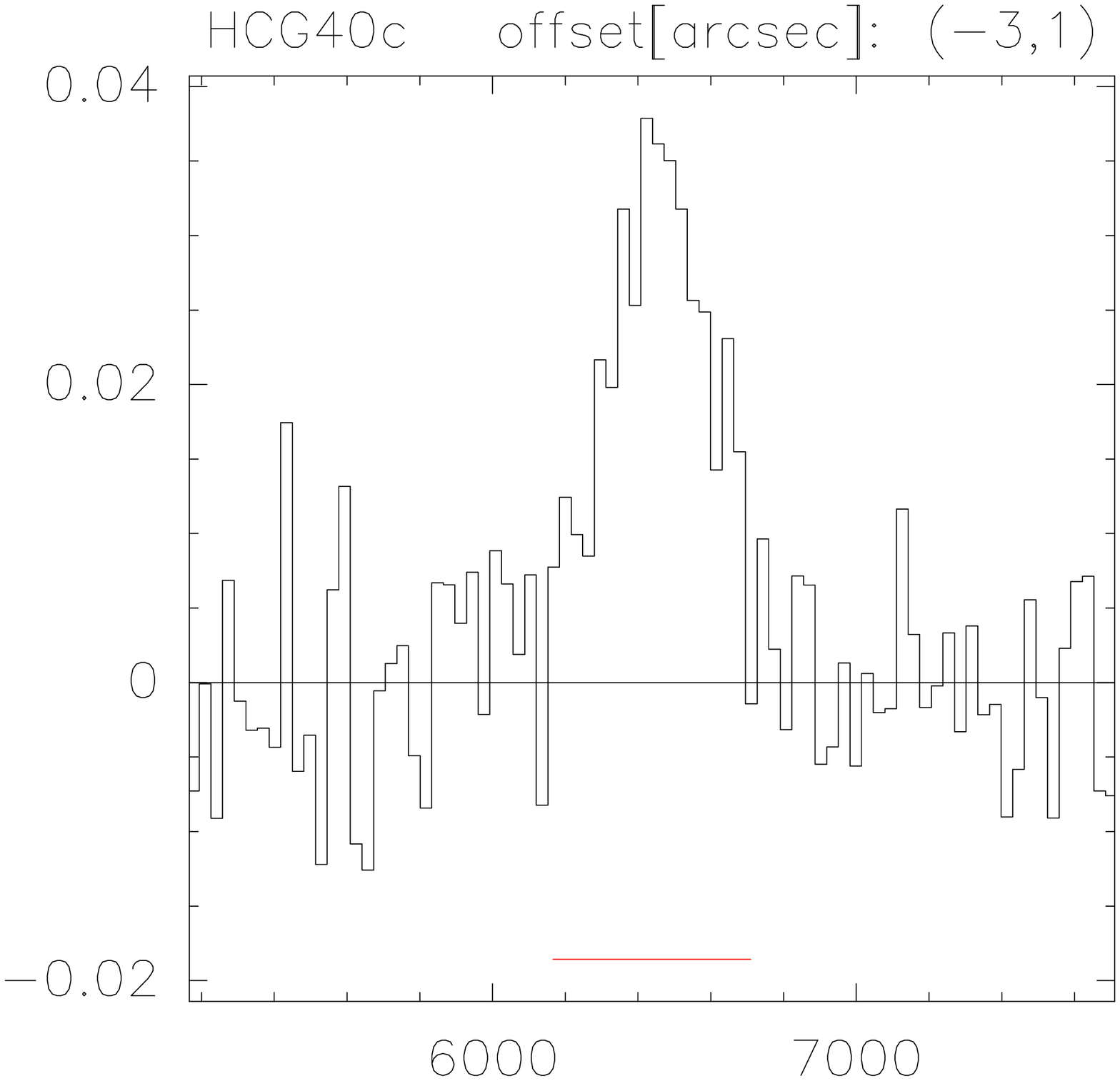}
\includegraphics[width=3.5cm,angle=-0]{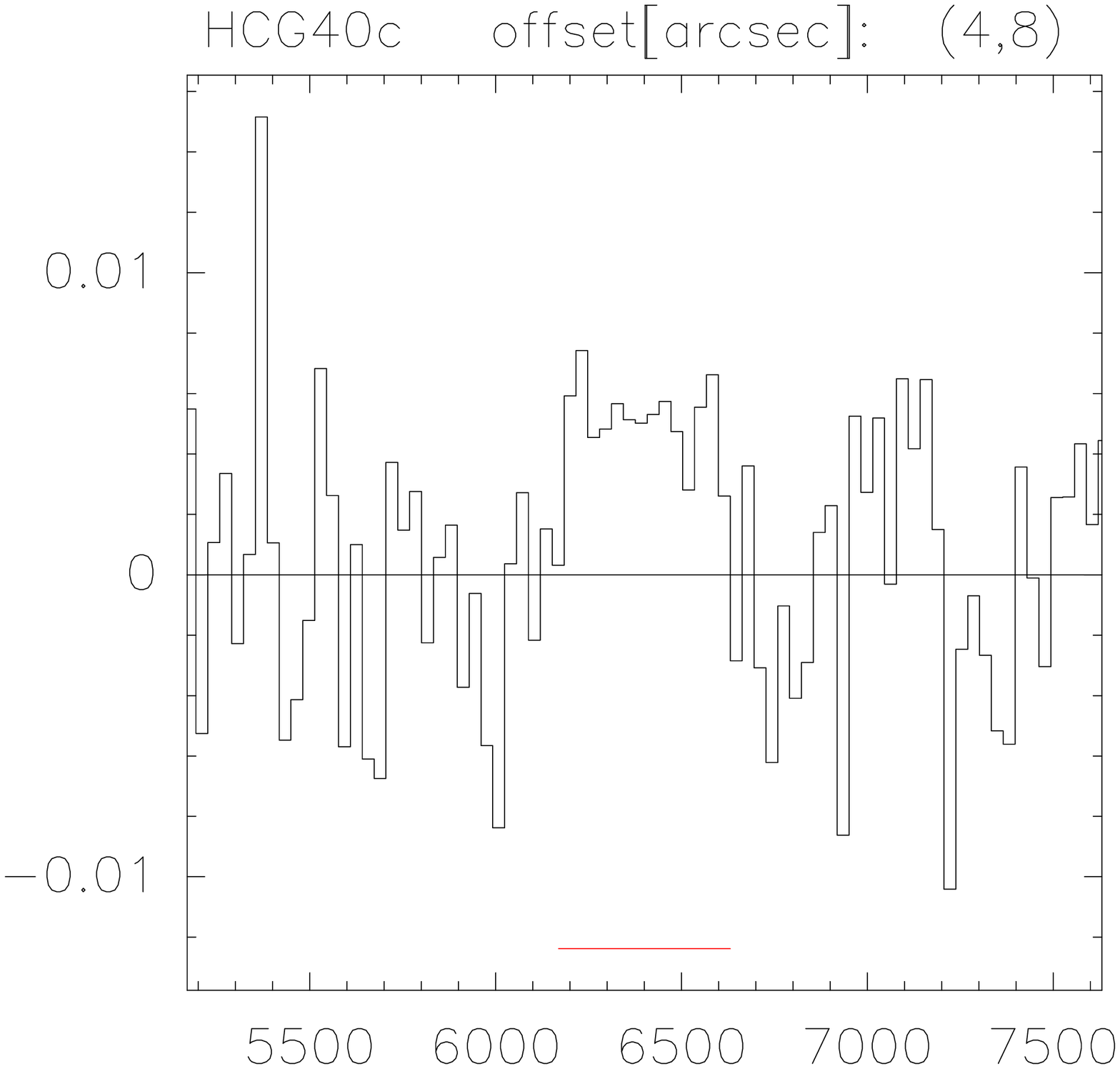}
\includegraphics[width=3.5cm,angle=-0]{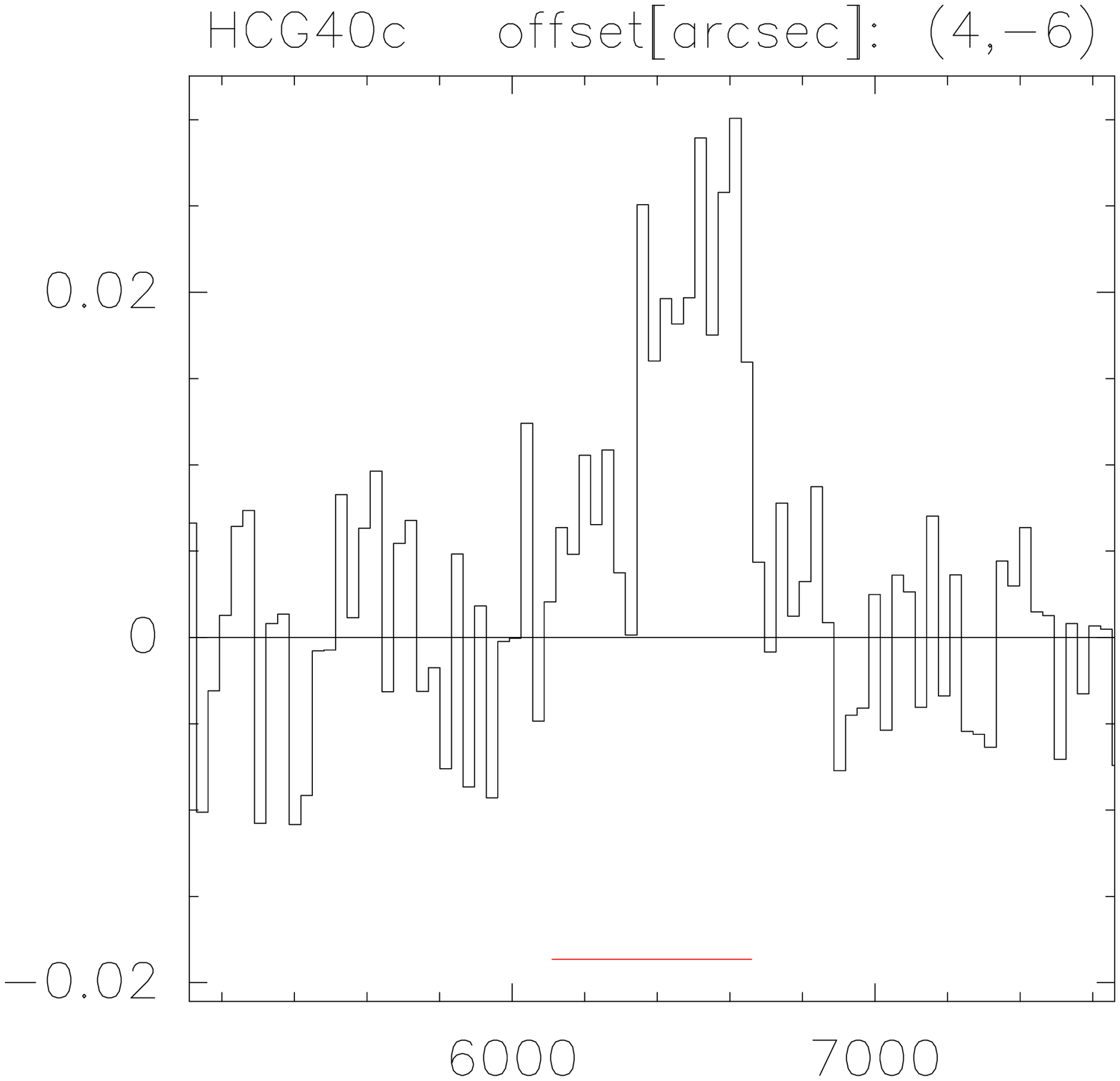}
\includegraphics[width=3.5cm,angle=-0]{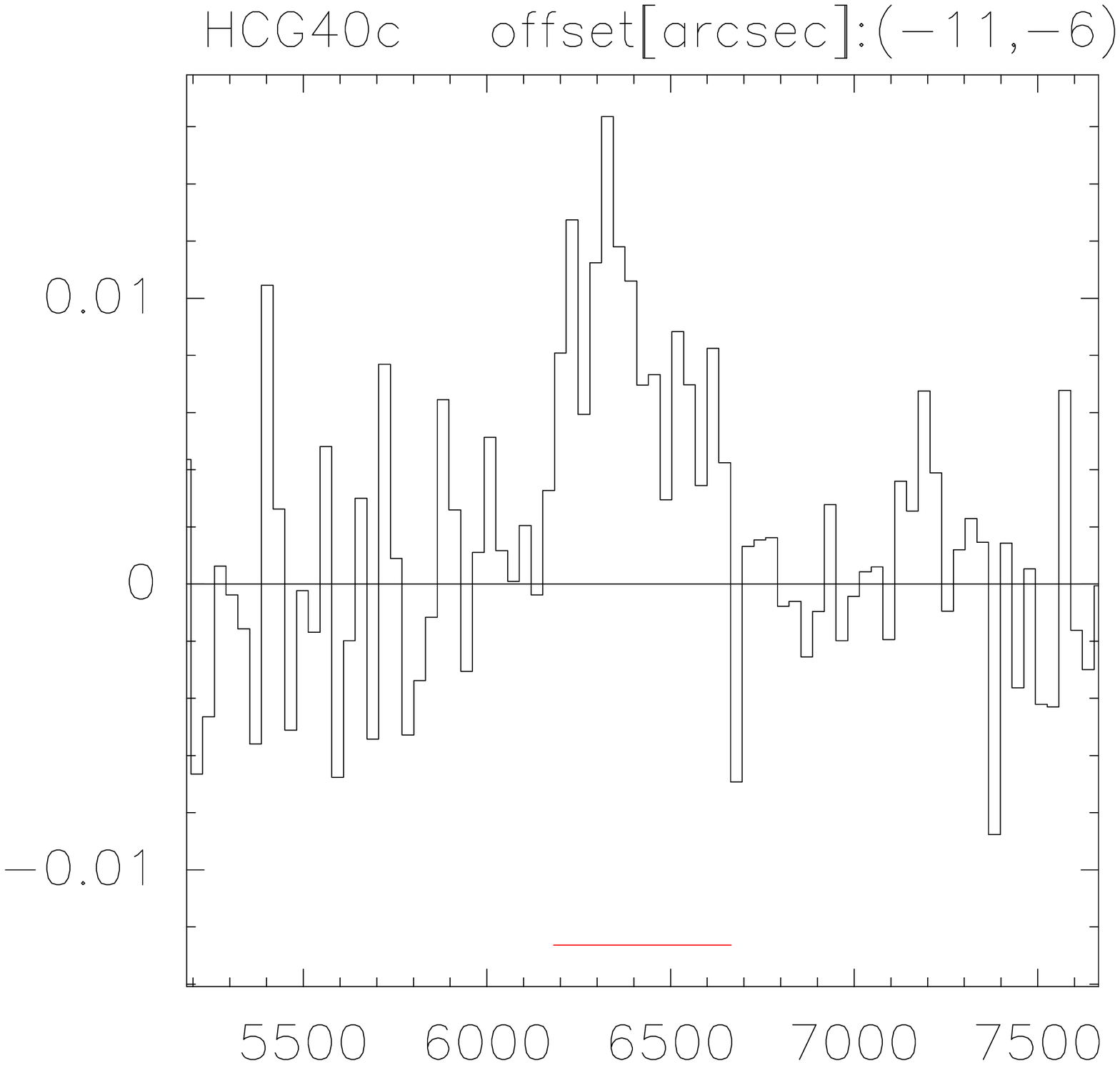}
\includegraphics[width=3.5cm,angle=-0]{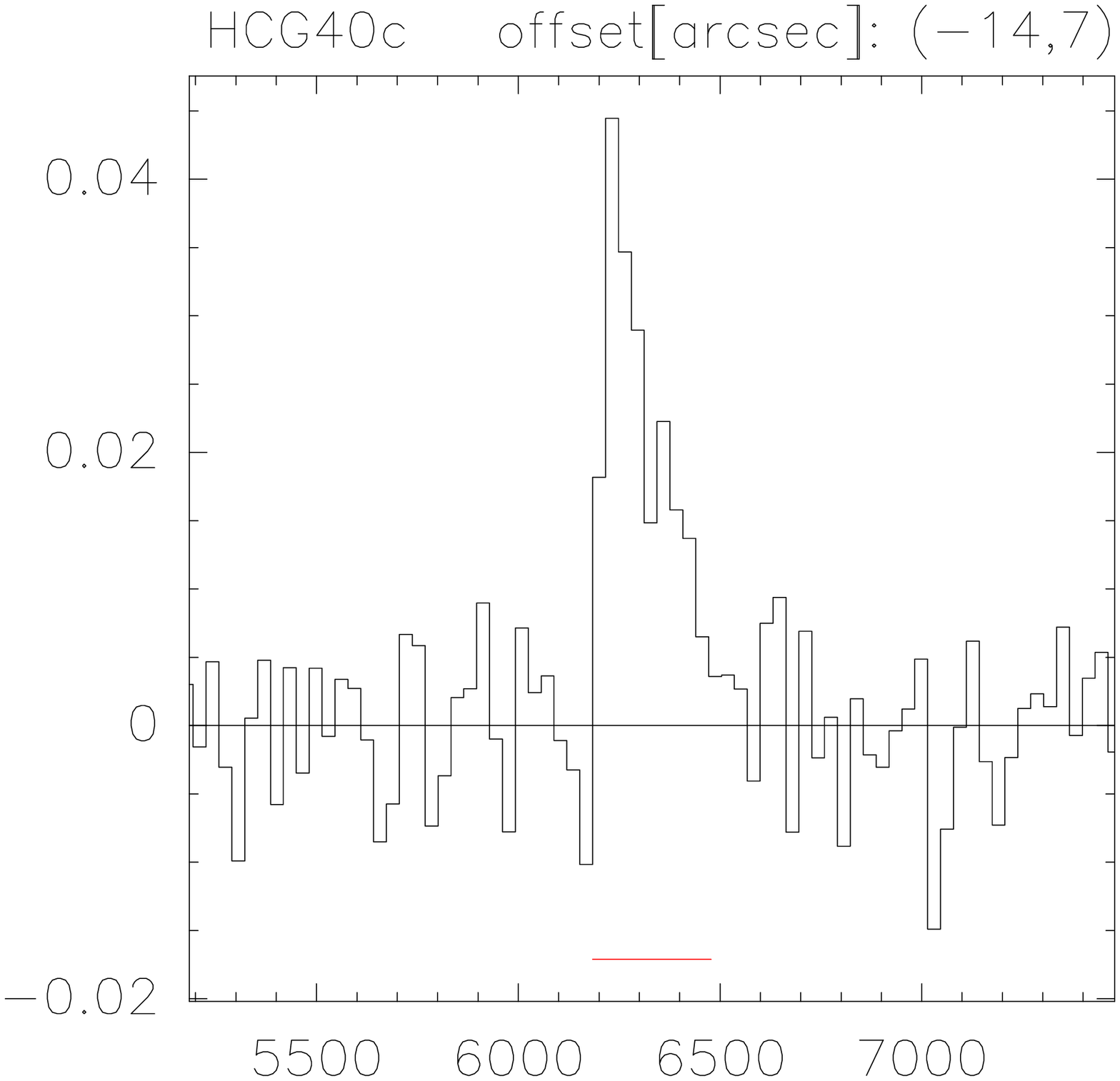}
}
\centerline{
\includegraphics[width=3.5cm,angle=-0]{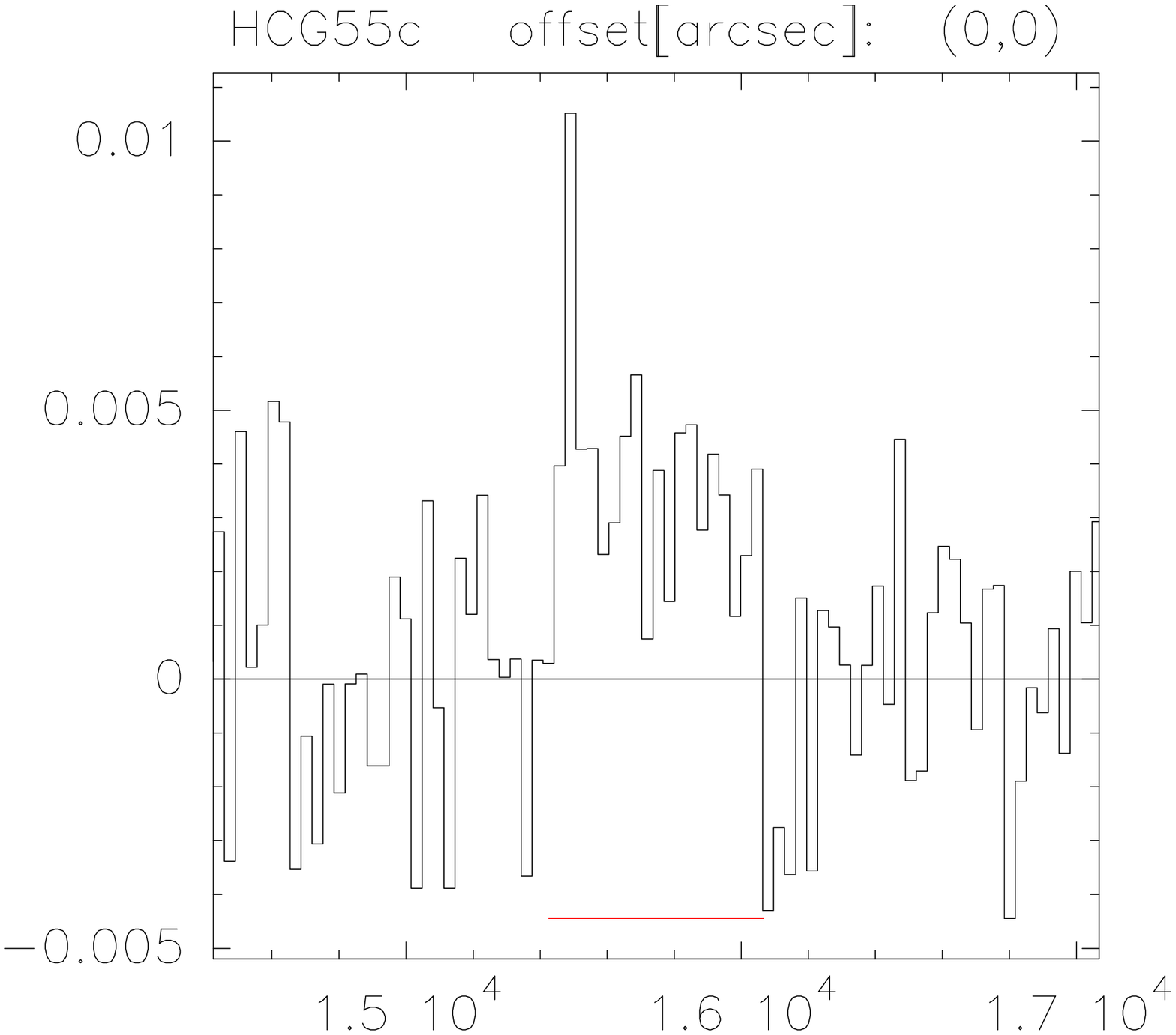}
\includegraphics[width=3.5cm,angle=-0]{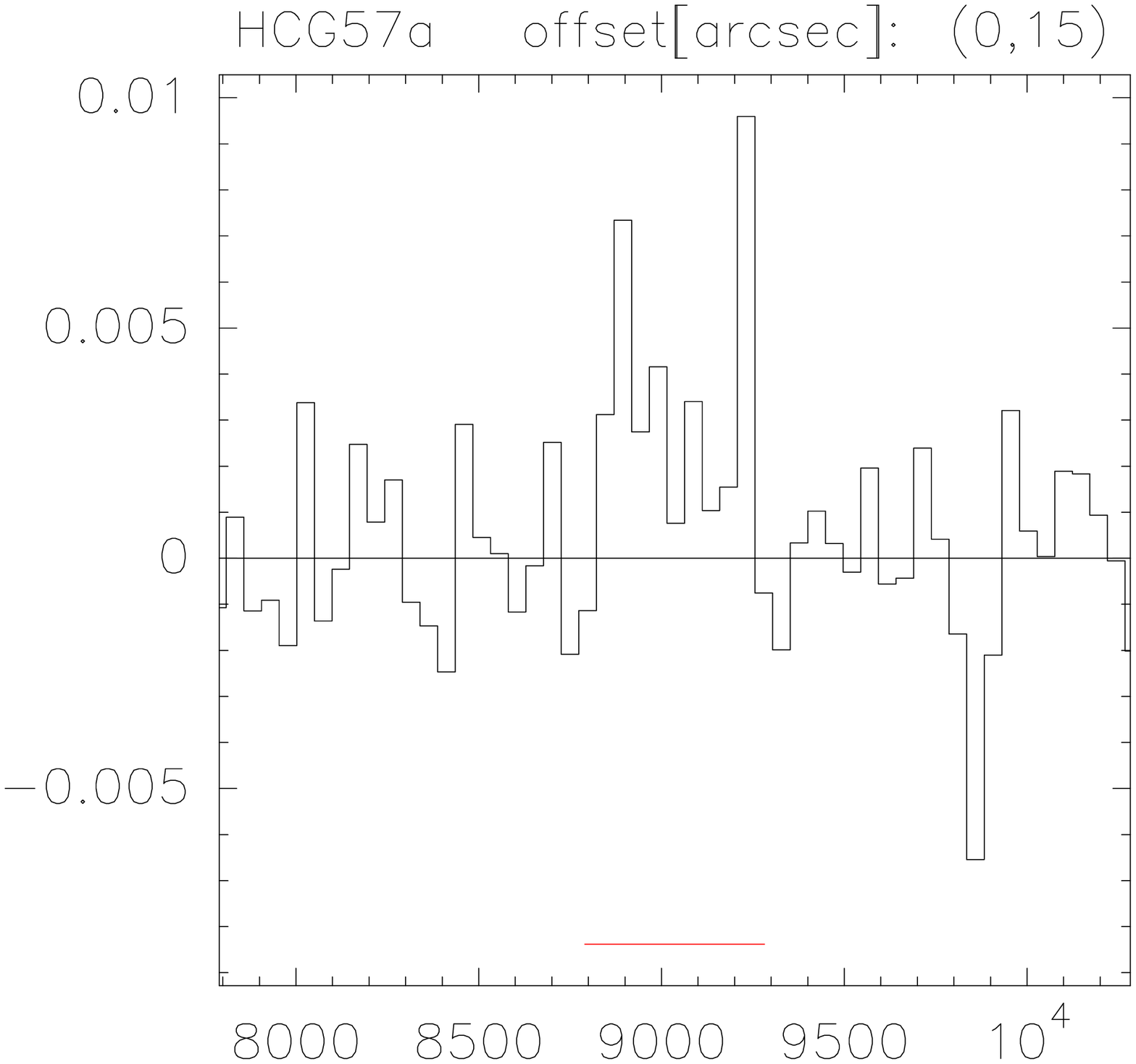}
\includegraphics[width=3.5cm,angle=-0]{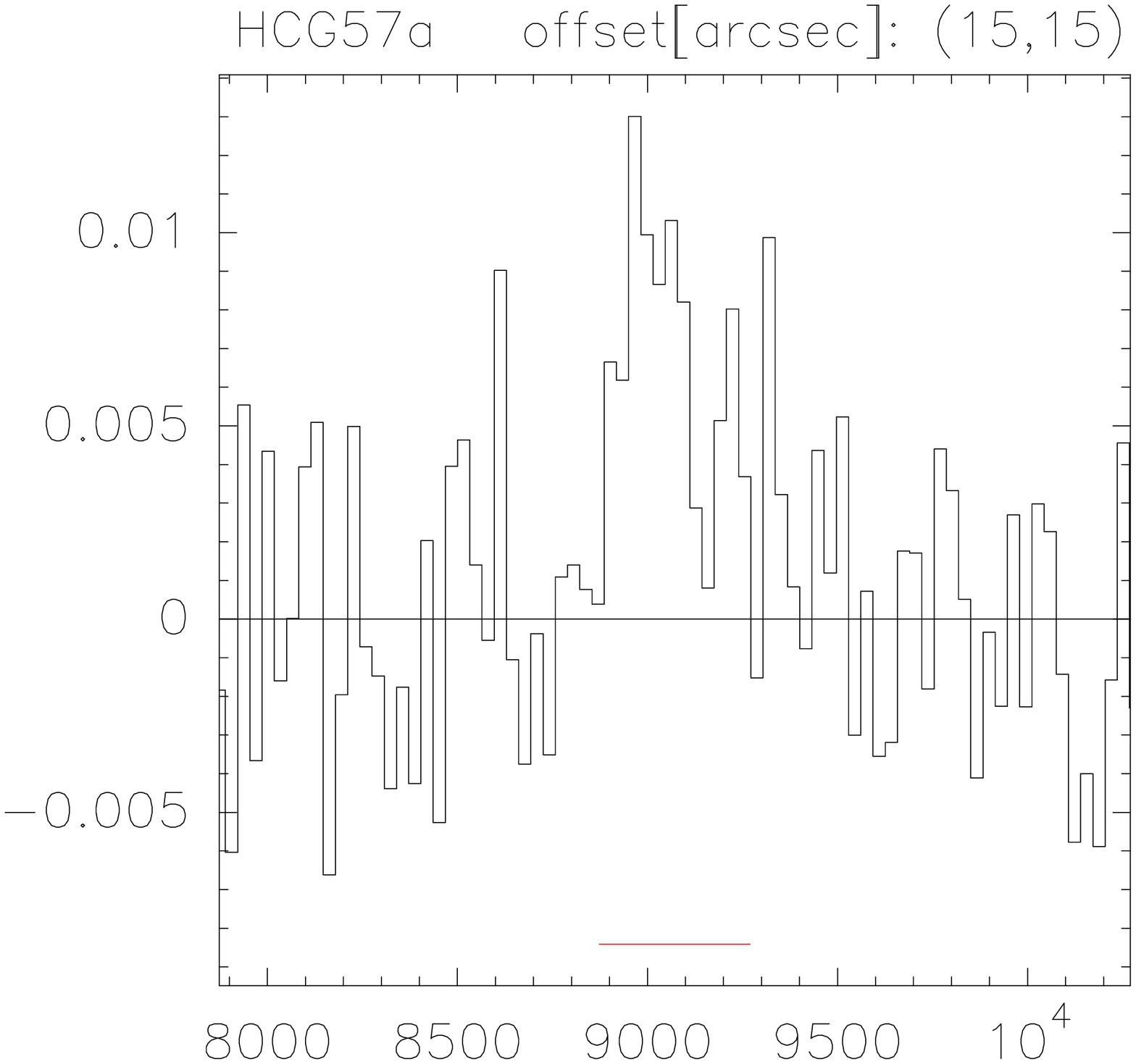}
\includegraphics[width=3.5cm,angle=-0]{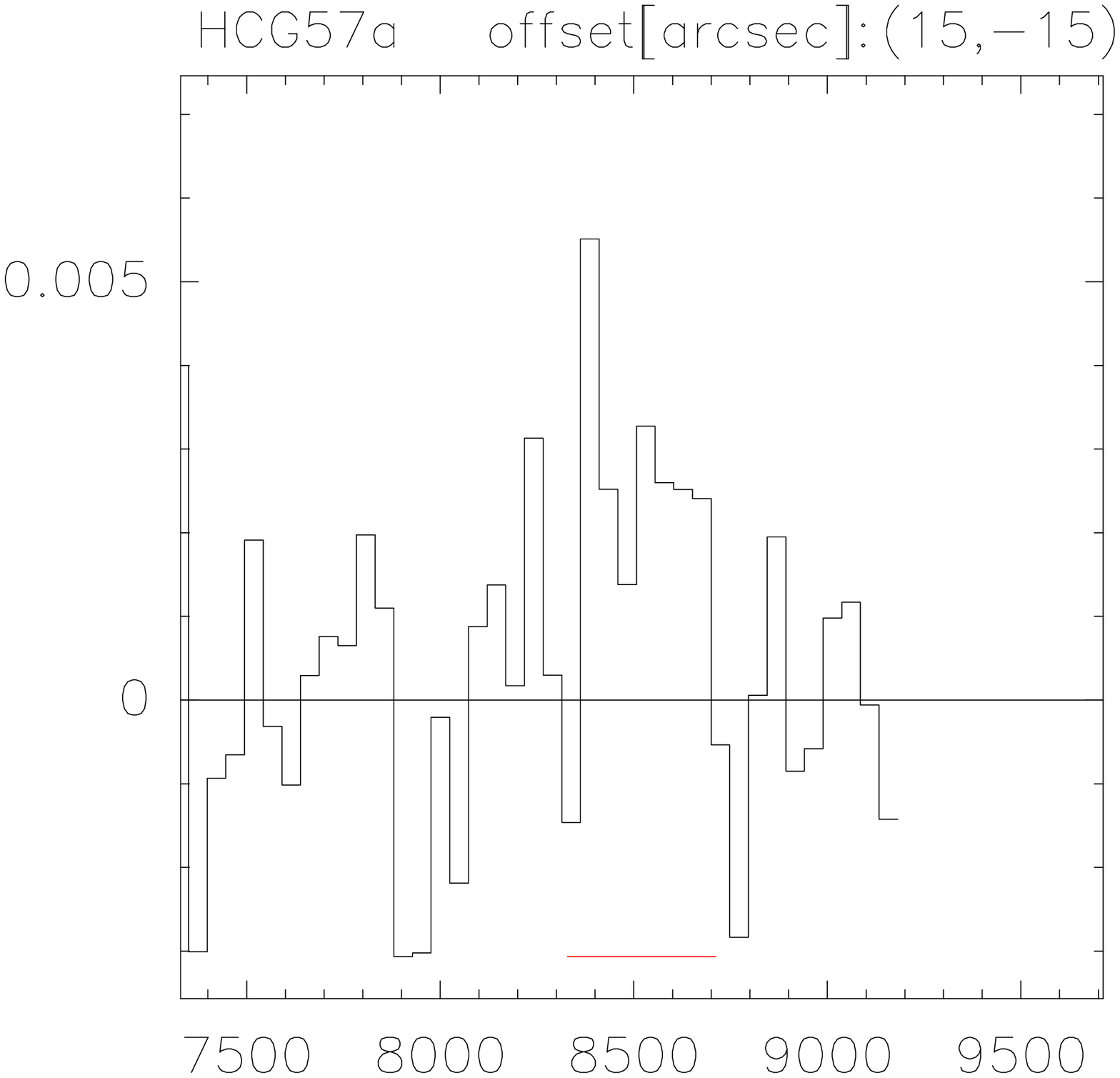}
\includegraphics[width=3.5cm,angle=-0]{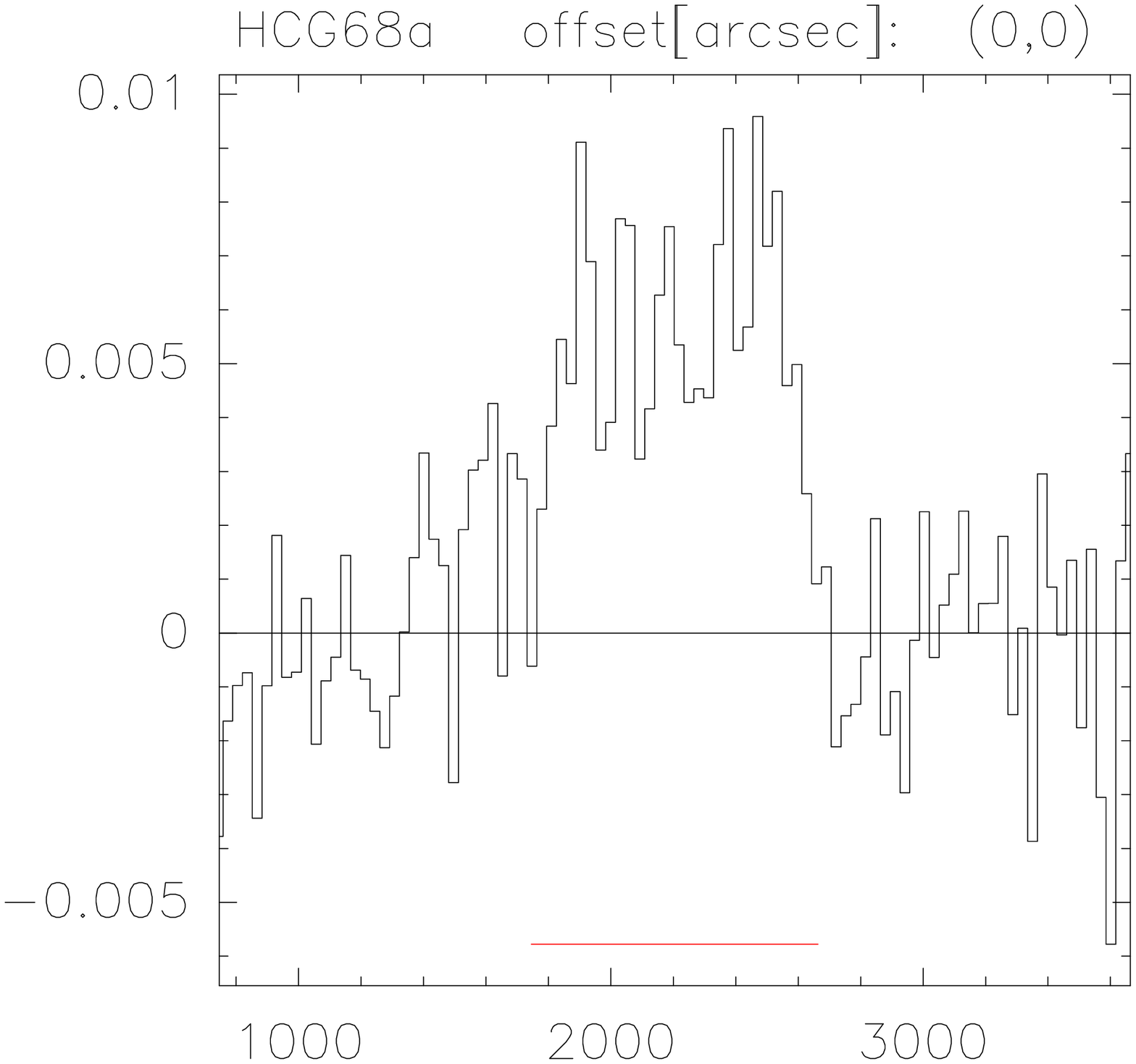}
}
\centerline{
\includegraphics[width=3.5cm,angle=-0]{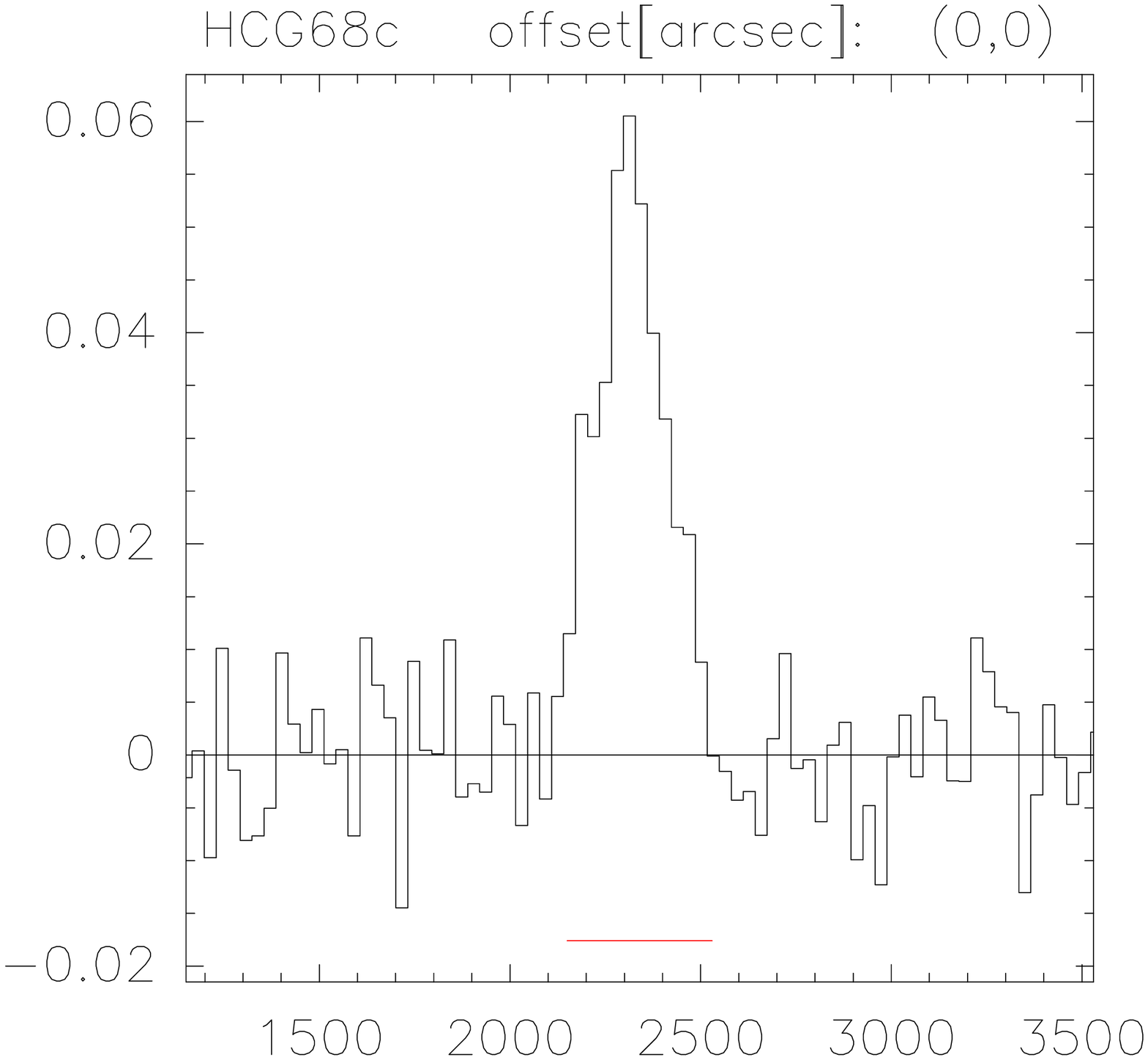}
\includegraphics[width=3.5cm,angle=-0]{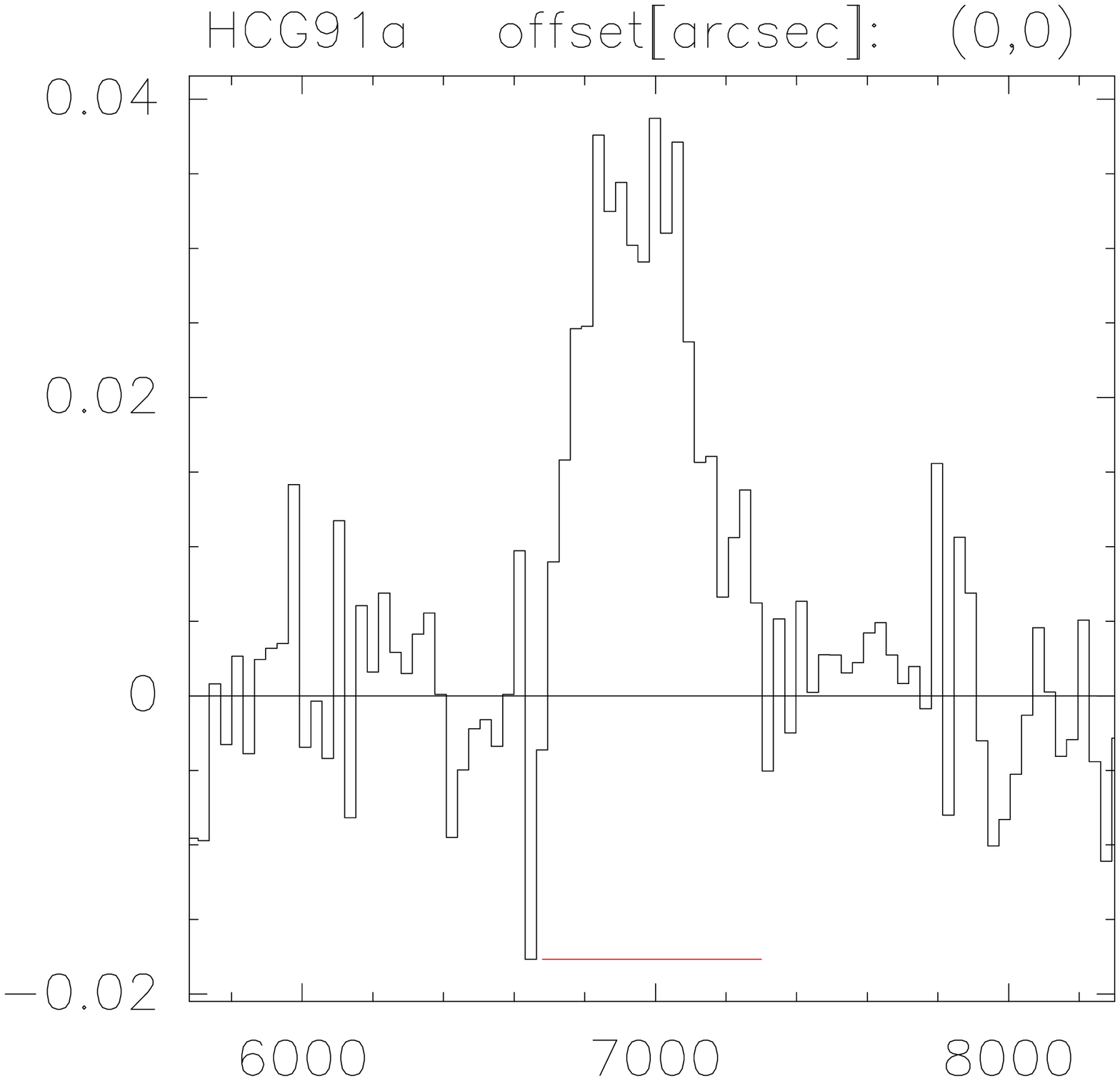}
\includegraphics[width=3.5cm,angle=-0]{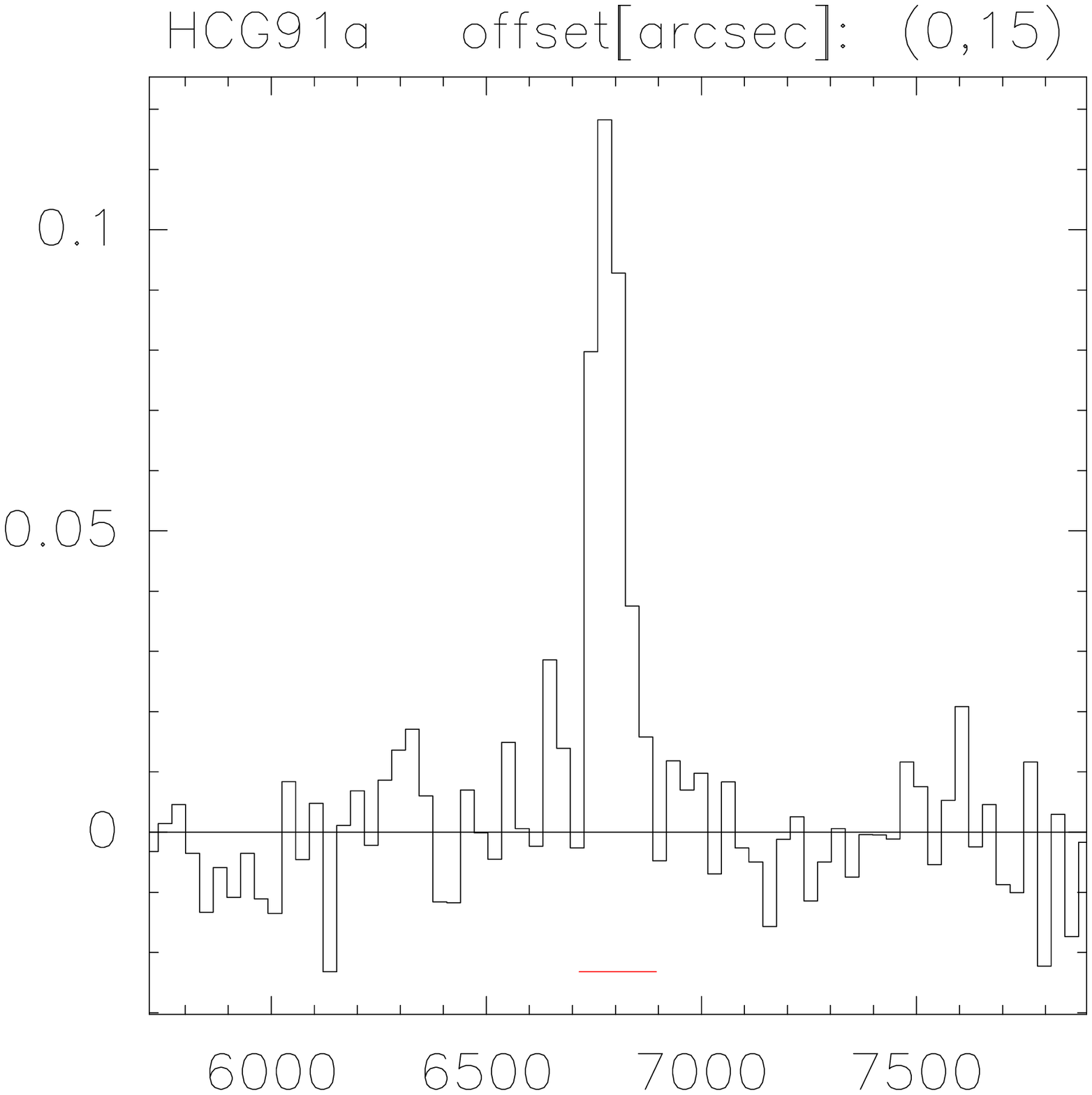}
\includegraphics[width=3.5cm,angle=-0]{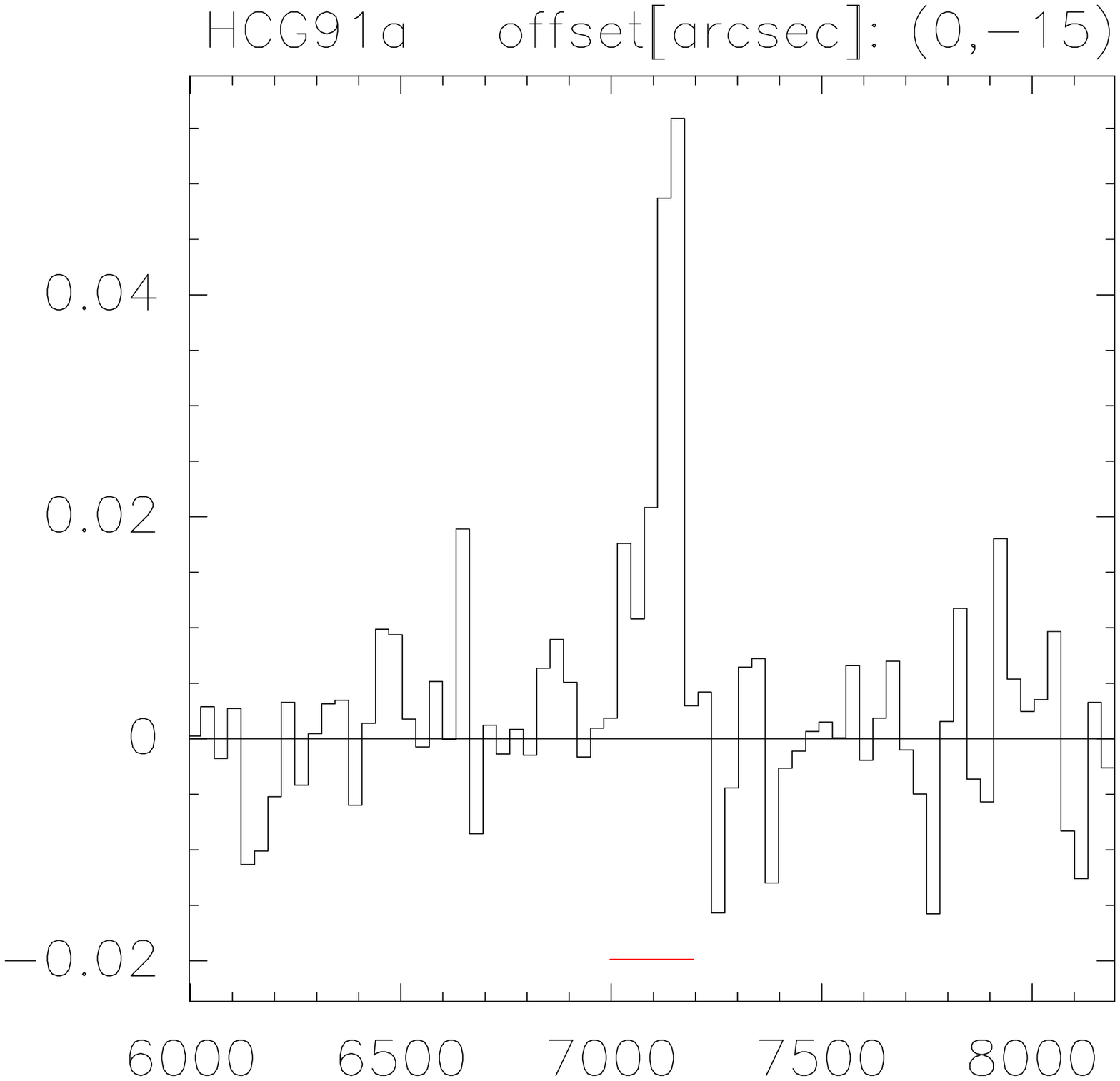}
\includegraphics[width=3.5cm,angle=-0]{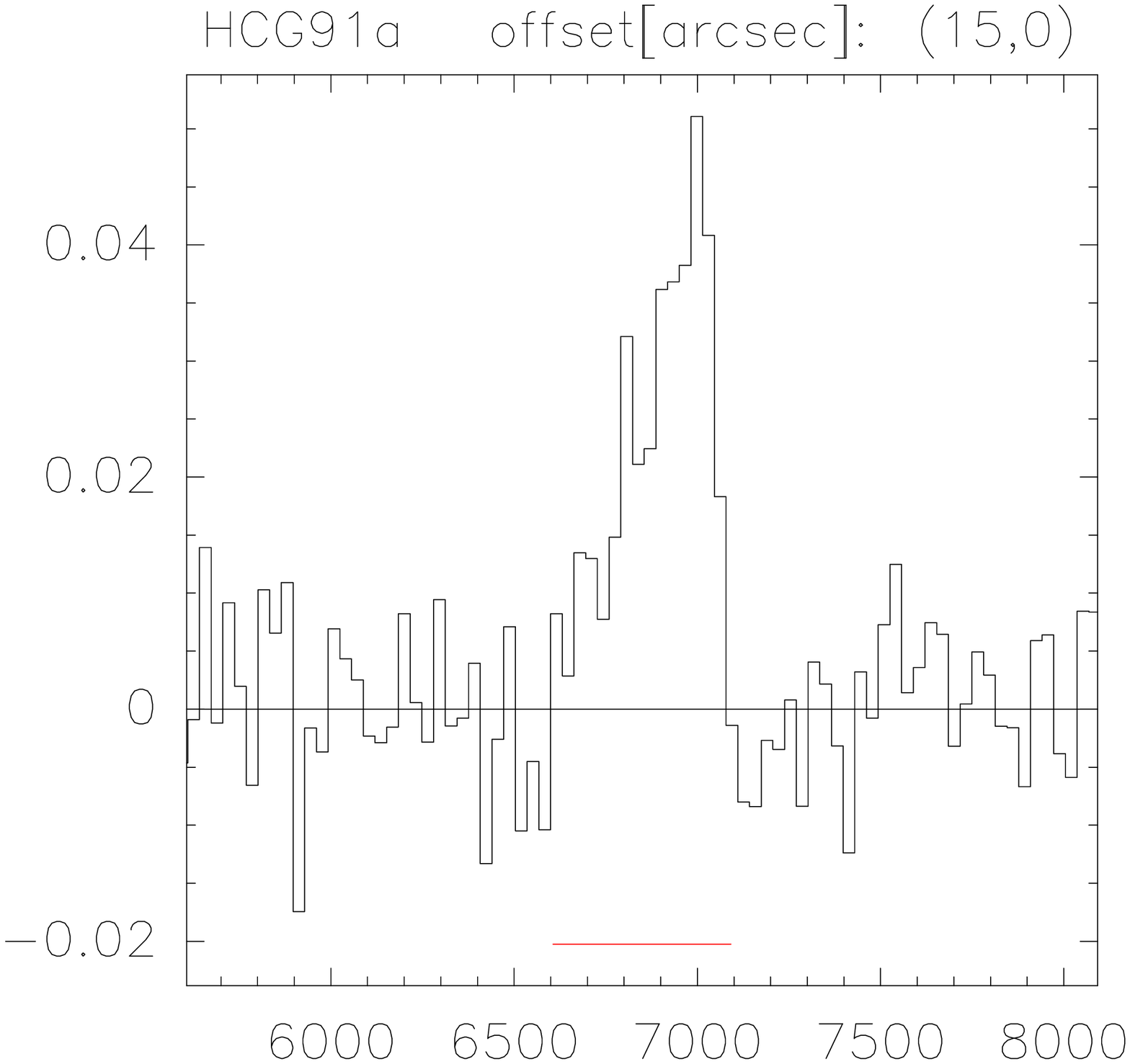}
}

\centerline{
\includegraphics[width=3.5cm,angle=-0]{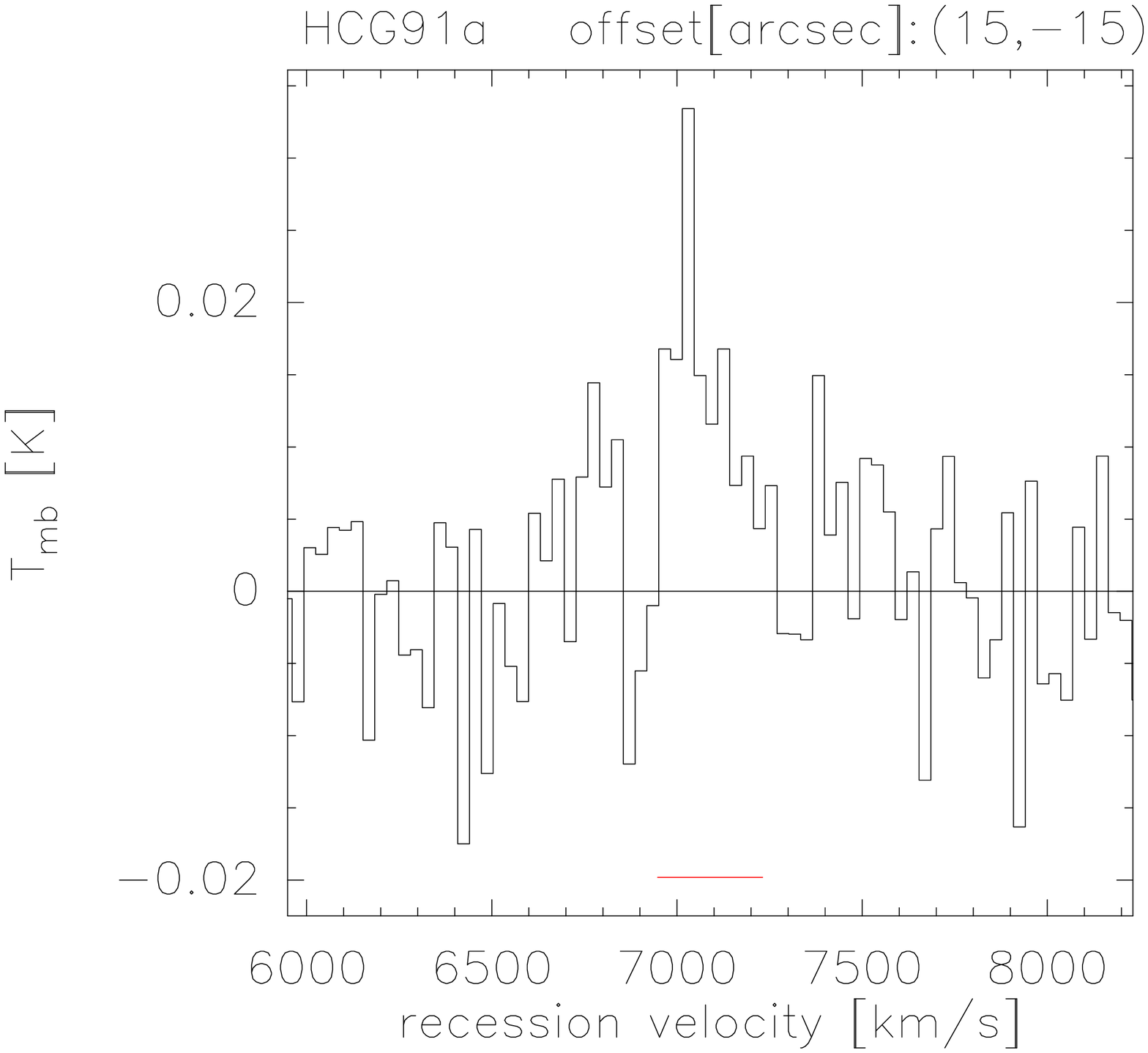}
\includegraphics[width=3.5cm,angle=-0]{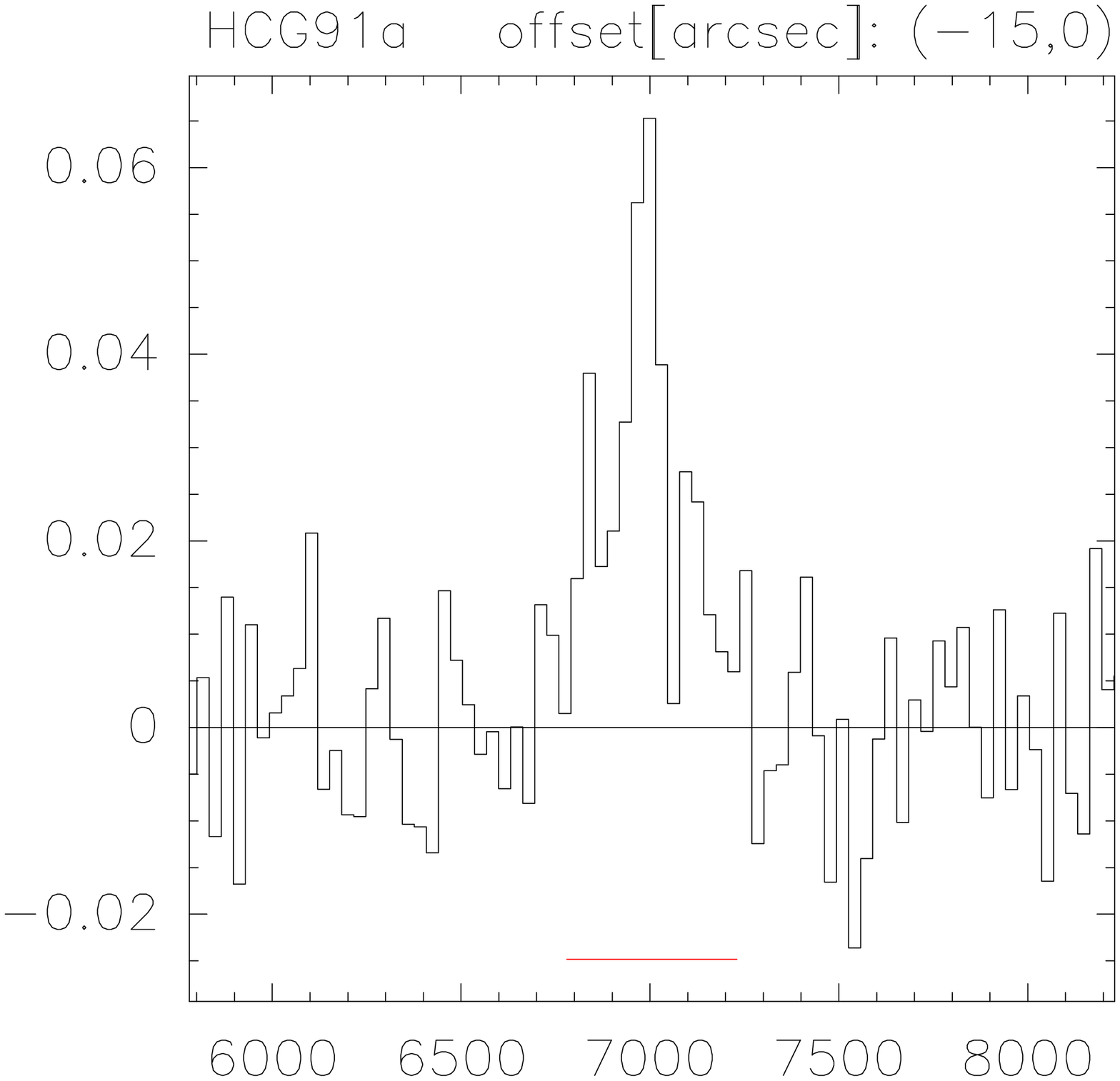}
\includegraphics[width=3.5cm,angle=-0]{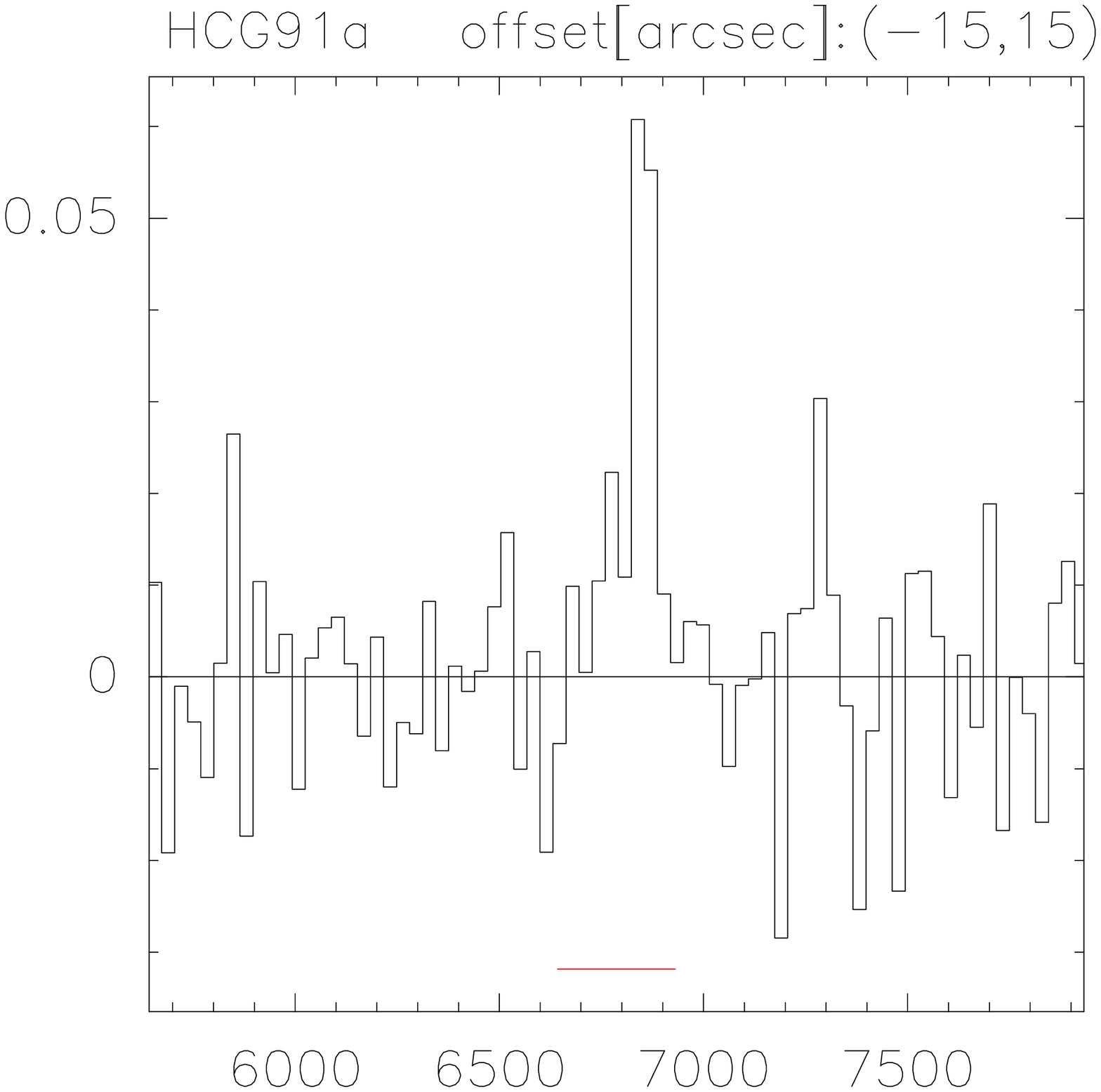}
\includegraphics[width=3.5cm,angle=-0]{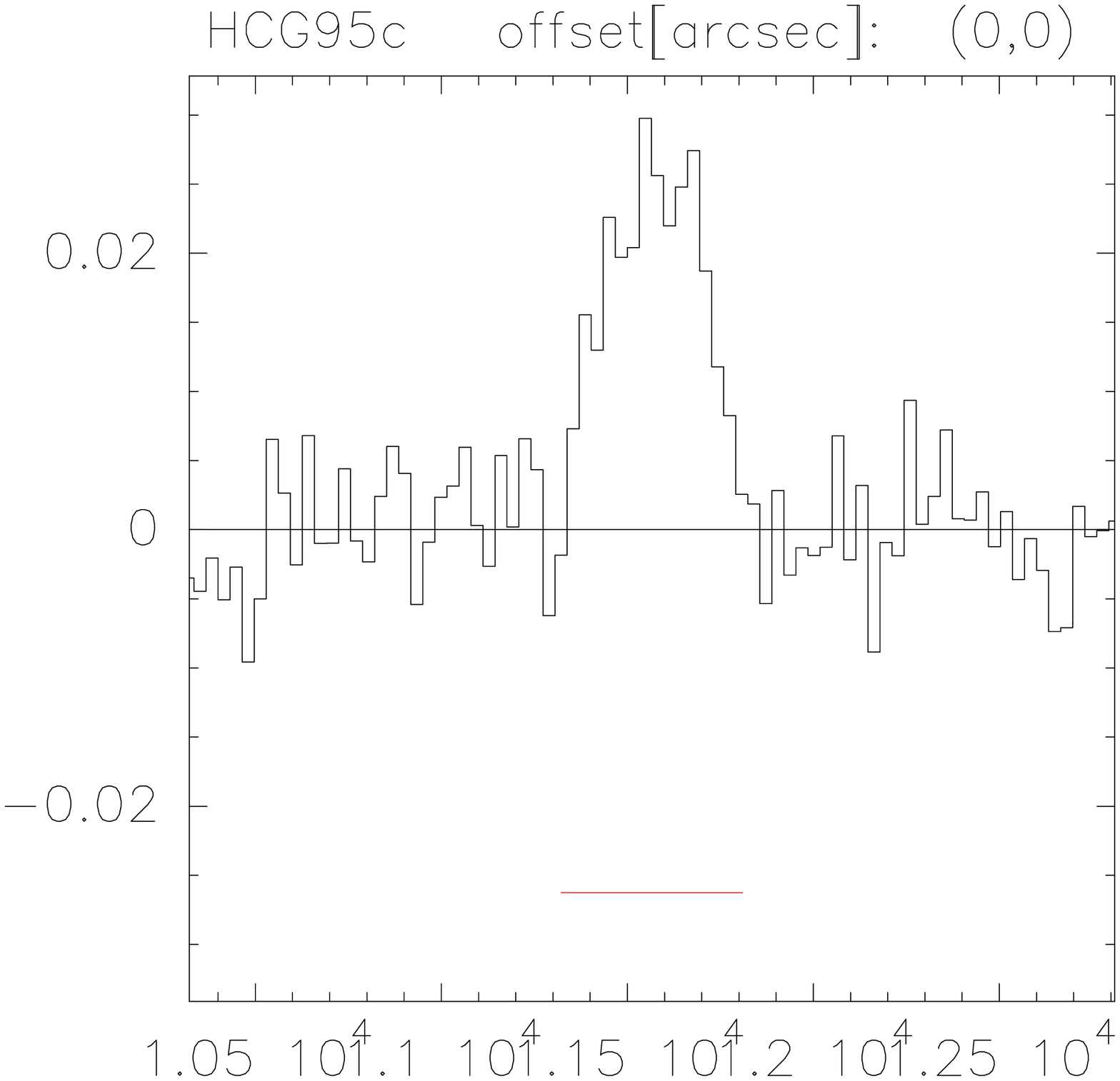}
\includegraphics[width=3.5cm,angle=-0]{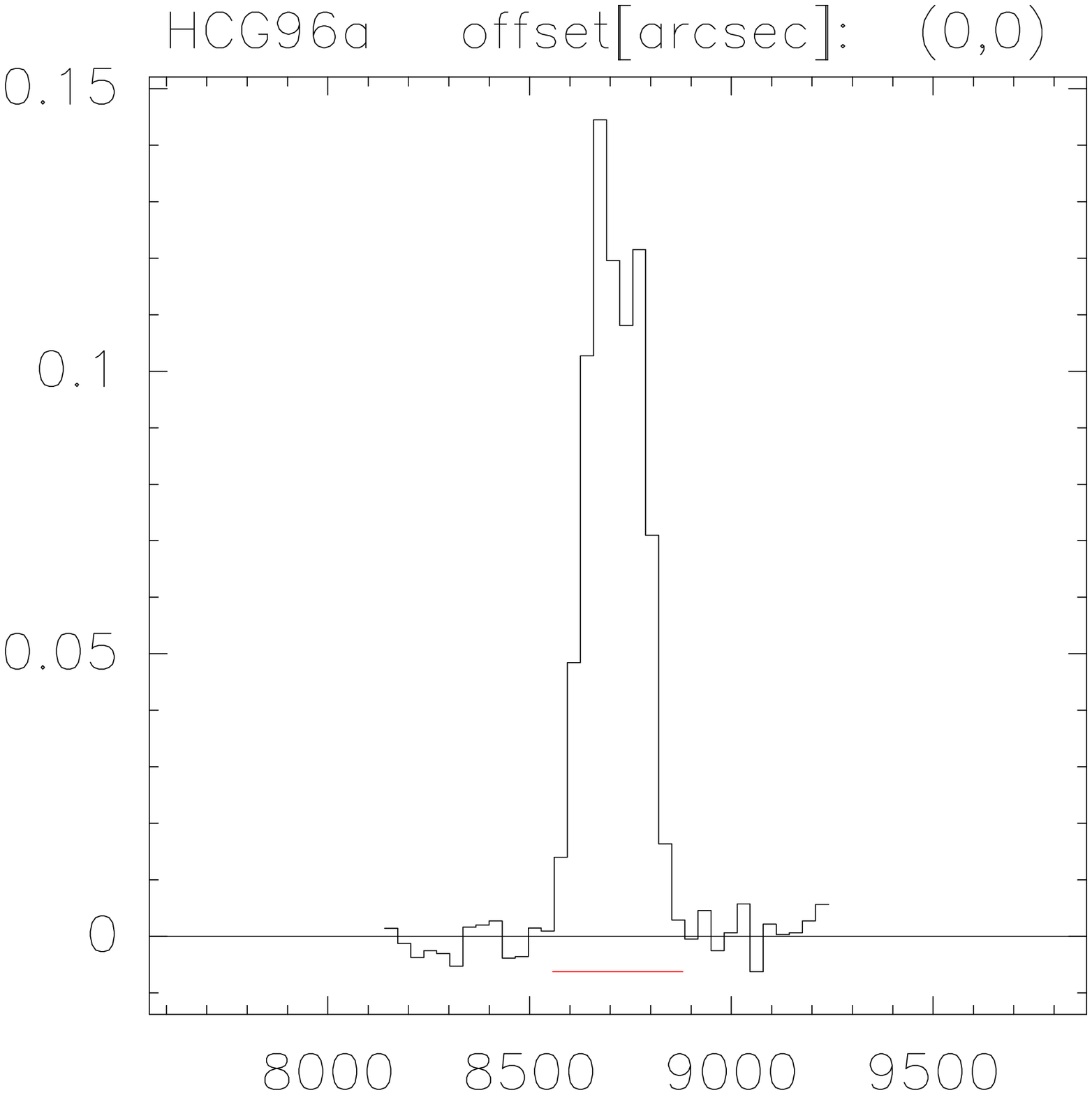}
}

\caption{CO(2-1) spectra of the detected (including tentative detections) spectra. The velocity resolution
is $\sim$ 32 \kms\ for most spectra and $\sim$  48 \kms\ for some cases where a lower resolution was required
to clearly see the line. The red line segment shows the zero-level line width of the 
CO line adopted for the determination of the velocity integrated intensity. 
An asterisk next to the name indicates a tentative detection.
}

\label{fig:spectra_co21}
\end{figure*}

\end{document}